\documentclass{JINST}
\usepackage{url}
\usepackage{lineno}
\usepackage{subfigure}
\usepackage{floatpag}
\floatpagestyle{empty}

\title{Validation of \textsc{Geant4} Monte Carlo Models with a Highly Granular Scintillator-Steel Hadron Calorimeter}

\author{\centering 
\LARGE\bf The CALICE Collaboration
}

\author{\centering
C.\,Adloff, 
J.\,Blaha, 
J.-J.\,Blaising, 
C.\,Drancourt,
A.\,Espargili\`{e}re, 
R.\,Gaglione, 
N.\,Geffroy, 
Y.\,Karyotakis, 
J.\,Prast,
G.\,Vouters
\\ \it
Laboratoire d'Annecy-le-Vieux de Physique des Particules, Universit\'{e} de Savoie,
CNRS/IN2P3,
9 Chemin de Bellevue BP110, F-74941 Annecy-le-Vieux CEDEX, France
}

\author{\centering
K.\,Francis,
J.\,Repond, 
J.\,Schlereth, 
J.\,Smith\footnote{Also at University of Texas, Arlington},
L.\,Xia 
\\ \it
Argonne National Laboratory,
9700 S.\ Cass Avenue,
Argonne, IL 60439-4815,
USA}

\author{\centering
E.\,Baldolemar, 
J.\,Li\footnote{Deceased}, 
S.\,T.\,Park, 
M.\,Sosebee, 
A.\,P.\,White, 
J.\,Yu
\\ \it
Department of Physics, SH108, University of Texas, Arlington, TX 76019, USA
}

\author{\centering
T.\,Buanes, G.\,Eigen
\\ \it
University of Bergen, Institute, of Physics, Allegaten 55, N-5007 Bergen, Norway
}

\author{\centering
Y.\,Mikami, 
N.\,K.\,Watson 
\\ \it
University of Birmingham,
School of Physics and Astronomy,
Edgbaston, Birmingham B15 2TT, UK
}
\author{\centering 
G.\,Mavromanolakis\footnote{Now at CERN}, 
M.\,A.\,Thomson, 
D.\,R.\,Ward, 
W.\,Yan\footnote{Now at Department of Modern Physics, University of Science and Technology of China, 96 Jinzhai Road, Hefei, Anhui, 230026, P.\, R.\, China}
\\ \it
University of Cambridge, Cavendish Laboratory, J J Thomson Avenue, CB3 0HE, UK
}

\author{\centering 
D.\,Benchekroun, 
A.\,Hoummada, 
Y.\,Khoulaki
\\ \it
Universit\'{e} Hassan II A\"{\i}n Chock, Facult\'{e} des sciences, B.P. 5366 Maarif, Casablanca, Morocco
}

\author{\centering 
J.\,Apostolakis, 
A.\,Dotti, 
G.\,Folger, 
V.\,Ivantchenko, 
V.\,Uzhinskiy
\\ \it 
CERN, CH-1211 Gen\`{e}ve 23, Switzerland
}

\author{\centering
M.\,Benyamna, 
C.\,C\^{a}rloganu, 
F.\,Fehr, 
P.\,Gay, 
S.\,Manen, 
L.\,Royer
\\ \it
Clermont Universit\'e, Universit\'e Blaise Pascal, CNRS/IN2P3, LPC, BP
10448, F-63000 Clermont-Ferrand, France
}

\author{\centering
G.\,C.\,Blazey,
A.\,Dyshkant, 
J.\,G.\,R.\,Lima, 
V.\,Zutshi
\\ \it
NICADD, Northern  Illinois University,
Department of Physics,
DeKalb, IL 60115,
USA
}

\author{\centering 
J.\,-Y.\,Hostachy, 
L.\,Morin
\\ \it
Laboratoire de Physique Subatomique et de Cosmologie, Universit\'{e} Joseph Fourier Grenoble 1,
CNRS/IN2P3, Institut Polytechnique de Grenoble,
53, rue des Martyrs,
F-38026 Grenoble CEDEX, France
}

\author{\centering 
U.\,Cornett, 
D.\,David, 
G.\,Falley, 
K.\,Gadow, 
P.\,G\"{o}ttlicher, 
C.\,G\"{u}nter,
B.\,Hermberg, 
S.\,Karstensen, 
F.\,Krivan,
A.\,-I.\,Lucaci-Timoce\footnotemark[3], 
S.\,Lu, 
B.\,Lutz, 
S.\,Morozov, 
V.\,Morgunov\footnote{Also at ITEP}, 
M.\,Reinecke, 
F.\,Sefkow, 
P.\,Smirnov,
M.\,Terwort,
A.\,Vargas-Trevino 
\\ \it
DESY, Notkestrasse 85,
D-22603 Hamburg, Germany
}

\author{\centering  
N.\,Feege, 
E.\,Garutti$^\spadesuit$, 
I.\,Marchesini\footnote{Also at DESY}, 
M.\,Ramilli
\\ \it
University of Hamburg,
Physics Department,
Institut f\"ur Experimentalphysik,
Luruper Chaussee 149,
D-22761 Hamburg, Germany
}

\author{\centering 
P.\,Eckert,
T.\,Harion, 
A.\,Kaplan,
 H.\,-Ch.\,Schultz-Coulon,
 W.\,Shen,
 R.\,Stamen
\\ \it
 University of Heidelberg, Fakult\"at fur Physik und Astronomie,
Albert Uberle Str. 3-5, 2.OG Ost,
D-69120 Heidelberg, Germany
}

\author{\centering 
B.\,Bilki, E.\,Norbeck, 
Y.\,Onel
\\ \it
University of Iowa, Department of Physics and Astronomy,
203 Van Allen Hall, Iowa City, IA 52242-1479, USA
}

\author{\centering 
G.\,W.\,Wilson
\\ \it
University of Kansas, Department of Physics and Astronomy,
Malott Hall, 1251 Wescoe Hall Drive, Lawrence, KS 66045-7582, USA
}

\author{\centering 
K.\,Kawagoe 
\\ \it
Department of Physics, Kyushu University, Fukuoka 812-8581, Japan
}

\author{\centering 
P.\,D.\,Dauncey,  
A.\,-M.\,Magnan
\\ \it
Imperial College London, Blackett Laboratory,
Department of Physics,
Prince Consort Road,
London SW7 2AZ, UK 
}

\author{\centering 
V.\,Bartsch\footnote{Now at University of Sussex, Physics and Astronomy Department, Brighton, Sussex, BN1 9QH, UK}, 
M.\,Wing
\\ \it
Department of Physics and Astronomy, University College London,
Gower Street,
London WC1E 6BT, UK
}

\author{\centering 
F.\,Salvatore\footnotemark[7]
\\ \it
Royal Holloway University of London,
Department of Physics,
Egham, Surrey TW20 0EX, UK
}

\author{\centering 
E.\,Calvo~Alamillo, 
M.-C.\, Fouz, 
J.\,Puerta-Pelayo 
\\ \it
CIEMAT, Centro de Investigaciones Energeticas, Medioambientales y Tecnologicas, Madrid, Spain 
}

\author{\centering 
B.\,Bobchenko, 
M.\,Chadeeva, 
M.\,Danilov\footnote{Also at MEPhI and at Moscow Institute of Physics and Technology}, 
A.\,Epifantsev, 
O.\,Markin,
R.\,Mizuk\footnotemark[8], 
E.\,Novikov, 
V.\,Popov, 
V.\,Rusinov, 
E.\,Tarkovsky
\\ \it
Institute of Theoretical and Experimental Physics, B. Cheremushkinskaya ul. 25,
RU-117218 Moscow, Russia
}

\author{\centering 
N.\,Kirikova,
V.\,Kozlov, 
P.\,Smirnov, 
Y.\,Soloviev 
\\ \it
P.\,N.\, Lebedev Physical Institute,
Russian Academy of Sciences,
117924 GSP-1 Moscow, B-333, Russia
}

\author{\centering 
P.\,Buzhan, A.\,Ilyin, V.\,Kantserov, V.\,Kaplin, A.\,Karakash, E.\,Popova, V.\,Tikhomirov 
\\ \it
Moscow Physical Engineering Institute, MEPhI,
Department of Physics,
31, Kashirskoye shosse,
115409 Moscow, Russia
}

\author{\centering 
C.\,Kiesling,
K.\,Seidel, 
F.\,Simon, 
C.\,Soldner, 
M.\,Szalay, 
M.\,Tesar,
L.\,Weuste
\\ \it
Max Planck Institute f\"ur Physik,
F\"ohringer Ring 6,
D-80805 Munich, Germany
}

\author{\centering 
M.\,S.\,Amjad, 
J.\,Bonis, 
S.\,Callier, 
S.\, Conforti\,di\,Lorenzo, 
P.\,Cornebise, 
Ph.\,Doublet, 
F.\,Dulucq, 
J.\,Fleury, 
T.\,Frisson, 
N.\,van der Kolk,  
H.\,Li\footnote{Now at LPSC Grenoble}, 
G.\,Martin-Chassard, 
F.\,Richard, 
Ch.\,de la Taille, 
R.\,P\"oschl, 
L.\,Raux, 
J.\,Rou\"en\'e, 
N.\,Seguin-Moreau
 \\ \it

Laboratoire de l'Acc\'{e}l\'{e}rateur Lin\'{e}aire, Centre
Scientifique d'Orsay, Universit\'{e} de Paris-Sud XI, CNRS/IN2P3, BP
34, B\^atiment 200, F-91898 Orsay CEDEX, France
}

\author{\centering 
M.\,Anduze,
V.\,Boudry, 
J-C.\,Brient, 
D.\,Jeans, 
P.\,Mora de Freitas, 
G.\,Musat, 
M.\,Reinhard, 
M.\,Ruan,  
H.\,Videau
\\ \it
 Laboratoire Leprince-Ringuet (LLR)  -- \'{E}cole Polytechnique, CNRS/IN2P3, F-91128 Palaiseau, France
}

\author{\centering 
B.\,Bulanek,
J.\,Zacek 
\\ \it
Charles University, Institute of Particle \& Nuclear Physics,
V Holesovickach 2,
CZ-18000 Prague 8, Czech Republic  
}

\author{\centering 
J.\,Cvach, 
P.\,Gallus, 
M.\,Havranek, 
M.\,Janata, 
J.\,Kvasnicka,
D.\,Lednicky,
M.\,Marcisovsky, 
I.\,Polak, 
J.\,Popule, 
L.\,Tomasek, 
M.\,Tomasek, 
P.\,Ruzicka, 
P.\,Sicho, 
J.\,Smolik, 
V.\,Vrba, 
J.\,Zalesak 
\\ \it
Institute of Physics, Academy of Sciences of the Czech Republic, Na Slovance 2,
CZ-18221 Prague 8, Czech Republic
}

\author{\centering 
B.\,Belhorma,
H.\,Ghazlane
\\ \it
Centre National de l'Energie, des Sciences et des Techniques Nucl\'{e}aires, 
B.P. 1382, R.P. 10001, Rabat, Morocco
}

\author{\centering              
T.\,Takeshita,
S.\,Uozumi
\\ \it
Shinshu University,
Department of Physics,
3-1-1 Asaki,
Matsumoto-shi, Nagano 390-861,
Japan \\
}

\author{{\centering 
M.\, G\"otze, 
O.\, Hartbrich, 
J.\,Sauer, 
S.\,Weber,
C.\,Zeitnitz
\\ \it
Bergische Universit\"{a}t Wuppertal,
Fachbereich 8 Physik,
Gaussstrasse 20,
D-42097 Wuppertal, Germany\\
}

\it
$^\spadesuit$ Corresponding author\newline
E-mail: \email{erika.garutti@desy.de}

}

\abstract{Calorimeters with a high granularity are a fundamental requirement of the Particle Flow paradigm. This paper focuses on the prototype of a hadron calorimeter with analog readout, consisting of thirty-eight scintillator layers alternating with steel absorber planes. The scintillator plates are finely segmented into tiles individually read out via Silicon Photomultipliers. The presented results are based on data collected with pion beams in the energy range from 8\,GeV to 100\,GeV. The fine segmentation of the sensitive layers and the high sampling frequency allow for an excellent reconstruction of the spatial development of hadronic showers. A comparison between data and Monte Carlo simulations is presented, concerning both the longitudinal and lateral development of hadronic showers and the global response of the calorimeter. The performance of several \textsc{Geant4} physics lists with respect to these observables is evaluated.}


\begin{document}

\tableofcontents

\clearpage
\section{Introduction}
\label{section:introduction}

The next generation of lepton colliders, such as the International Linear Collider (ILC)~\cite{Brau:2007zza}, will require unprecedented detector performances in order to fully exploit the physics potential. A clean separation between the hadronic decays of the $W$ and $Z$ bosons sets the goal for the jet energy resolution. Ideally, the di-jet mass resolution should be comparable to the decay widths of the weak vector bosons. This requires a jet energy resolution $\sigma_E/E_j$ of the order of 3-4\% at the $Z$ peak, which translates approximately into $\sigma_E/E_j \sim 30\%/\sqrt{E_j/\mathrm{GeV}}$. Within the ILC community, Particle Flow calorimetry has been identified  as the most promising way to achieve this level of precision~\cite{Brient:2002gh,Morgunov:2002pe,Thomson:2009rp},  as successfully tested on Monte Carlo simulations. For instance, the Pandora Particle Flow algorithm, which has been developed in the context of the detector optimization studies for the ILC, has demonstrated the feasibility of jet energy resolutions of $\sigma_E/E_j \lesssim 3.8\%$ for $40 - 400$\,GeV jets~\cite{Thomson:2009rp}.

The key factor in Particle Flow calorimetry is the single particle separability within a jet, which requires a high granularity of the calorimeters. The CALICE collaboration is performing studies of several designs of highly granular calorimeters, with sensitive layers finely segmented into cells, which are individually read out. The stability over a large number of channels of new readout technologies has been demonstrated and reliable procedures of cell equalization and calibration have been established. Furthermore, the high spatial resolution allowed by the fine granularity provides valuable input to the validation of the shower models used for Monte Carlo simulations. Such studies are of broader interest, which goes beyond the support of simulation studies of Particle Flow algorithms. Uncertainties in the shower modeling still contribute significantly to energy scale uncertainties in high energy physics experiments, for instance at the Large Hadron Collider detectors~\cite{Aad:2012vm}.

Since 2006 several beam tests of the CALICE prototypes have been conducted at the Deutsches Elektronen-Synchrotron (DESY), at the European Organization for Nuclear Research (CERN) and at the Fermi National Accelerator Laboratory (FNAL). This paper focuses on data collected at CERN in 2007, when the experimental set-up consisted of a silicon-tungsten electromagnetic sampling calorimeter (SiW-ECAL)~\cite{Anduze:2008hq}, a scintillator-steel hadron sampling calorimeter with analog readout (AHCAL)~\cite{collaboration:2010hb} and a scintillator-steel tail catcher and muon tracker (TCMT)~\cite{:2012kx}.

The analyzed data were collected when the prototypes were exposed to pion beams in the energy range from 8\,GeV to 100\,GeV, provided by the CERN SPS H6 beam line. The overall response of the calorimeter and the lateral and longitudinal profiles of hadronic showers are compared to the predictions made by several Monte Carlo simulations, which make use of different \textsc{Geant4} physics lists (Sec.~\ref{section:simulations}).

The study is completed by an update on the validation of the calibration of the AHCAL performed with electromagnetic showers, already described in a previous publication~\cite{collaboration:2010rq}. This study uses data collected both at FNAL and CERN, exposing the prototypes to electron and positron beams in the energy range from 1\,GeV to 50\,GeV.

\section{Experimental Setup}
\label{section:experimentalsetup}

During the beam tests at CERN the detectors were exposed to muon, positron and pion beams. The SiW-ECAL and the AHCAL were mounted on a movable stage, providing the possibility to translate and rotate the calorimeters with respect to the beam. However, the studies discussed in the following only make use of data collected with the beam incident in the center of the calorimeters, along their center line.

\begin{figure}
\centering
\includegraphics[width=0.95\textwidth]{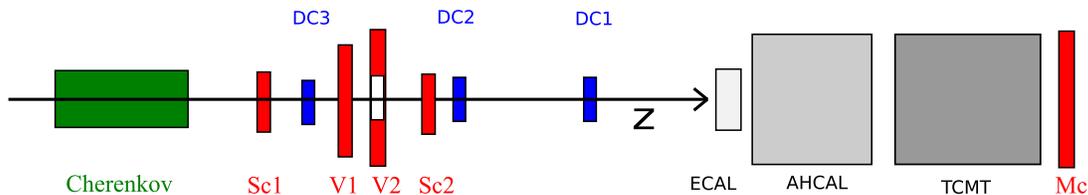}
\caption[Experimental setup]{\sl Schematic layout (not to scale) of the CALICE experimental setup at CERN, with calorimeters and beam instrumentation. Sc1 and Sc2 are the scintillators for the beam trigger, V1 is the multiplicity counter, V2 is the veto trigger and Mc is the scintillator for the muon trigger.}
\label{figure:calicesetup}
\end{figure}

A schematic overview of the experimental setup is shown in Fig.~\ref{figure:calicesetup}. Three sets of wire chambers were operated upstream of the detectors in order to measure the beam coordinates. The coincidence signal of two upstream $10 \times 10$\,cm$^2$ scintillator counters (Sc1 and Sc2 in Fig.~\ref{figure:calicesetup})  triggered the readout of the detector (\textit{beam trigger}). One scintillator (V1 in Fig.~\ref{figure:calicesetup}) with an area of $20 \times 20$\,cm$^2$ and analog readout tagged multi-particle events (\textit{multiplicity counter}). A $100 \times 100$\,cm$^2$ scintillator with a $20 \times 20$\,cm$^2$ hole in the center (V2 in Fig.~\ref{figure:calicesetup}) rejected beam halo events (\textit{veto trigger)}. Muons were identified by a downstream $100 \times 100$\,cm$^2$ scintillator (Mc in Fig.~\ref{figure:calicesetup}). A Cherenkov counter was also operated in threshold mode to discriminate between electrons and negative pions or between positive pions and protons.

\subsection{The SiW-ECAL}

The SiW-ECAL is divided into three stacks, each composed of 10 modules of alternating tungsten and silicon layers. Each stack has tungsten layers with different thicknesses: 1.4\,mm (0.4 radiation lengths $X_{0}$) per layer in the first stack, 2.8\,mm or 0.8\,$X_{0}$ in the second and 4.2\,mm or 1.2\,$X_{0}$ in the third one. The silicon layers are segmented into PIN diodes of 1\,$\times$\,1\,cm$^2$, for a total number of about 9700 read-out cells.

The overall thickness of the prototype is about 20\,cm, corresponding to 24.6\,$X_{0}$ or 0.9 nuclear interaction lengths $\lambda_{\rm I}$. The lateral size of the SiW-ECAL is $18 \times 18$\,cm$^2$.

\subsection{The AHCAL}

\begin{figure}[]
\centerline{
\includegraphics[height=0.30\textheight,width=0.45\columnwidth]{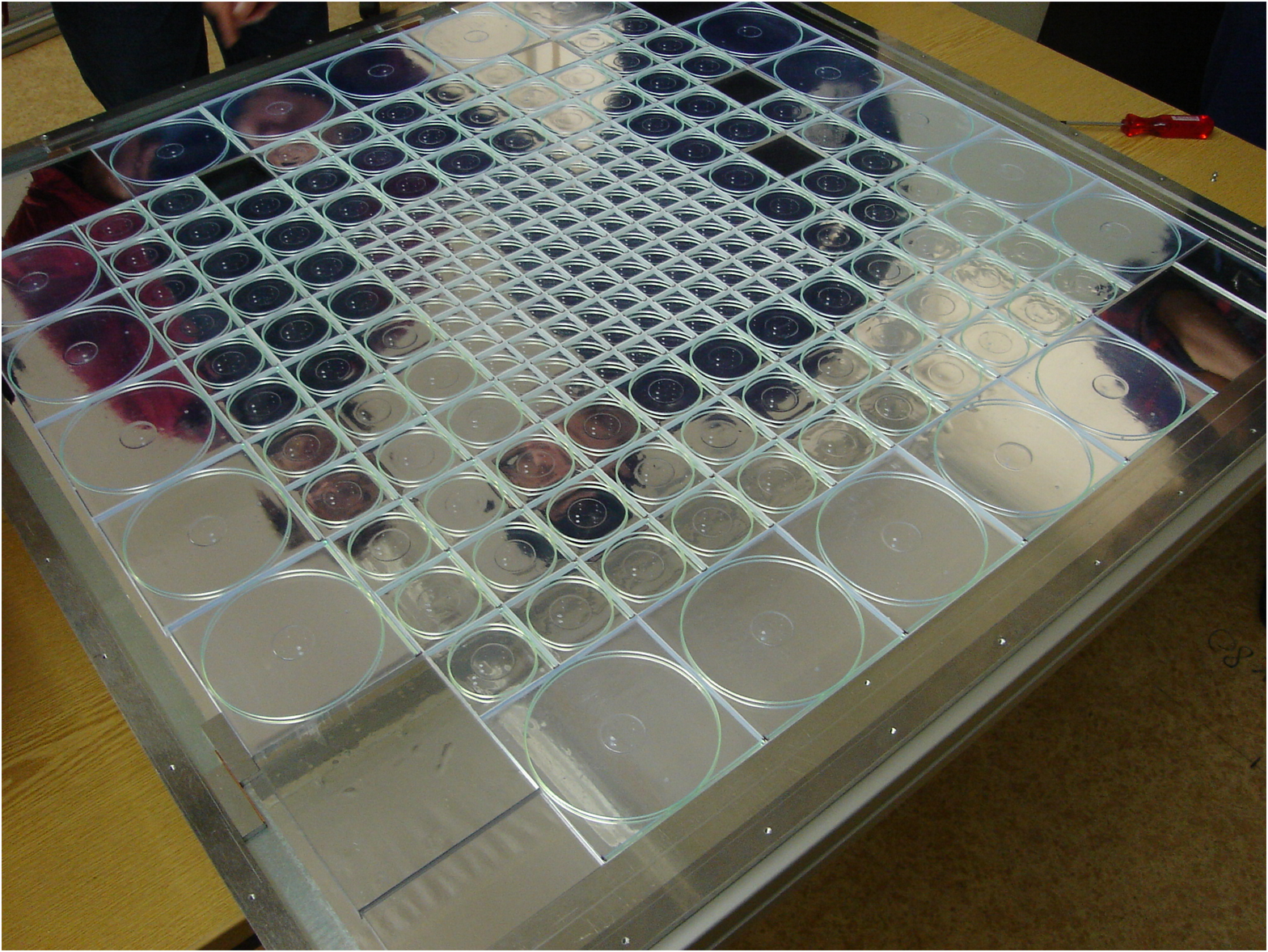}
\includegraphics[height=0.30\textheight,width=0.45\columnwidth]{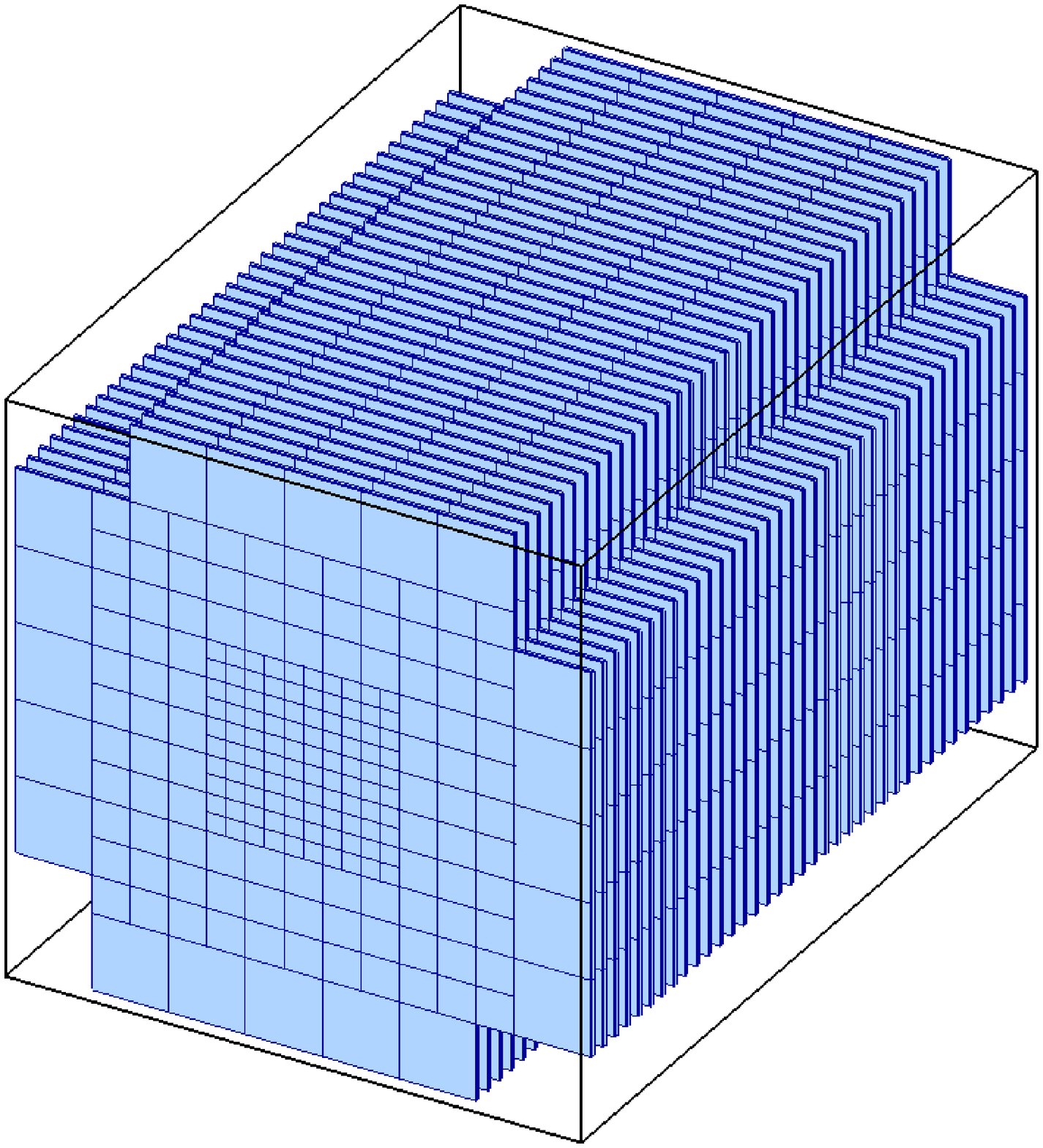}
}
\caption[Segmentation of an AHCAL sensitive layer and AHCAL steel stack support structure]{\sl Left: Segmentation of a sensitive layer of the AHCAL. The size of the scintillator tiles increases towards the outer region. Right: schematic view of the active layers of the AHCAL.}
\label{figure:segmentation}
\end{figure}

The AHCAL is a sampling structure of 38 modules, each consisting of a $\sim$2\,cm thick steel absorber plate and a sensitive layer instrumented with 0.5\,cm thick scintillator tiles. The total depth of the prototype is 1.2\,m, translating into about 5.3\,$\lambda_{\rm I}$, while the lateral dimensions are approximately $1 \times 1$\,m$^2$. Each sensitive layer is composed of scintillator tiles of different sizes (Fig.~\ref{figure:segmentation}, left). The 30\,$\times$\,30\,cm$^2$ core has a granularity of 3\,$\times$\,3\,cm$^2$, while the outer region is equipped with tiles of increasing sizes (6\,$\times$\,6\,cm$^2$ and 12\,$\times$\,12\,cm$^2$). In the last 8 layers the highly granular core is replaced by $6 \times 6$\,cm$^2$ tiles. The scintillation light from each tile is read out individually by a Silicon PhotoMultiplier (SiPM)~\cite{Bondarenko:2000in,Buzhan:2003ur}, coupled to the scintillator via a WaveLength Shifting fiber (WLS). The SiPMs employed in the AHCAL have a photosensitive surface of 1.1\,mm$^2$, which is divided into 1156 pixels. These pixels are individually equipped with a quenching resistor and mounted on a common substrate, in such a way that the charge signal is proportional to the number of pixels fired.

The readout system of the SiPMs is operated in two different modes, called \textit{calibration} and \textit{physics} modes. For calibration purposes single photons need to be resolved in the SiPM spectrum, since the distance between two consecutive peaks in the spectrum determines the gain of the SiPM. Therefore, a short shaping time of the signal (40\,ns) and a high amplification are needed. In contrast, during physics runs large signals are produced and the amplification needs to be reduced by approximately a factor 10 to minimize saturation effects. Furthermore, a longer shaping time of about 180\,ns is used to provide sufficient latency for the beam trigger decision.

The response of the SiPMs is measured both in physics and calibration modes, in order to determine the gain and saturation level of the SiPMs and the electronics intercalibration between the two operation modes. The response is monitored using LEDs with tunable intensities.

Temperature sensors placed inside the detector in correspondence to each module monitor temperature changes with an accuracy better than 0.6\,$^\circ$C .

\subsection{The TCMT}

The TCMT is positioned downstream with respect to the AHCAL in order to absorb the tails of the showers leaking out of the AHCAL. The TCMT has a lateral size of 109\,$\times$\,109\,cm$^2$ and is 142\,cm in depth, corresponding to 5.8\,$\lambda_{\rm I}$. It consists of two sections, a {\it fine} one and a {\it coarse} one. Each section has 8 sensitive layers, alternating with steel absorbers. The absorber plates are 2\,cm thick in the fine section and 10\,cm thick in the coarse section. The sensitive layers are 0.5\,cm thick and are segmented into 5\,cm wide and 1\,m long scintillator strips, with alternating horizontal and vertical orientation in adjacent layers. The scintillation light is collected by WLS fibers and detected by SiPMs, as for the AHCAL prototype.

\section{Calibration}
\label{section:calibration}

As typical of semiconductor devices, SiPMs are sensitive to temperature and bias voltage changes that affect most of their performance parameters~\cite{collaboration:2010hb}. Moreover, SiPMs show saturation effects, due to the finite number of pixels and the finite pixel recovery time. One of the main technical aims of the AHCAL prototype is to show that these effects can be handled over a large number of channels.

The calibration chain is described in~\cite{collaboration:2010rq}. It proceeds through the following steps:

\begin{itemize}

\item inter-cell equalization of the response;

\item calibration of the SiPM signal and correction for the non-linear response;

\item conversion of the calibrated signal to the GeV scale.
\end{itemize}
The equalization of the response of the 7608 AHCAL cells is performed using the reference signal from muons~\cite{D'Ascenzo:2009zz}. Muons represent the best approximation of the behavior of a Minimum Ionizing Particle (MIP). The signal measured in a cell expressed in ADC counts can be converted to MIPs, using as unit the Most Probable Value (MPV) of the response to the muon beam.

In order to calibrate the SiPM signals, the gain is measured periodically during data taking using the LED system. A correction is also applied to account for the non-linear response of the SiPMs, based on the specific response curve of each SiPM, measured during dedicated laboratory tests. Since the SiPM properties depend on the temperature and the voltage, a given set of calibration constants is only valid for measurements at the same operation conditions and needs to be extrapolated to the operating conditions to account for possible changes, such as temperature fluctuations~\cite{Feege:2008zz}.

Finally, the conversion of the signal to the GeV scale is derived from the response of the detector to electromagnetic showers (Sec.~\ref{section:calibratiovalidation}).

\section{Monte Carlo Simulations}
\label{section:simulations}

The Monte Carlo simulations for the CALICE prototypes are carried out in the framework of \textsc{Geant4}~\cite{Simone:1995sz}. The version 9.4 patch 3 of \textsc{Geant4} is used as default in the following, apart from explicit comparisons with previous versions. The geometries of the calorimeters are simulated within \textsc{Geant4} using Mokka~\cite{mokka-note,mokka-www}. The detectors upstream of the calorimeters are simulated as well. The origin of the simulated particles is located upstream of the full system, such that any interactions with the Cherenkov counter, the scintillator-triggers, the tracking chambers and the air volumes in between are taken into account. Sensitive and passive materials, gaps and support structures are also considered in detail. The simulated AHCAL active layers have a uniform granularity of 1~$\times$~1\,cm$^2$ cell size. The realistic geometry of the AHCAL is obtained during the digitization procedure, as described below.

\subsection{Digitization}

The simulated events are processed further in the so-called digitization step, which models the response of the detector and its electronics. This allows the same treatment for data and Monte Carlo in the subsequent analysis steps, such as calibration and reconstruction. This procedure takes into account several factors:

\begin{itemize}
\item the detector granularity. The signal amplitude of the 1\,$\times$\,1\,cm$^2$ virtual cells simulated is summed up to obtain the real geometry with 3\,$\times$\,3, 6\,$\times$\,6 and 12\,$\times$\,12\,cm$^2$ cells.

\item the light sharing between neighboring tiles (known also as \textit{cross-talk}). A 2.5\% light sharing from each tile edge is used in simulations~\cite{collaboration:2010rq}.

\item non-linearity effects of the SiPMs, based on their specific saturation curves.

\item Poissonian fluctuations of the photo-electron statistics.

\item the noise contribution. The noise is overlaid on the simulation, using real noise measured in dedicated triggers without beam.
\end{itemize}

Shielding effects in the scintillator material saturate the scintillation process at high ionization densities. This causes a non-linearity of the light yield, which has been simulated in Monte Carlo according to the Birks Law~\cite{Birks:1964zz}.

The read out electronics have a defined time window. Late energy depositions, for instance from neutrons, might escape this time window.  In order to reproduce this effect, a time cut of 150\,ns is applied in simulations.

The effects of gaps between the calorimeter tiles, as well as the non-uniform response of the tiles have been studied in Monte Carlo events in ~\cite{Sefkow:2010rt}, showing that these type of effects do not have a significant influence on the measurement of hadron showers.

\subsection{Physics Lists}

The interactions of hadrons with matter cannot be modeled from first principles alone; several phenomenological models, working with different approximations, exist. An overview of the different models is given in~\cite{eudet-memo}. The main features are summarized here:

\begin{itemize}

\item {\bf String Parton Models}. These models are employed to simulate the interaction of medium or high energy hadrons with nuclei. As a first stage the interaction of the particles with at least one nucleon of the nucleus is modeled using a string excitation model. Two different approaches are available in \textsc{Geant4}: the Fritiof ({\bf FTF}) and the Quark Gluon String ({\bf QGS}) model. In the FTF approach the diffractive scattering of the primary particle with the nucleons is realized only via momentum exchange. In the QGS model the hadron-nucleon interaction is mediated via pomerons. The products of an interaction between the primary particle and the nucleus are one or several excited strings and a nucleus in an excited state. The fragmentation of the excited strings into hadrons is handled by a longitudinal string fragmentation model, with differences between FTF and QGS. The interaction of secondaries with the excited nucleus is handled by a shower model or by a precompound model (see below). In the latter case the string parton models get the suffix ``P'' (FTFP and QGSP). The de-excitation of the excited nucleus is further simulated by nuclear fragmentation, precompound and nuclear de-excitation models.

\item {\bf Parameterized Models}. Low Energy Parametrized ({\bf LEP}) and High Energy Parametrized ({\bf HEP}) models are based on fits to experimental data, with little theoretical guidance. They do not conserve energy, momentum, charge or other quantum numbers on an event-by-event basis, but instead seek to conserve these quantities on average. The approach of \textsc{Geant4} is to limit the use of these models and replace them by more sophisticated ones where possible.

\item {\bf Cascade Models}. The Bertini cascade {\bf BERT} and the binary cascade {\bf BIC} models are employed at medium and low energies, where the quark-substructure of individual nuclei can be neglected. The BERT model describes a nucleus as a sphere with constant nucleon density. Incident hadrons collide with protons and neutrons in the target nucleus and produce secondaries which in turn collide with other nucleons generating a so-called intra-nuclear cascade. At the end of the cascade the excited nucleus is represented as a sum of particle-hole states which is then decayed by pre-equilibrium, nucleus explosion, fission and evaporation models. The BIC model considers the nucleus as a sum of discrete nucleons, at defined positions and with defined momenta. The propagation through the nucleus of the incident hadron and the secondaries produced are modeled by a cascading series of two-particle collisions. Secondaries are created during the decay of resonances formed during the collisions. The decay of the excited nucleus is handled by the precompound model.

\item {\bf Precompound Model}. The precompound model generates the final state for hadron inelastic scattering. It describes the emission of protons, neutrons, and light ions before the equilibrium of a nuclear system is reached. The final products are passed to de-excitation models.

\item {\bf \texttt{CHIPS} Model}. The Chiral Invariant Phase Space (\verb=CHIPS=) model is a nuclear fragmentation model acting at the quark level. The model is based on the concept of a quasmon, which is an excited intermediate state of massless quarks that are asymptotically free, formed from the quarks of the projectile hadron and of the target nucleon. The quasmons decay via internal quark fusion or by double quark exchange with neighboring quasmons. This model is still at an experimental phase and several parameters are still being optimized.

\end{itemize}

\begin{figure}
\centering
\includegraphics[width=0.95\textwidth]{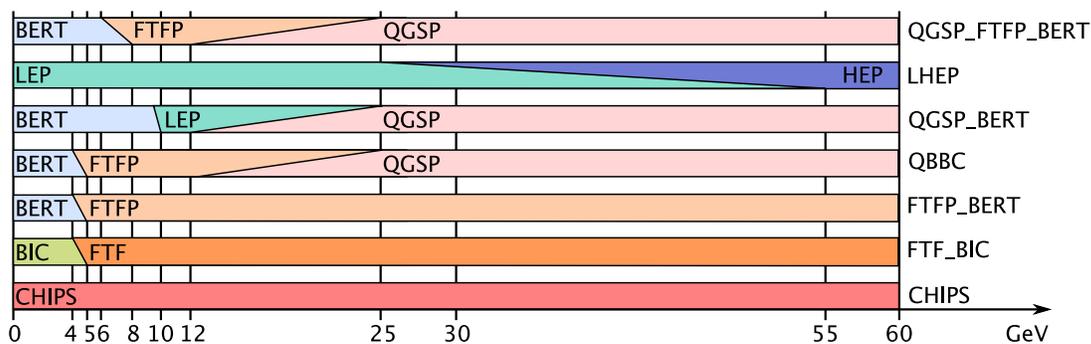}
\caption[Physics lists]{\sl Models applied at different energies in several physics lists for the simulation of pions in \textsc{Geant4}, version 9.4 patch3.}
\label{figure:physicslists}
\end{figure}

Several ``physics lists'' are available in \textsc{Geant4}, which combine different models in different energy ranges, with a random choice of which model is used for overlapping energy regions. A summary of the physics lists considered in this paper and of their composition at different energies is given in Fig.~\ref{figure:physicslists}.

\section{Event Selection}
\label{section:eventselection}

This paper presents electron (or positron\footnotemark\footnotetext{In the following, by ``electron beam'' we refer to either an electron or positron beam, since no distinction between the two beam particles is relevant for the performed studies.}) and negative pion data collected with the experimental set-up described in Sec.~\ref{section:experimentalsetup}. While performing measurements with an electron beam, the SiW-ECAL is moved out of the beam line. Though the beam is highly enriched with the nominal particle, some contamination is present and the recorded event samples need to be purified for analysis. The selection applied is summarized in the following and has been tested on Monte Carlo simulations, comparing the selection efficiencies for simulated muons, electrons and pions. Some of the selection criteria use information from the AHCAL itself. Applying the same requirements to simulated events confirms that no bias is introduced in the energy distributions. More details on the event selection are given in~\cite{feegethesis}.

Beam events are selected requiring the beam trigger. Events where particles generate a beam trigger signal but fail to reach the calorimeters are rejected by requiring a minimum energy deposition of 4\,MIPs in the $3 \times 3$\,cm$^2$ cells of the first five layers of the AHCAL. Signals of less than 0.5\,MIPs in single cells are rejected in all calorimeters to reduce the noise contribution. This requirement has been tested on events collected without incoming beam, since they have a topology similar to that given by spurious trigger events (i.e. only noise). About 96\% of the beam-off events are rejected. Some events contain additional particles in the beam halo or particles that initiate a shower before reaching the AHCAL. These events are excluded by requiring no signal from the veto trigger and less than 15 hits in the $6 \times 6$\,cm$^2$ cells of the first five layers of the AHCAL. In order to exclude events with more than one particle depositing energy in the AHCAL at the same time, only events with a multiplicity counter amplitude of less than 1.4\,MIPs are kept. This selection yields a multi-particle contamination of less than 0.1\%, which has been evaluated using the estimated distribution of the signal from double particle events in the multiplicity counter. Such a distribution has been simulated from the known distribution of noise, measured during triggers without beam, and the signal from single particle events. The signal from single particles has been obtained fitting the distribution of the measured amplitude in the multiplicity counter in a low amplitude region, where no significant contribution from double particle events is expected.

In order to improve the purity of the electron data, a signal is required in the Cherenkov detector. In addition, the center of gravity of the energy deposits in the beam direction is required to be in the front part of the AHCAL (less than 360\,mm beyond the front face of the AHCAL). At least one cluster of neighboring hits with a total energy of 18\,MIPs or higher has to be found in the AHCAL and an energy of less than 5\,MIPs has to be deposited in the last 10 layers of the AHCAL.

For minimizing the electron contamination in pion data, events with a signal in the Cherenkov detector are rejected. The contamination from muons is suppressed by requiring the identification in the AHCAL of the position where the pion shower is initiated by the first hard interaction, since muons generally do not produce showers. The position of the first hard interaction is identified using a cluster-based algorithm, based on the three-dimensional distribution of hits~\cite{Lutz:2010zz}. Seed hits with visible energies of more than 1.65\,MIPs are sorted by their $z$-positions in ascending order. Starting with the first seed hit according to this ordering, each seed hit and all neighboring hits are assigned to a cluster. As long as one or more of the cells added to a cluster meet the seed hit requirement, the clustering continues and all hits adjacent to these cells are assigned to the same cluster. The cluster closest to the point the pion enters the calorimeter consisting of at least 4 hits and having a minimum energy of 16\,MIPs is identified as the beginning of the shower. The end of the cluster axis pointing in the direction of the incoming hadron is used as location of the first inelastic scattering. According to Monte Carlo simulations this procedure allows rejection of muons with an efficiency of more than 80\%. Since muons can produce shower-like clusters along the track due to Bremsstrahlung, the muon contamination is further reduced by requiring additionally more than 60 hits in the AHCAL, yielding a muon rejection efficiency of more than 95\%.

In order to minimize the systematic uncertainties arising from combining the information from different detectors, pion events for which the shower starting point is found in the SiW-ECAL are rejected. The algorithm used to identify the beginning of the hadronic shower in the SiW-ECAL is described in~\cite{:2011ha}. The beginning of the shower is found by imposing thresholds on the energy measured in the different layers and on the number of hits deposited.  This requirement further minimizes the contamination from electrons, which start to shower in the SiW-ECAL.

\section{Calibration Validation}
\label{section:calibratiovalidation}

Although the AHCAL is designed to measure hadrons, during beam tests it has also been exposed to electron beams. The study of the electromagnetic response serves to prove the understanding of the detector and to validate the calibration procedure, since electromagnetic showers can be precisely modeled. Additionally, the AHCAL is a non-compensating calorimeter and the energy response to hadrons may be non-linear. In contrast, apart from saturation effects, the response to electromagnetic showers is linear and allows for the determination of the conversion of the calibrated signal to the energy scale in units of GeV (referred to as electromagnetic energy scale).

The study of the electromagnetic response of the AHCAL is described in detail in~\cite{collaboration:2010rq}. Electron data of energies between 10\,GeV and 50\,GeV are compared to Monte Carlo simulations and the systematic uncertainties associated to the different steps of the calibration procedure are estimated.

Overall, the studies in~\cite{collaboration:2010rq} validate the simulation of electromagnetic showers at a level sufficient for the present analysis of the hadronic response. A residual disagreement of the order of 2\,mm is found between data and simulation, when comparing the radial development.

In order to confirm that the results in~\cite{collaboration:2010rq} are up-to-date with the most recent calibration and reconstruction software tools and to extend them over a broader energy range, the measurement of the energy response to electromagnetic showers is repeated here. This new study focuses on the determination of the electromagnetic energy scale, since it is applied in the following to pion data. The validation of the simulation of electromagnetic showers for the observables discussed in~\cite{collaboration:2010rq} is not repeated. The data set has been extended with respect to~\cite{collaboration:2010rq} to lower energies down to 1\,GeV, using data collected at FNAL exposing the AHCAL to electron beams, with an experimental set-up similar to the one employed at CERN (Sec.~\ref{section:experimentalsetup}).

\begin{figure}
\centerline{
\includegraphics[width=0.5\columnwidth]{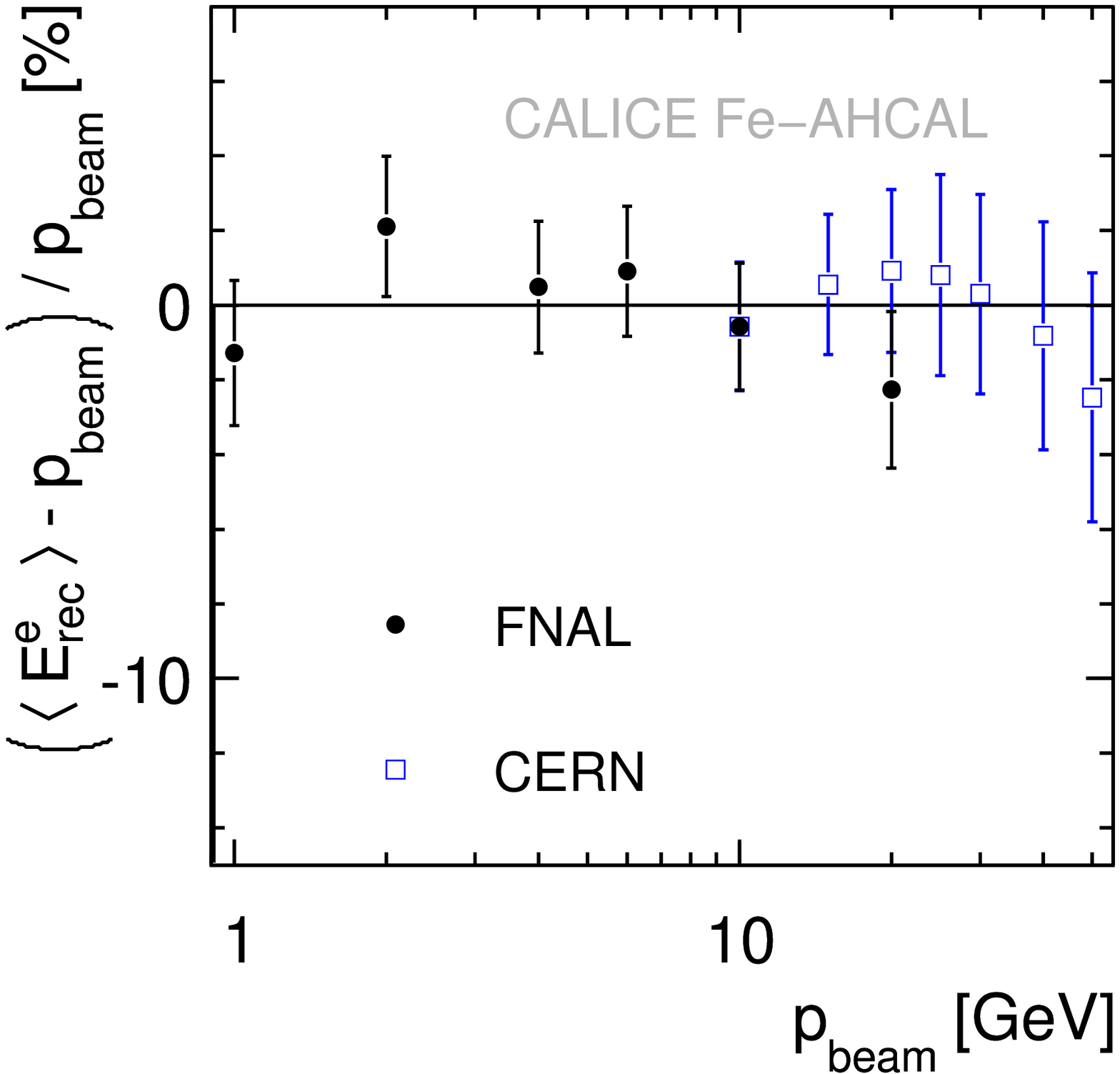}
\includegraphics[width=0.5\columnwidth]{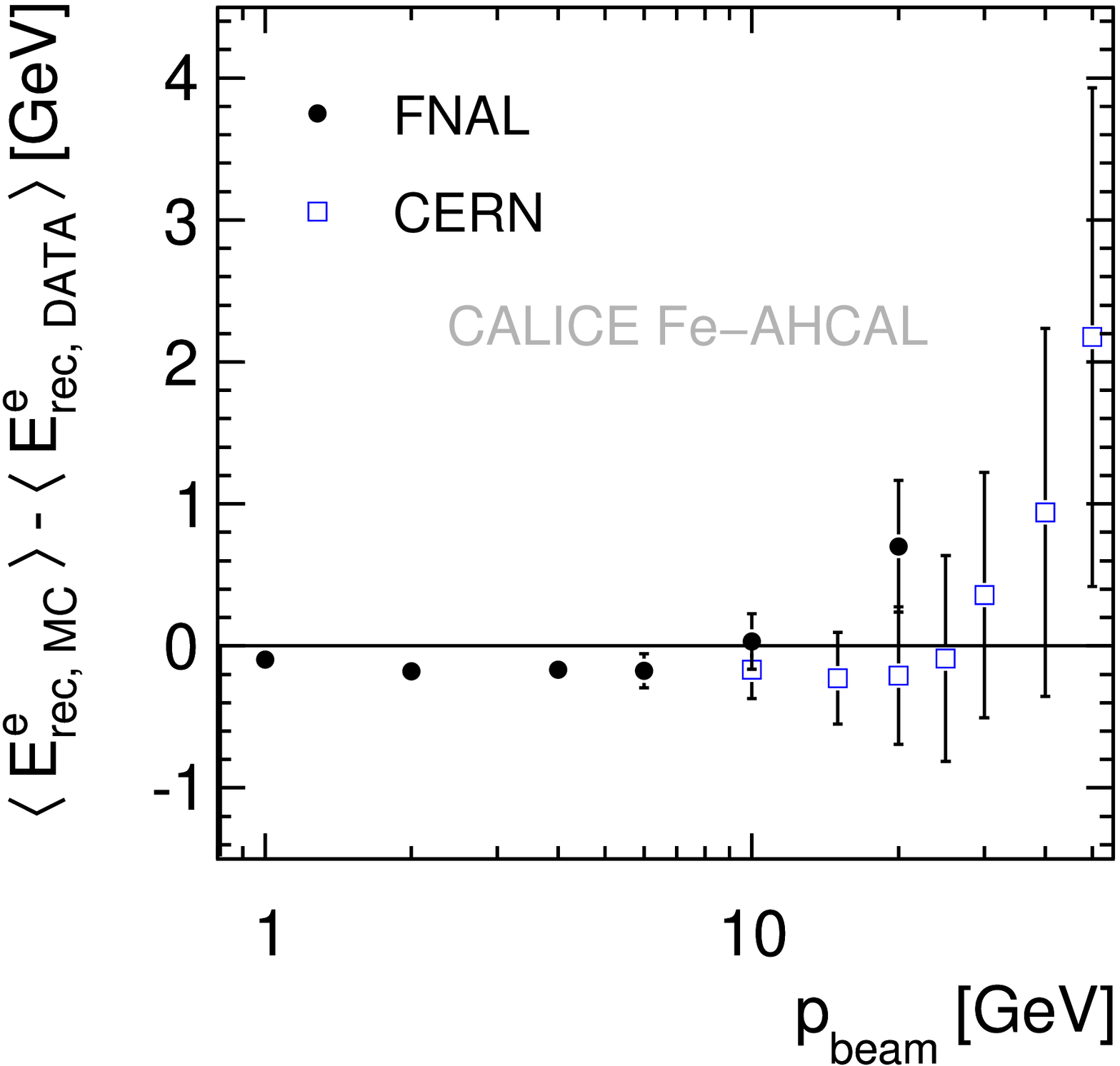}
}
\caption[Electromagnetic validation]{\sl Left: Residuals from linearity for electromagnetic data collected at FNAL (circles) and CERN (squares). Right: Difference between the simulated and the measured electromagnetic response. The error bars take into account both the statistical and the systematic uncertainties, as described in the text.}
\label{figure:electromagneticvalidation}
\end{figure}

The expected linear response to electrons is described by the following equation:
\begin{equation}
\langle E_{\mathrm{rec}}^{\mathrm{e}} \rangle = p_{\mathrm{beam}} \cdot u + v ,
\label{equation:electrom}
\end{equation}
where $\langle E_{\mathrm{rec}}^{\mathrm{e}} \rangle$ is the mean calibrated response to electrons expressed in MIPs, $p_{\mathrm{beam}}$ is the beam momentum, and the parameters $u$ and $v$ are obtained from a linear fit to the dependence of $\langle E_{\mathrm{rec}}^{\mathrm{e}} \rangle$ on $p_{\mathrm{beam}}$. The factor $u$ represents the calibration to the GeV scale of the electromagnetic response. The offset $v$ accounts for the combined effect of the 0.5\,MIPs threshold applied to all events (Sec.~\ref{section:eventselection}) and of the residual noise above the threshold. The residuals to the fit are shown in Fig.~\ref{figure:electromagneticvalidation} (left). The linearity is better than about 2\% at all energies. The errors taken into account in the fit and shown in the figure are both the statistical and the systematic uncertainties. The systematic uncertainty includes a calibration scale uncertainty of 1.6\% and the SiPMs saturation uncertainty, which ranges between 0.5\% at 1\,GeV and 3\% at 50\,GeV.

The same procedure is applied to Monte Carlo, yielding the results summarized in Tab.~\ref{table:electromagneticvalidation}, which are compatible with the results obtained previously in~\cite{collaboration:2010rq}. The factor $u$ agrees with the results obtained for data within the fit uncertainties, while the offset $v$ disagrees by about 5 standard deviations (5\,MIPs or $\sim$100\,MeV). This remaining discrepancy is attributed to an imperfect implementation of noise, light-sharing and threshold effects. An imperfect modeling of the beam line and of the energy lost in the upstream material could also contribute. Since in electromagnetic showers the signal cells below threshold concentrate on the edges, an inaccurate simulation of the lateral extension of electromagnetic showers, as observed in~\cite{collaboration:2010rq}, could also contribute to this effect. 

The difference between the simulated response and the measured response is shown in Fig.~\ref{figure:electromagneticvalidation} (right). The discrepancy in $v$ is visible at low energies, while at high energies the Monte Carlo slightly underestimates the SiPMs saturation.

\begin{table}
 \centering
 \begin{tabular}{|c|c|c|}
	\hline
	& $u$ [MIP$/$GeV] & $v$ [MIP] \\
	\hline
	Data & 42.4 $\pm$ 0.3 & -1.1 $\pm$ 0.9 \\
	\hline
	MC & 42.8 $\pm$ 0.1 & -6.6 $\pm$ 0.5 \\
	\hline
\end{tabular}
 \caption[Electromagnetic response]{\sl Measured parameters of the linear fit to the electromagnetic response, as defined in Eq.~\ref{equation:electrom}, for data and Monte Carlo. The errors reported are the fit uncertainties.}
 \label{table:electromagneticvalidation}
\end{table}

The electromagnetic analysis allows the conclusion that the detector calibration and simulation are sufficiently understood, in order to carry on studies based on hadronic data. The calibration to the GeV scale obtained for electromagnetic data is applied in the following to pion data and to the digitized Monte Carlo.

\subsection{Systematic Uncertainties}
\label{section:systematics}

The results shown in the following focus on the response of the calorimeter to hadrons. The systematic uncertainty on the calibration scale of 1.6\%, quoted above for the electromagnetic study, affects the hadronic analysis in the same way. This uncertainty is relevant for the study of the hadronic energy response discussed in Sec.~\ref{section:energyresponse}. The saturation uncertainty, which ranges between 0.5\% at 1\,GeV and 3\% at 50\,GeV for elecromagnetic showers, is expected to have a reduced impact on hadronic showers, due to the lower energy density. This is discussed in~\cite{collaboration:2010rq}, where it is shown that electromagnetic showers have more hits with high energy deposition that pion showers, even when the beam energy is only half that of the pion. Therefore, this source of uncertainty is neglected during the measurement of the hadronic energy response.

The measurement of the pion interaction length is dominated by the fit uncertainty, as discussed in Sec.~\ref{section:hadinterlength}.

The individual response of each layer is relevant for the longitudinal shower profiles discussed in Sec.~\ref{section:longitudinal}. For this study a dedicated evaluation of the systematics has been performed, as described in Sec.~\ref{section:longitudinal}.  A similiar procedure has been performed for radial profiles (Sec.~\ref{section:radial}), yielding negligible uncertainties.

Systematic uncertainties are assumed to affect in a negligible way the observables concerning average global quantities of the calorimeter, such as the center of gravity and the standard deviation of longitudinal profiles (Sec.~\ref{section:longitudinal}) , the mean energy-weighted shower radius and the standard deviation of radial energy distributions (Sec.~\ref{section:radial}). As a cross-check, the uncertainties obtained for longitudinal profiles have been propagated to the measurement of the longitudinal center of gravity of showers, obtaining uncertainties lower than 1\%, and to the standard deviation of the longitudinal profiles, yielding uncertainties lower than 0.5\%. Such a level of accuracy does not affect comparisons between data and Monte Carlo, which involve larger effects.

\section{Energy Response}
\label{section:energyresponse}

The calibrated response to pion showers in units of MIPs is converted to the GeV scale according to the following equation:

\begin{eqnarray}
E_{\mathrm{rec}}\,[{\rm GeV}] = (E_{\mathrm{ECAL1}} + E_{\mathrm{ECAL2}} \cdot 2 + E_{\mathrm{ECAL3}} \cdot 3) \cdot 0.002953 \nonumber \\
+ ( E_{\mathrm{AHCAL}} -v ) / u,
\label{equation:energyconv}
\end{eqnarray}
where $E_{\mathrm{ECAL1}}$ is the energy in MIPs measured in the first section of the SiW-ECAL. $E_{\mathrm{ECAL2}}$, $E_{\mathrm{ECAL3}}$ are defined analogously. $E_{\mathrm{AHCAL}}$  is the energy in MIPs measured in the AHCAL. The factors 2 and 3 account for the different sampling structure in the three sections of the SiW-ECAL, with different absorber thicknesses. The factor $0.002953$ is obtained from simulations. It has been evaluated using simulated muons, since the selection criteria require pions to start showering in the AHCAL and so to traverse the SiW-ECAL losing energy only by ionization, similarly to muons. The parameters $u$ and $v$ have been defined in Sec.~\ref{section:calibratiovalidation} and refer to the electromagnetic energy scale. Additional corrections to account for the different response to electromagnetic and hadronic showers need to be applied, if one is interested in the absolute scale of the hadronic response. This topic is discussed in~\cite{:2012gv} and is not relevant for this paper, which only concerns comparisons between the measured and the simulated response to pions.

In addition to the requirements discussed in Sec.~\ref{section:eventselection}, only events for which the hadronic shower starts in the first five AHCAL layers are considered for the studies presented in this section. This requirement minimizes the leakage of the showers out of the AHCAL and reduces the energy collected by the TCMT. This additional selection can be applied since this paper focuses on the validation of Monte Carlo models, rather than on the estimate of the energy resolution. Studies of leakage for non-contained showers have been performed in~\cite{IMarchesini:2011,:2012kx}.

\begin{figure}
\centerline{
\includegraphics[width=0.8\columnwidth,height=0.25\textheight]{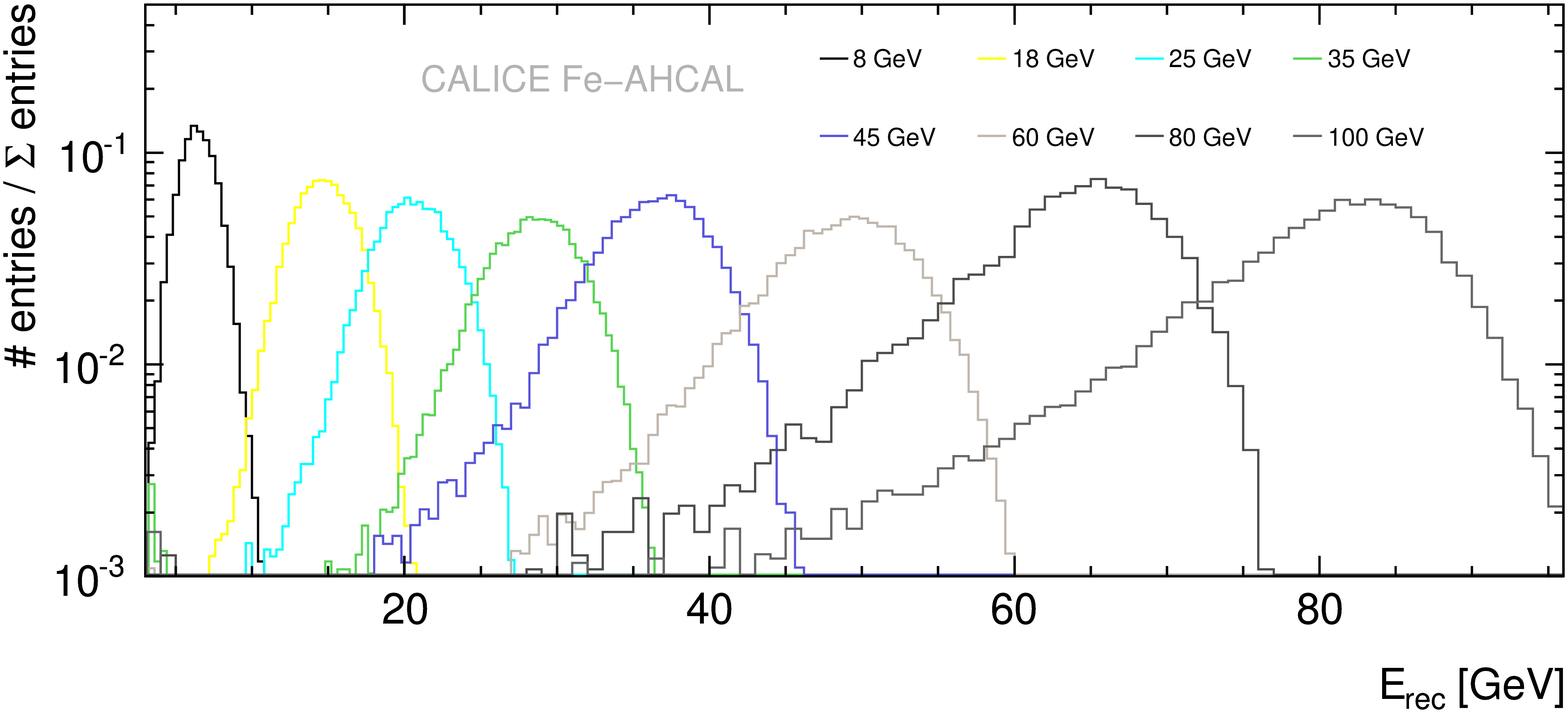}
}
\centerline{
\includegraphics[width=0.4\columnwidth,height=0.25\textheight]{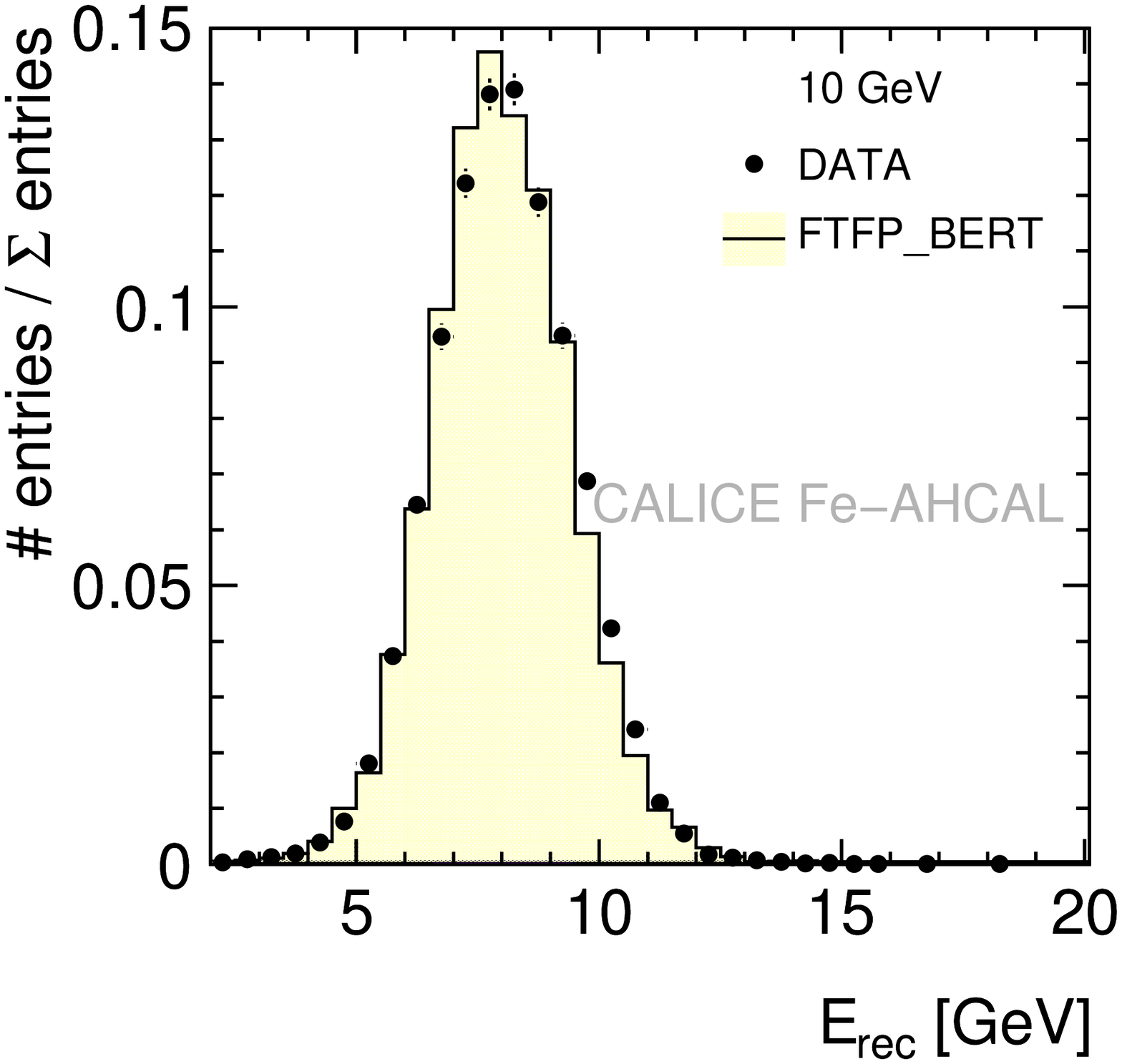}
\includegraphics[width=0.4\columnwidth,height=0.25\textheight]{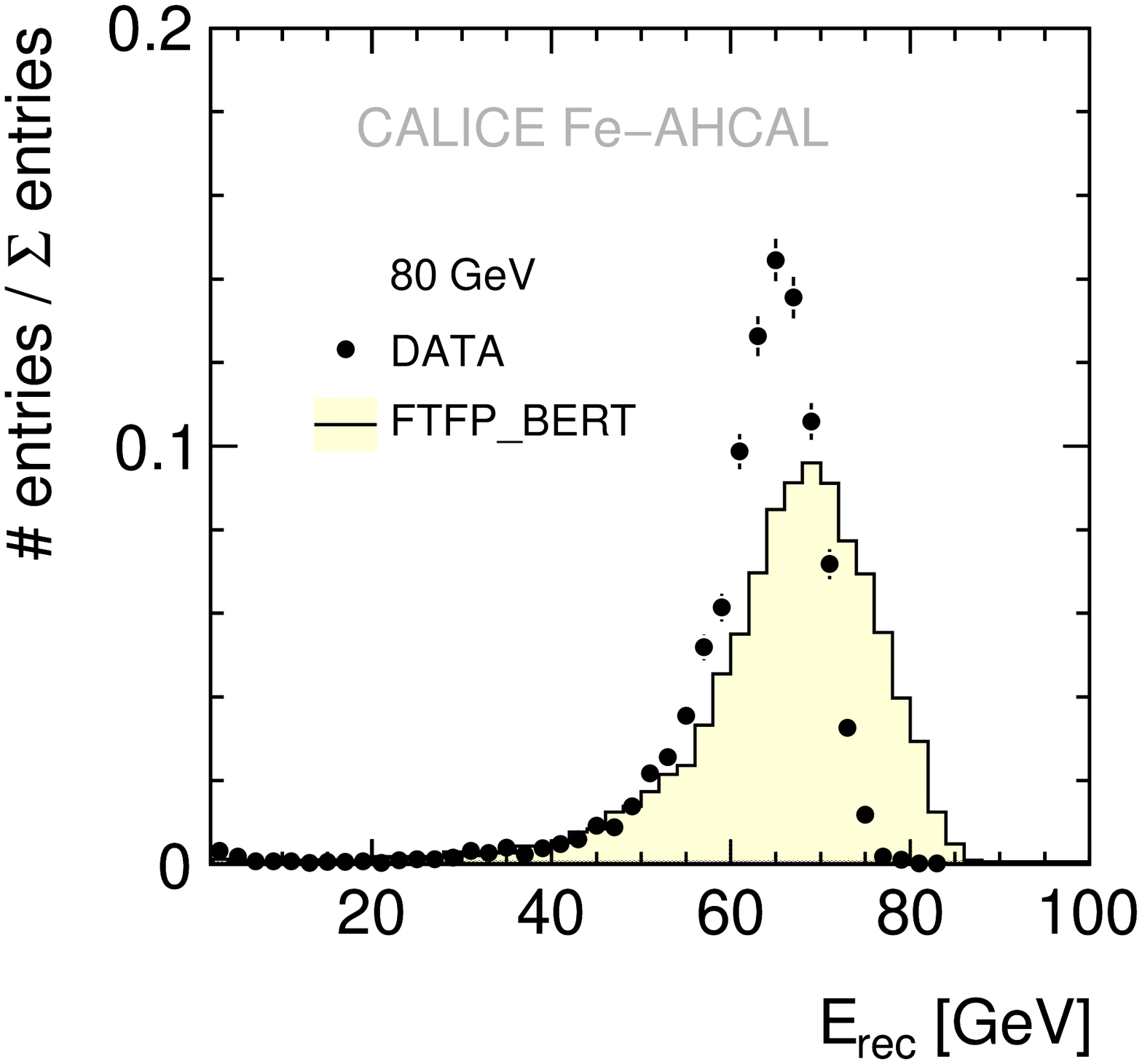}
}
\caption[Energy response distributions]{\sl Top: Total energy distribution for selected pion events at different energies. Bottom, left (right): Response to 10\,GeV (80\,GeV) pions for data (points) and for digitized Monte Carlo simulations using the \texttt{FTFP\_BERT} physics list.}
\label{figure:responseplots}
\end{figure}

The total energy distribution for pion runs at different energies is shown in Fig.~\ref{figure:responseplots} (top).  Figure~\ref{figure:responseplots} (lower left) presents the response to 10\,GeV pions for data and for digitized Monte Carlo simulations based on the \texttt{FTFP\_BERT} physics list. The shape of the response as well as the position of the peak are rather well simulated by the Monte Carlo. At higher energies, as exemplified by an 80\,GeV run in Fig.~\ref{figure:responseplots} (lower right), the agreement gets worse. This is due to a worse simulation of the energy response and to an imperfect description of the shower length by the Monte Carlo, which results in a different impact of leakage in data and Monte Carlo. The longitudinal development of showers is described in Sec.~\ref{section:longitudinal}.

\begin{figure}
\centerline{	
     \subfigure{ \label{} \includegraphics[trim=0 0 45 0, clip, height=8cm]{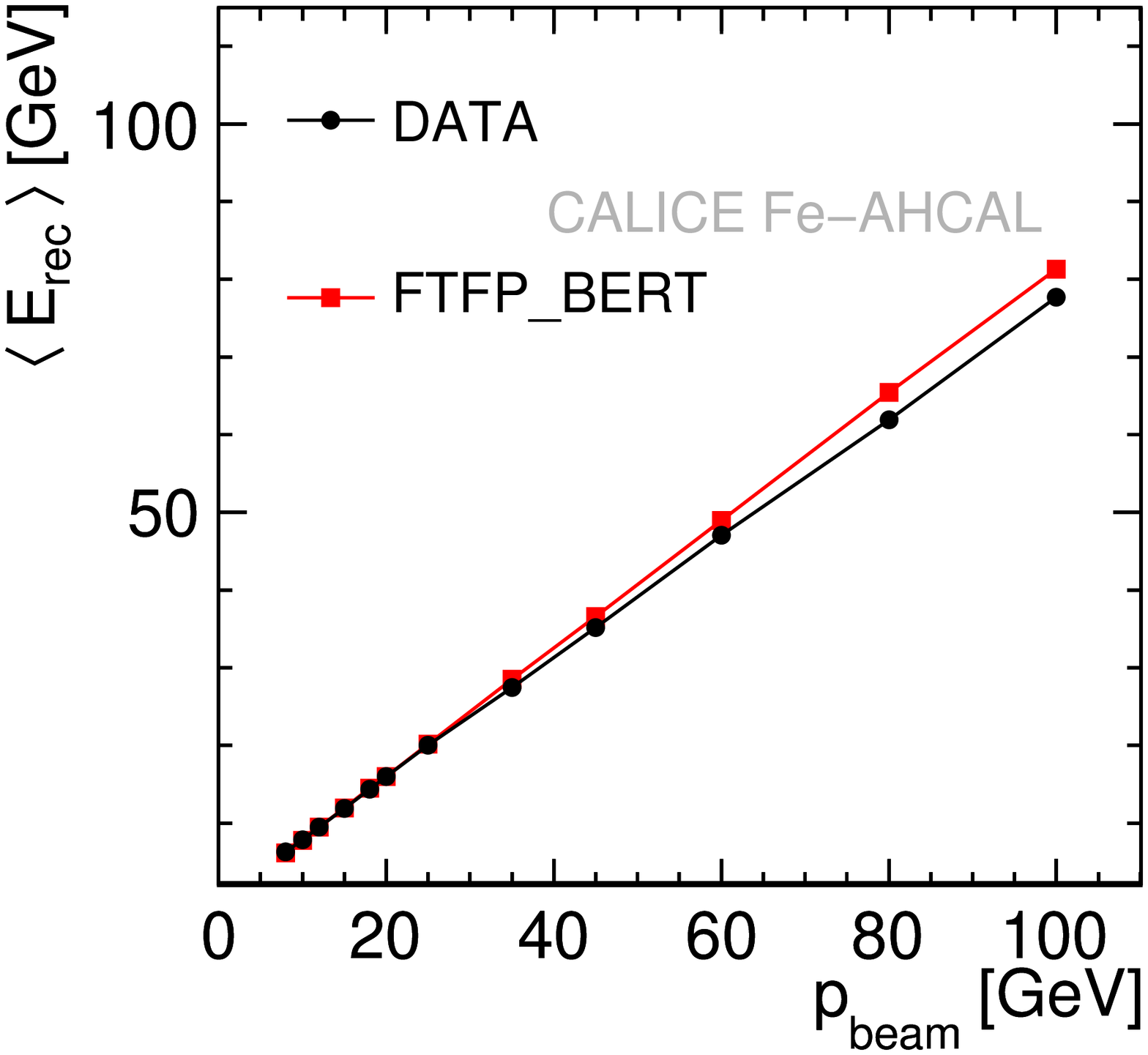} } \hspace{-4.5mm}
     \subfigure{ \label{} \includegraphics[trim=0 0 45 0, clip, height=8cm]{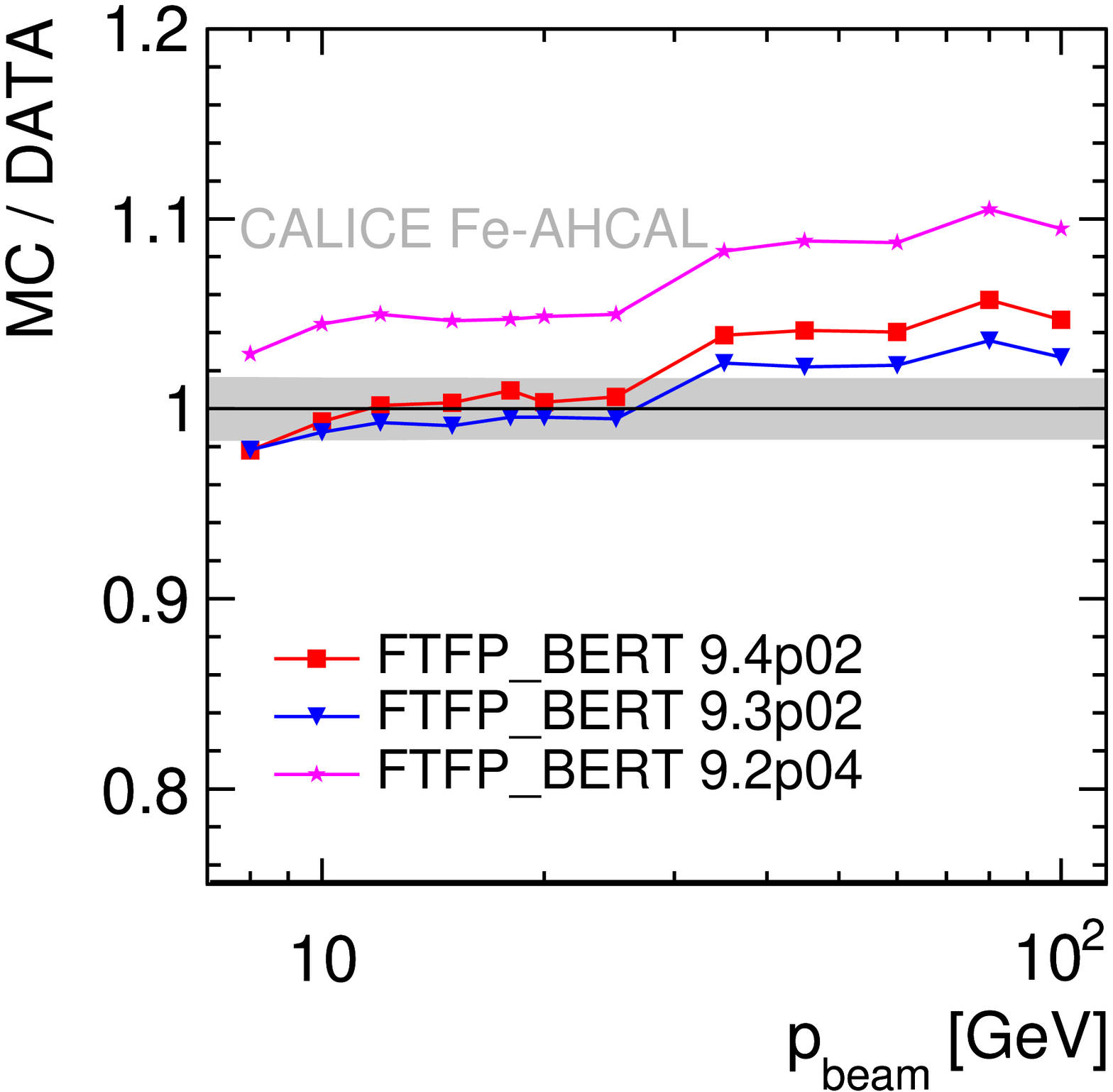} } \hspace{5mm}	
}
\centerline{
     \subfigure{ \label{} \includegraphics[trim=0 0 45 0, clip, height=6cm]{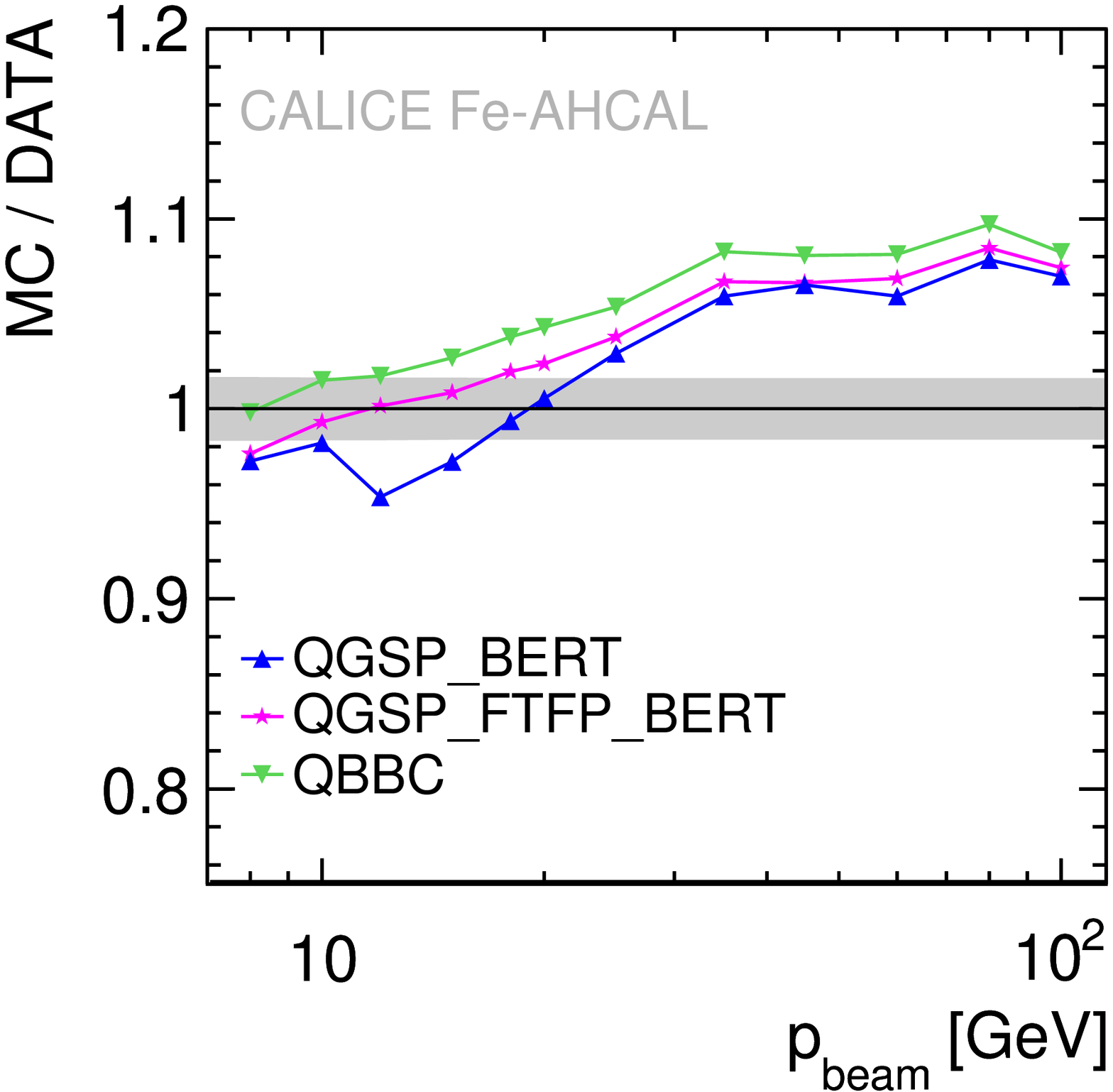} } \hspace{-4.5mm}
     \subfigure{ \label{} \includegraphics[trim=100 0 45 0, clip, height=6cm]{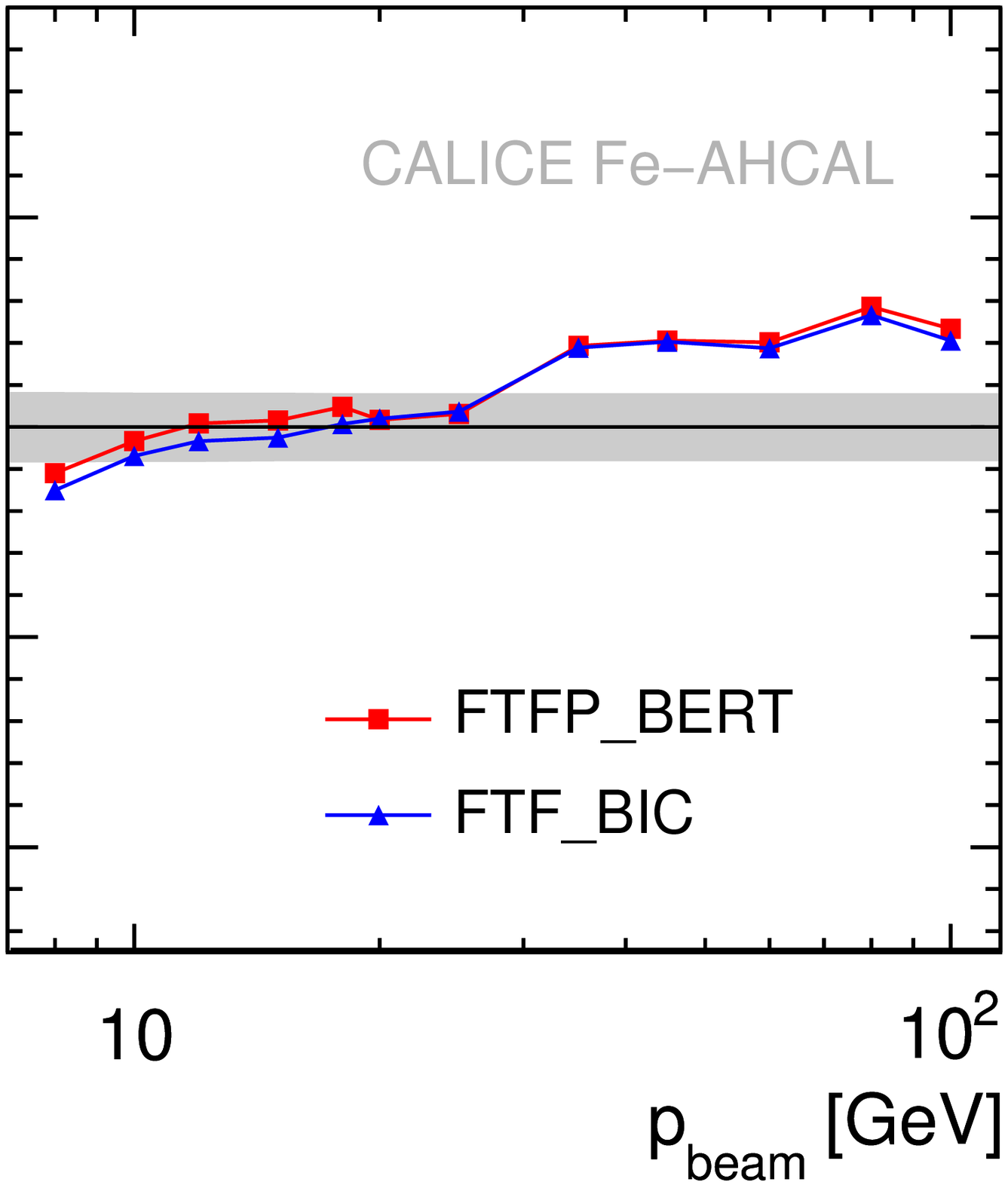} } \hspace{-4.5mm}
     \subfigure{ \label{} \includegraphics[trim=100 0 45 0, clip, height=6cm]{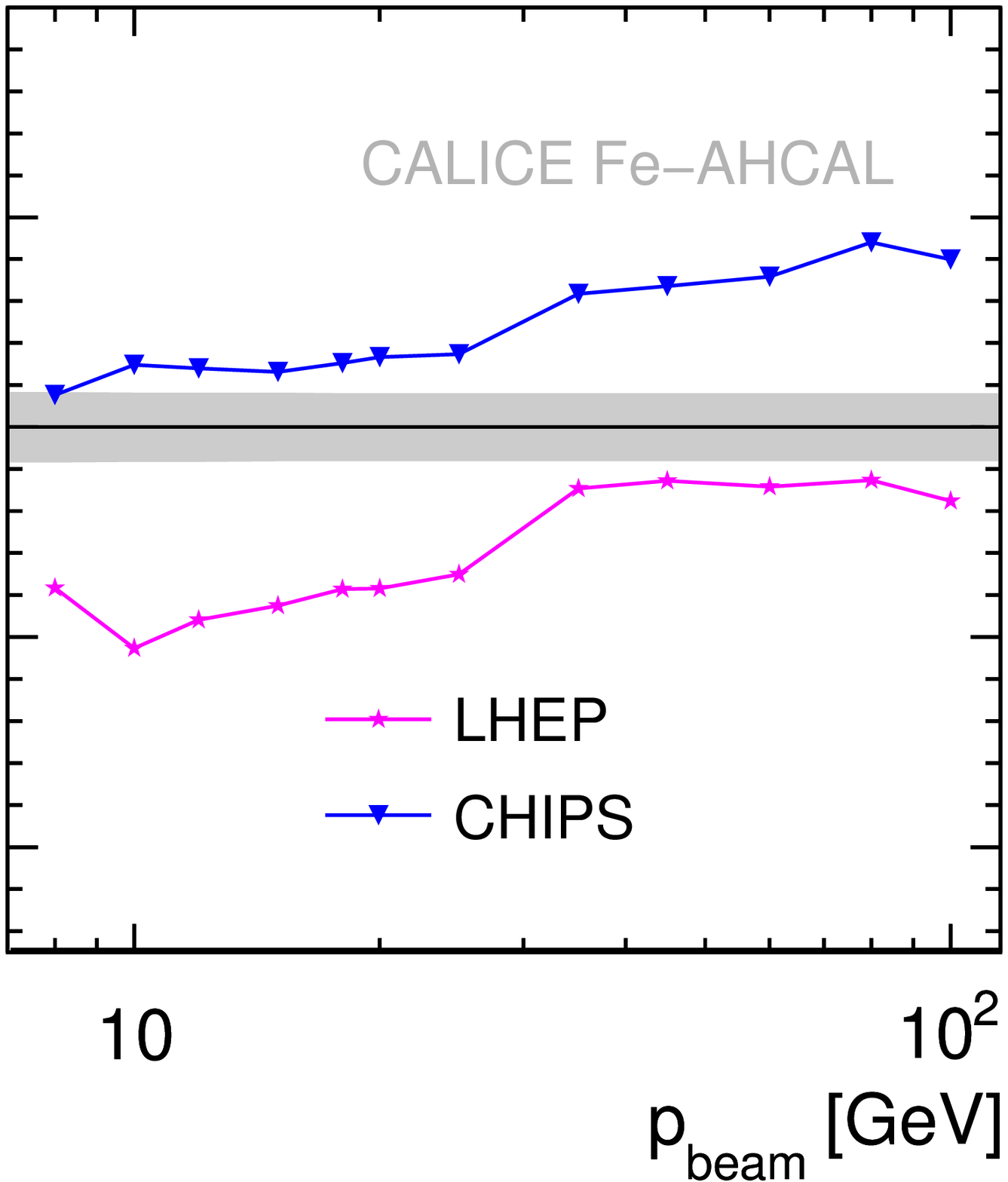} }
}
\caption[Energy response in data and Monte Carlo]{\sl Summary of the energy response to pions. Top, left: For data and for the \texttt{FTFP\_BERT} physics list. Top, right: Ratio between Monte Carlo and data using the \texttt{FTFP\_BERT} physics list with different versions of \textsc{Geant4}. Bottom: Ratio between Monte Carlo and data for several physics list. The gray band in the ratios represents the uncertainty on data (statistical plus a systematic calibration uncertainty of 1.6\%).}
\label{figure:responsemontecarlo}
\end{figure}

In order to quantify the agreement between data and Monte Carlo models, their mean response as function of beam momentum is compared in Fig.~\ref{figure:responsemontecarlo}. The average energy response $\langle E_{\mathrm{rec}} \rangle$ for the different beam energies is given by the arithmetic mean of the energy distributions, after subtraction of the average contribution of the noise above the 0.5\,MIPs threshold. The \verb=QGSP_BERT=, the \verb=QGSP_FTFP_BERT= and the \verb=QBBC= physics lists agree within 4\% with data at 8\,GeV, where the Bertini model is employed. At higher energies the disagreement with data increases, by about 7-10\% at 100\,GeV. The difference between \verb=QGSP_FTFP_BERT= and \verb=QBBC= at low energies, where they apply the same model, is attributed to the different treatment of the inelastic scattering of secondary protons and neutrons at very low energies, which is described using the Bertini model in \verb=QGSP_FTFP_BERT=, while \verb=QBBC= uses the BIC model. The Fritiof-based physics lists underestimate the response compared to the data at 8\,GeV by up to 2-3\%, while they agree with data in the range 10-30\,GeV. At higher energies, between 40\,GeV and 100\,GeV, they overestimate data, though the deviations are below 6\%. The performance of the \texttt{FTFP\_BERT} physics list for previous versions of \textsc{Geant4} is shown in Fig.~\ref{figure:responsemontecarlo} (top, right). The performance has significantly improved from the version 9.2 to the version 9.3. In the most recent version the response has slightly increased with respect to the version 9.3, worsening the agreement with data at high energies. \verb=CHIPS= overestimates data at all energies, by up to 10\% at 80\,GeV. At low energies, below 20\,GeV the description of the energy response improves and the disagreement reduces to 2\%. The \verb=LHEP= physics list, based on parameterized models, largely underestimates the response by up to 8-10\% at 8-10\,GeV. At high energies above 40\,GeV the disagreement between data and Monte Carlo reduces to about 3\%.

\section{Pion Interaction Length}
\label{section:hadinterlength}

As long as a pion traversing a material does not interact strongly with a nucleus, it loses energy mainly by ionization. The probability $P_I$ of having an inelastic hadron-nucleus interaction before a distance $x$ is given by:
\begin{equation}
P_I = 1 - e^{-x/\lambda_\pi},
\label{equation:lambdaexp}
\end{equation}
where $\lambda_\pi$ is the pion interaction length. 

\begin{figure}
\centerline{
\includegraphics[width=0.5\columnwidth]{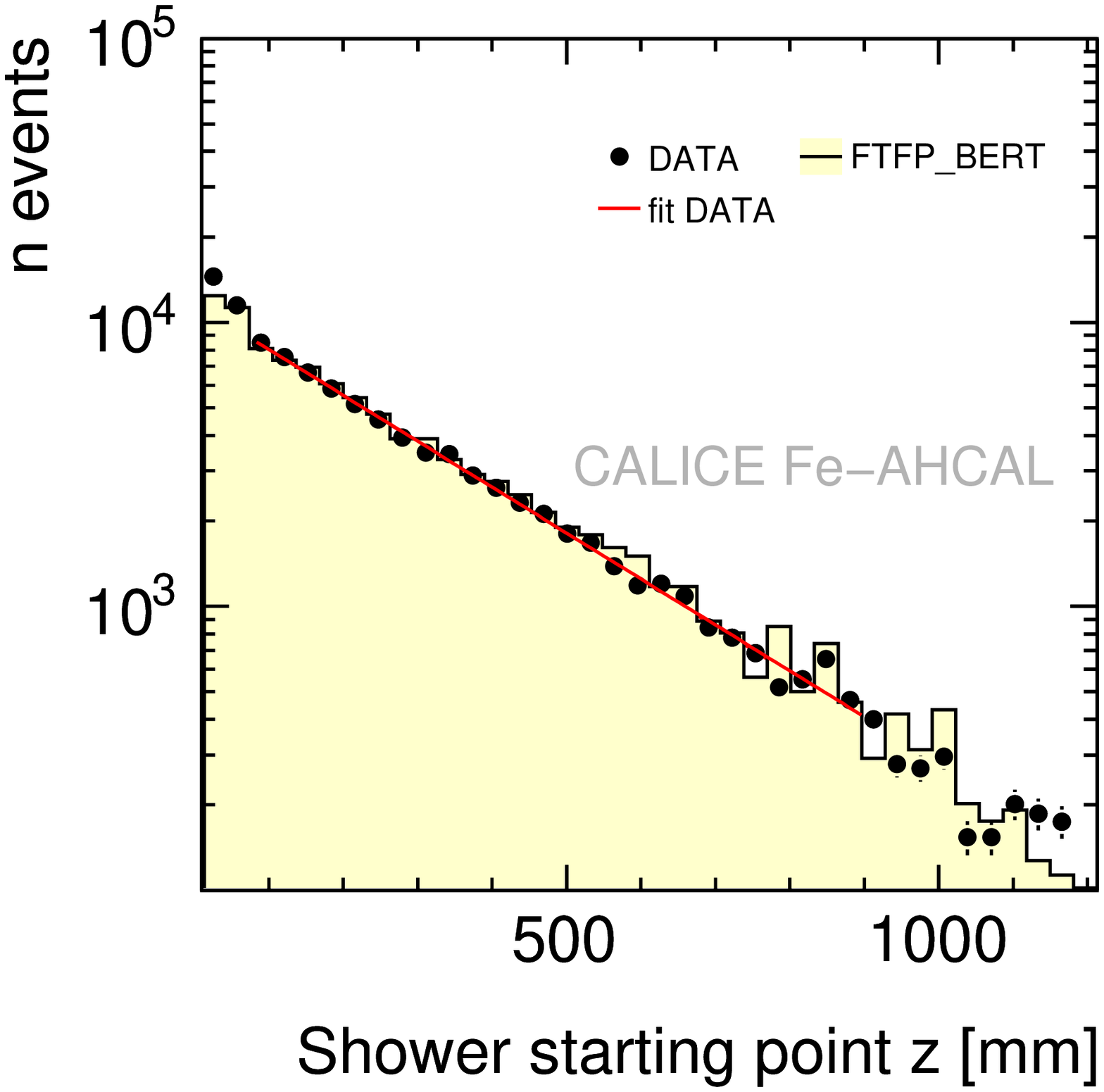}
\includegraphics[width=0.5\columnwidth]{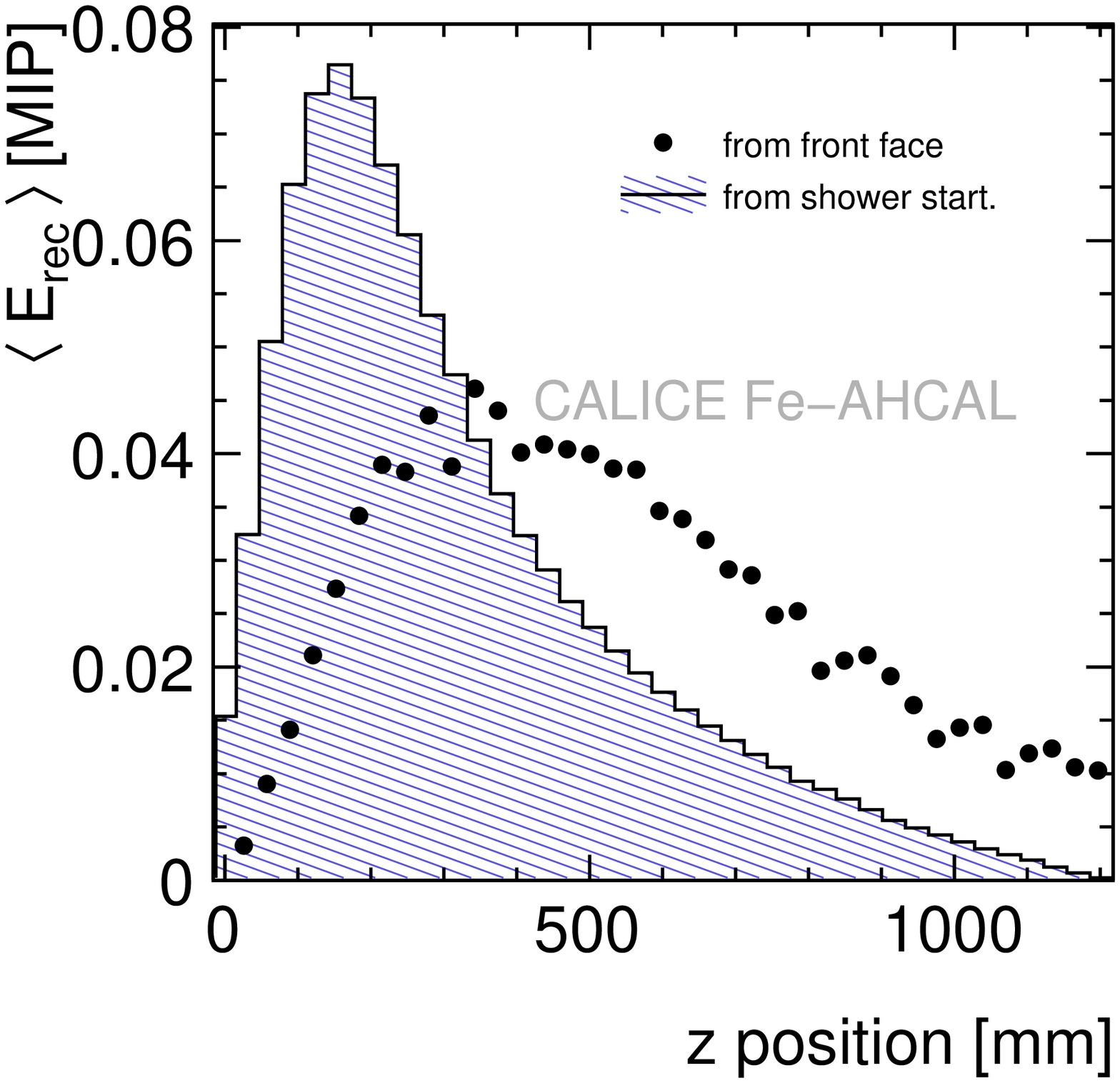}
}
\caption[First hard interaction and longitudinal profile]{\sl Left: Distribution of the measured layer of the first hard interaction for 45\,GeV pions, for data (circles) and for the \texttt{FTFP\_BERT} physics (histogram). The distributions are normalized to an arbitrary number. The fit to the data distribution is also drawn. Right: Longitudinal profile of 45\,GeV pion showers relative to the calorimeter front face (circles) and relative to the first hard interaction (histogram). The profiles are normalized to unity.}
\label{figure:showerstartdistr}
\end{figure}

The position of the first hard interaction of the primary beam particle is approximated by the reconstructed layer where the shower starts. The measured distribution of the shower starting point follows the exponential behavior expected from Eq.~\ref{equation:lambdaexp}, as shown in Fig.~\ref{figure:showerstartdistr} (left) for 45\,GeV pion showers. 

From this distribution it is possible to derive the effective interaction length of a pion $\lambda_\pi$ in the AHCAL. The interaction length is obtained fitting an exponential to the distribution. The first layers are excluded from the fit, since the uncertainty of the algorithm used to determine the layer of the first hard interaction is larger there, due to the contamination from showers that start to develop in the last layers of the SiW-ECAL. The last layers are also not considered when performing the fit, since the fluctuations are higher due to the lower statistics and the algorithm used to identify the shower starting point has not been optimized for the coarser granularity of the last layers of the AHCAL. Hence, the fit is performed in the range from 120\,mm to 800\,mm of the longitudinal $z$ coordinate.

\begin{figure}
\centerline{	
     \subfigure{ \label{} \includegraphics[trim=0 0 45 0, clip, height=8cm]{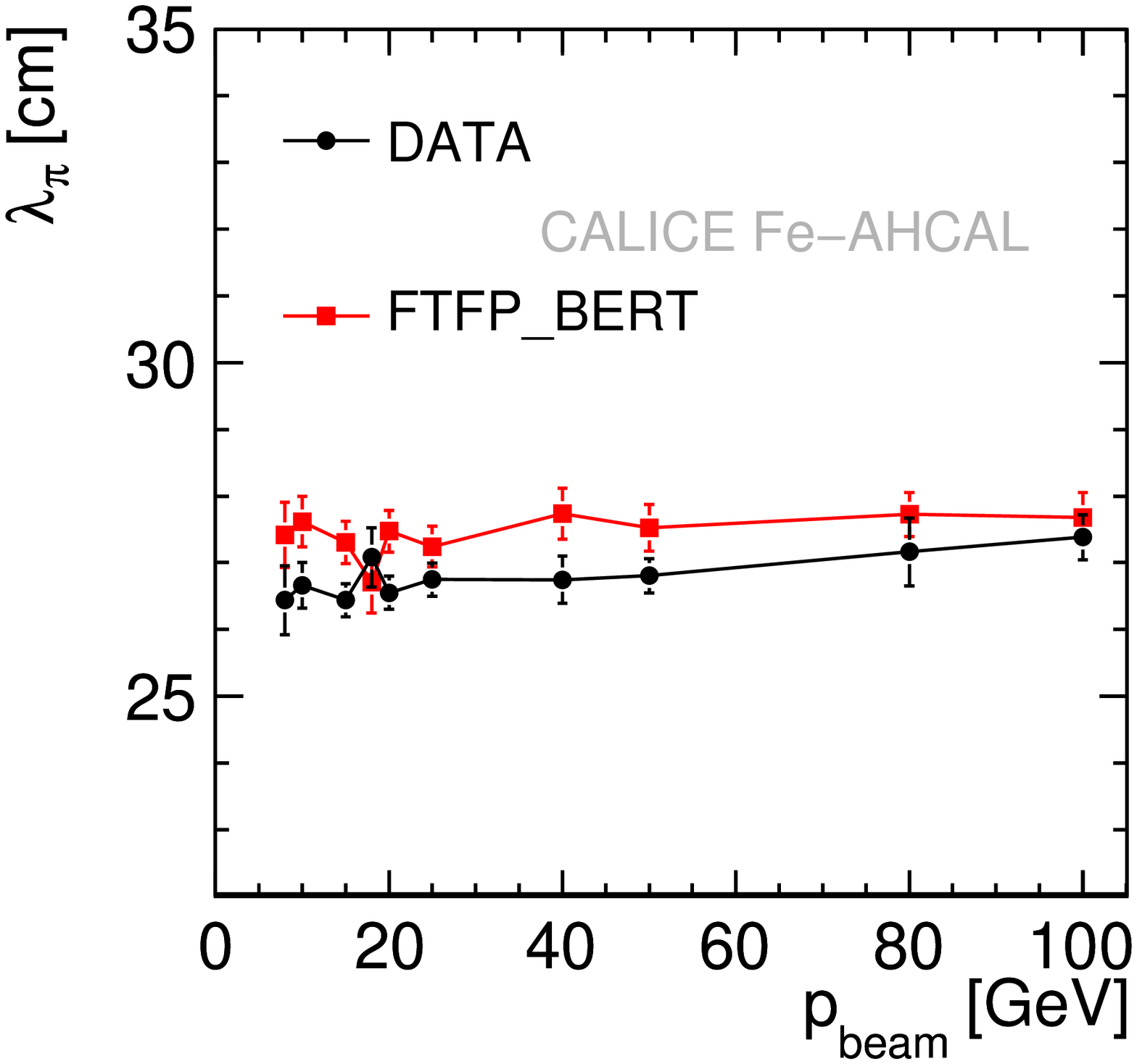} } \hspace{-4.5mm}
     \subfigure{ \label{} \includegraphics[trim=0 0 45 0, clip, height=8cm]{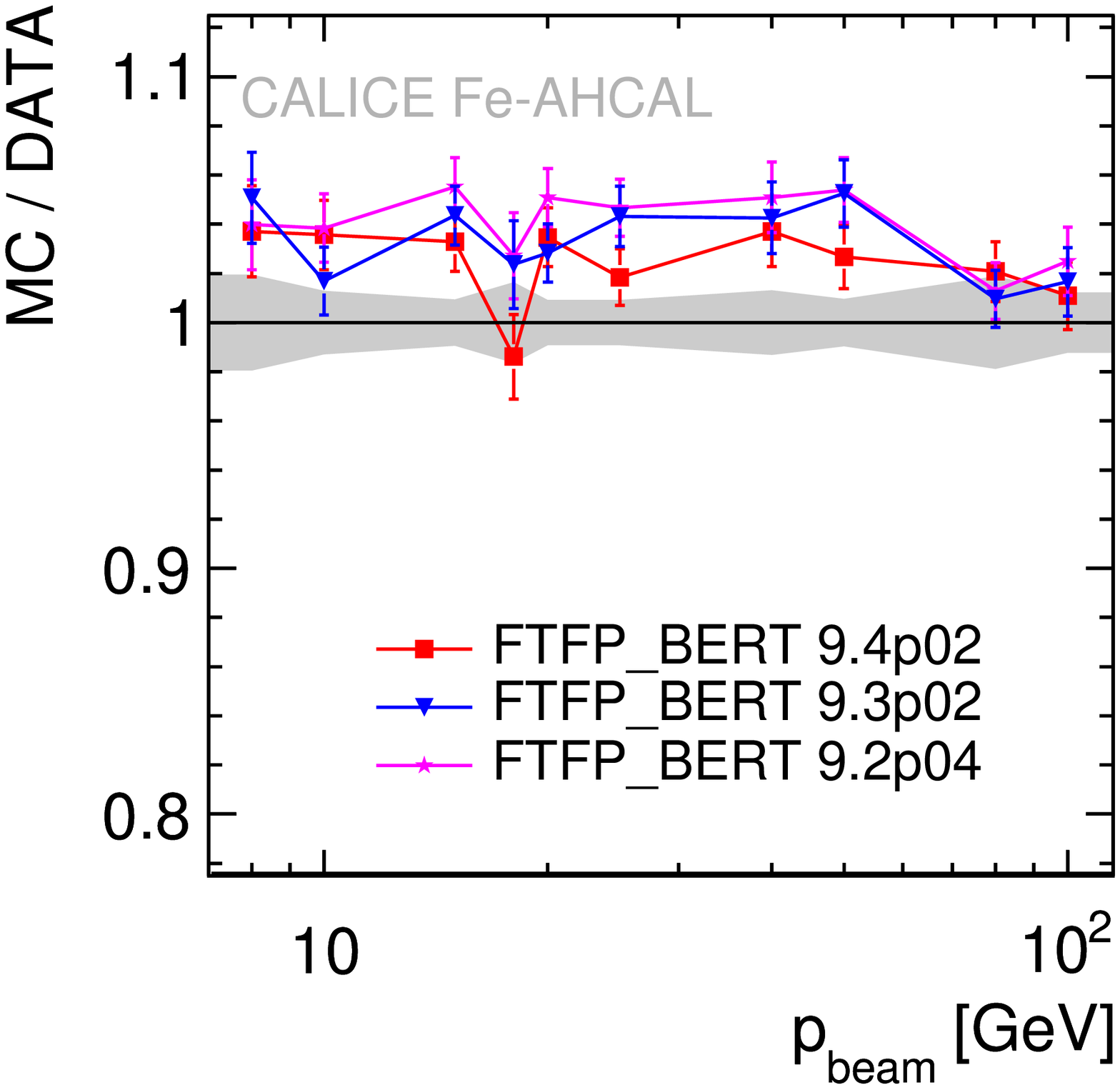} } \hspace{5mm}	
}
\centerline{
     \subfigure{ \label{} \includegraphics[trim=0 0 45 0, clip, height=6cm]{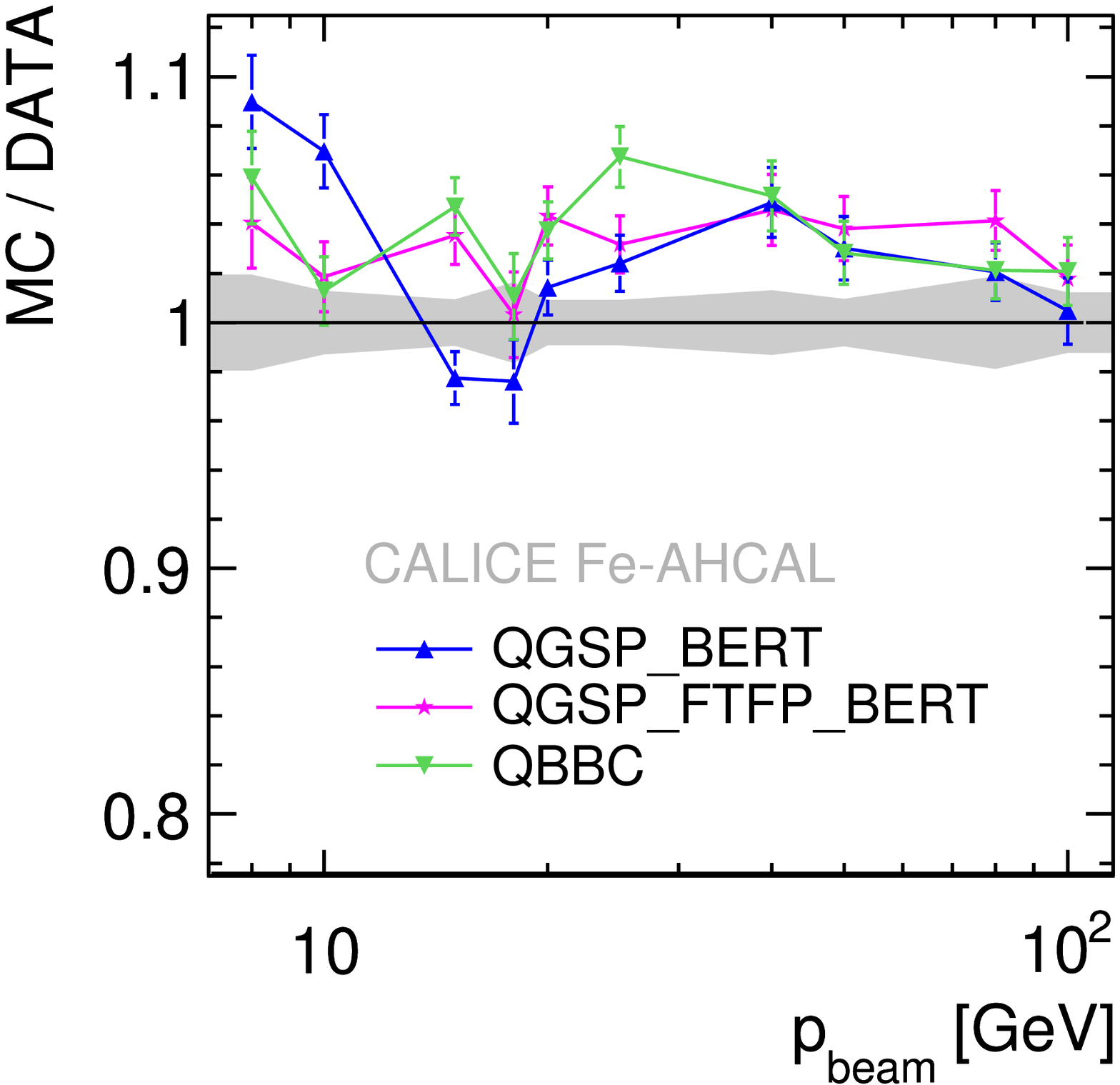} } \hspace{-4.5mm}
     \subfigure{ \label{} \includegraphics[trim=100 0 45 0, clip, height=6cm]{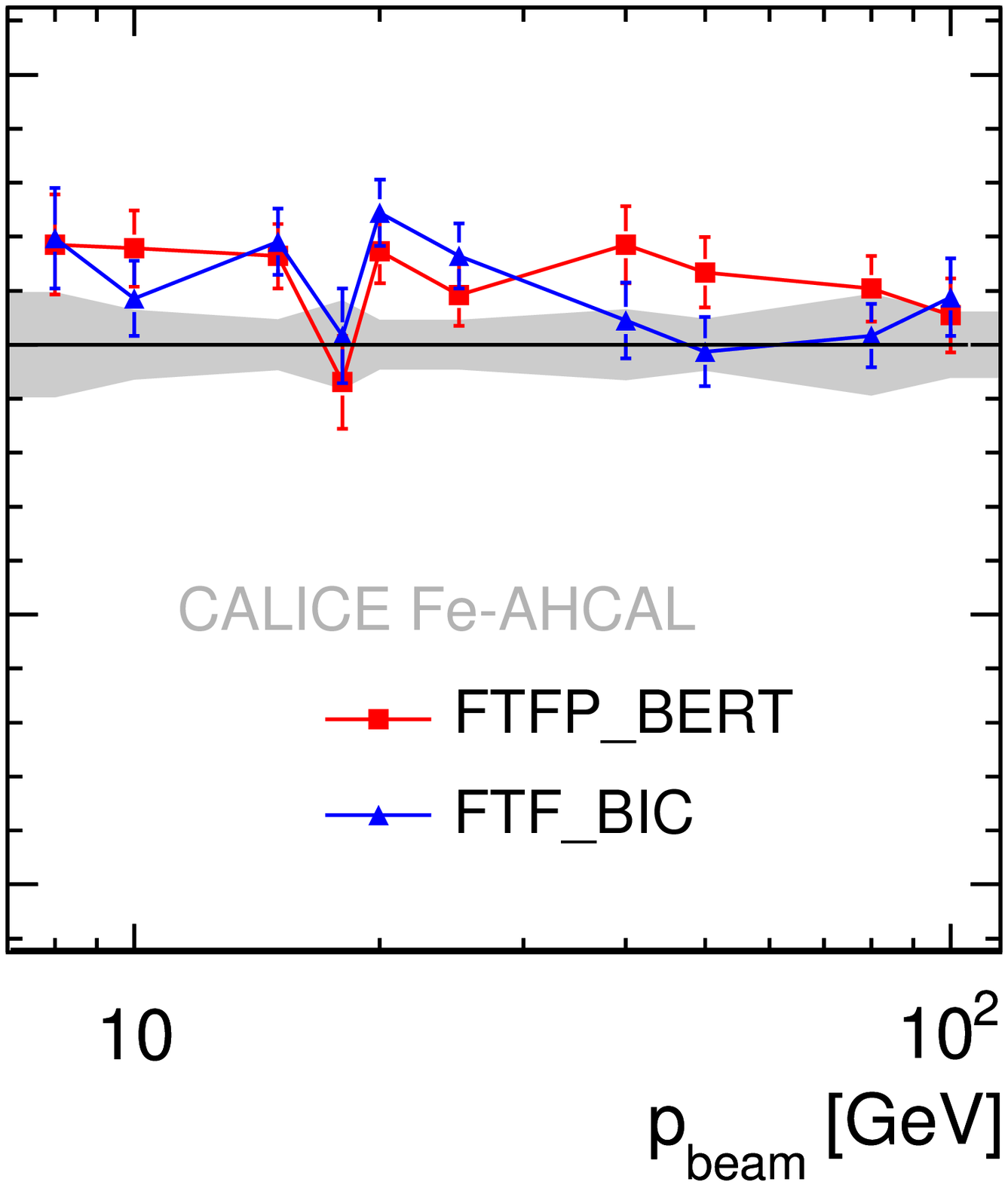} } \hspace{-4.5mm}
     \subfigure{ \label{} \includegraphics[trim=100 0 45 0, clip, height=6cm]{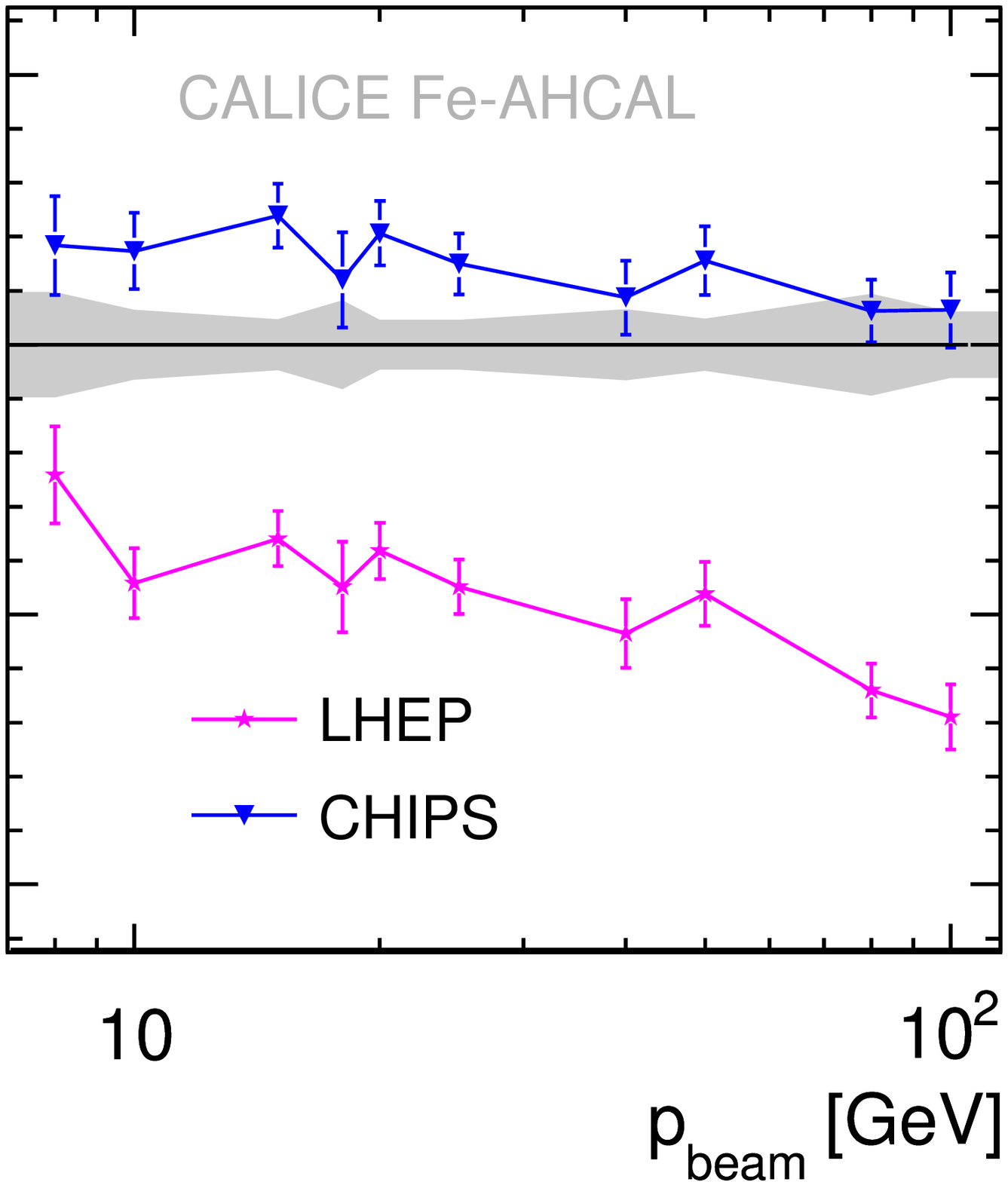} }
}
\caption[Interaction length]{\sl Summary of the measurement of the interaction length of pions in the AHCAL. Top, left: For data and for the \texttt{FTFP\_BERT} physics list. Top, right: Ratio between Monte Carlo and data using the \texttt{FTFP\_BERT} physics list with different versions of \textsc{Geant4}. Bottom: Ratio between Monte Carlo and data for several physics lists. The gray band in the ratios represents the uncertainty on data (statistical plus systematic fit uncertainty). The error bands take into account both the statistical uncertainty and the systematic uncertainty due to the choice of the fit range.}
\label{figure:interactionlength}
\end{figure}

The extracted values of the interaction length for data and Monte Carlo are shown in Fig.~\ref{figure:interactionlength}. No energy dependence of the interaction length is observed in data. The measurements yield an average of $\lambda_\pi = (26.8 \pm 0.46)$\,cm. The error includes the statistical error, determined from the standard deviation of the measurements performed at different energies, and the systematic error. The only source of systematics taken into account is the fit uncertainty due to choice of the fit range. Such an uncertainty has been determined by comparing the fit results for two different $z$ ranges, namely [120, 800]\,mm and [90, 900]\,mm.  The $\lambda_\pi$ value obtained is in good agreement with the expectations from the material composition of the detector, which yield a $\lambda_\pi$ of approximately 28\,cm (from~\cite{collaboration:2010hb}: 4.28\,$\lambda_\pi$ for an AHCAL thickness of 120.26\,cm, which yield a $\lambda_\pi$ of 28.1\,cm). This serves as a cross-check of the detector modeling used for Monte Carlo simulations, which assumes the material composition of the detector described in~\cite{collaboration:2010hb}, and of the algorithm developed for the identification of the shower starting point, used in the event selection. 

All Monte Carlo models except \verb=CHIPS= and \verb=LHEP= use the same cross-sections for hadron-nucleus interactions, which is reflected in a general agreement between the models themselves within fit uncertainties. The \verb=QGSP_FTFP_BERT= and the Fritiof-based physics lists agree with data at approximately the 4\% level at all energies, while \verb=QBBC= agrees with data within 6\%. The \verb=QGSP_BERT= physics list is consistent with data within 4\% for energies greater than 10\,GeV, while at 10\,GeV and 8\,GeV the disagreement with data increases by up to $\sim$9\%. \verb=CHIPS= shows a behavior similar to the other models, with a good agreement with data, better than 4\% at all energies. Only \verb=LHEP= has an energy-dependent trend and systematically underestimates data by up to 14\% at 100\,GeV.

\section{Longitudinal Development}
\label{section:longitudinal}

\begin{figure}
\centerline{
\includegraphics[trim=0 0 10 10, clip, height=6cm]{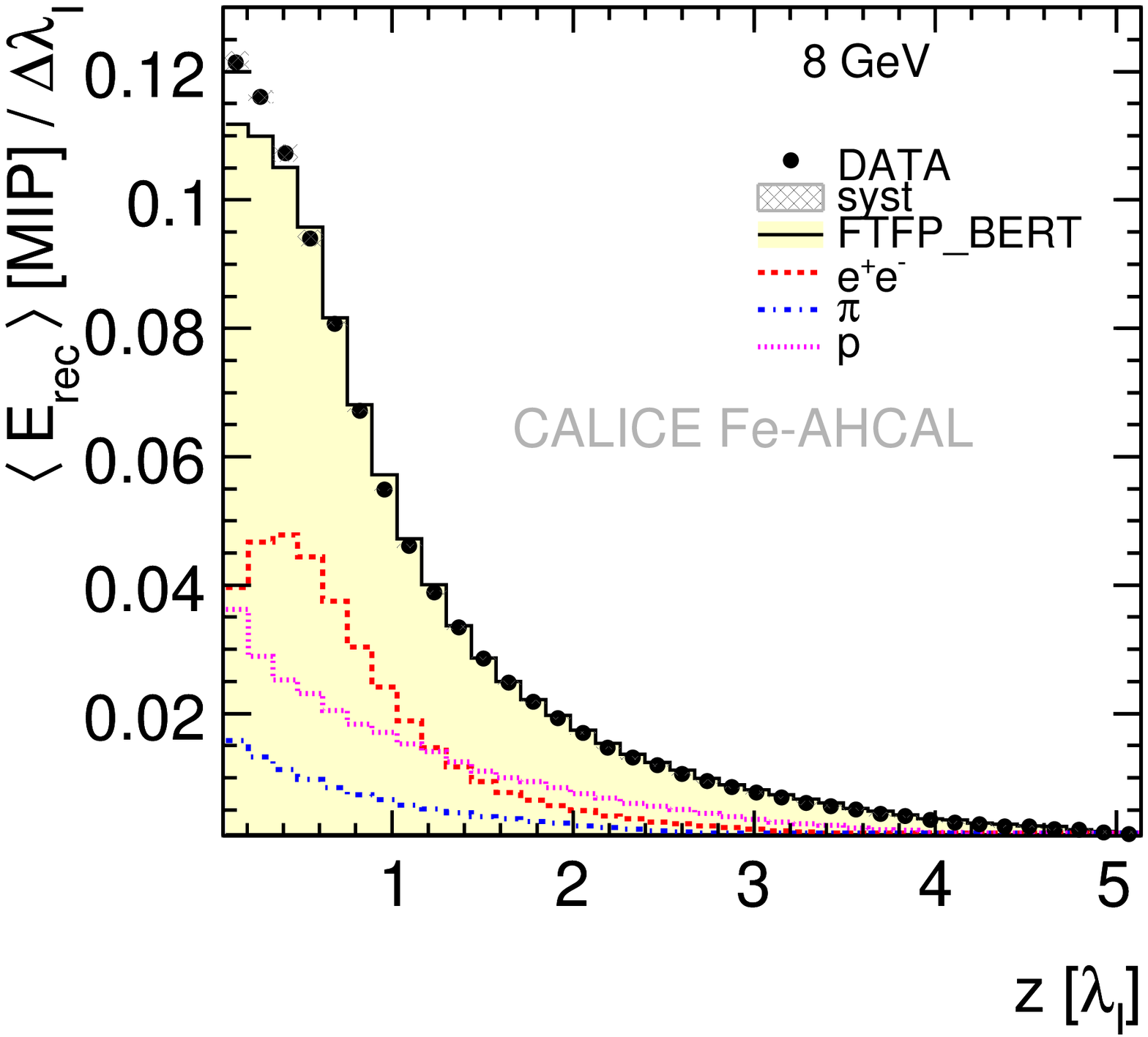}\hspace{-4.mm}
\includegraphics[trim=80 0 10 10, clip, height=6cm]{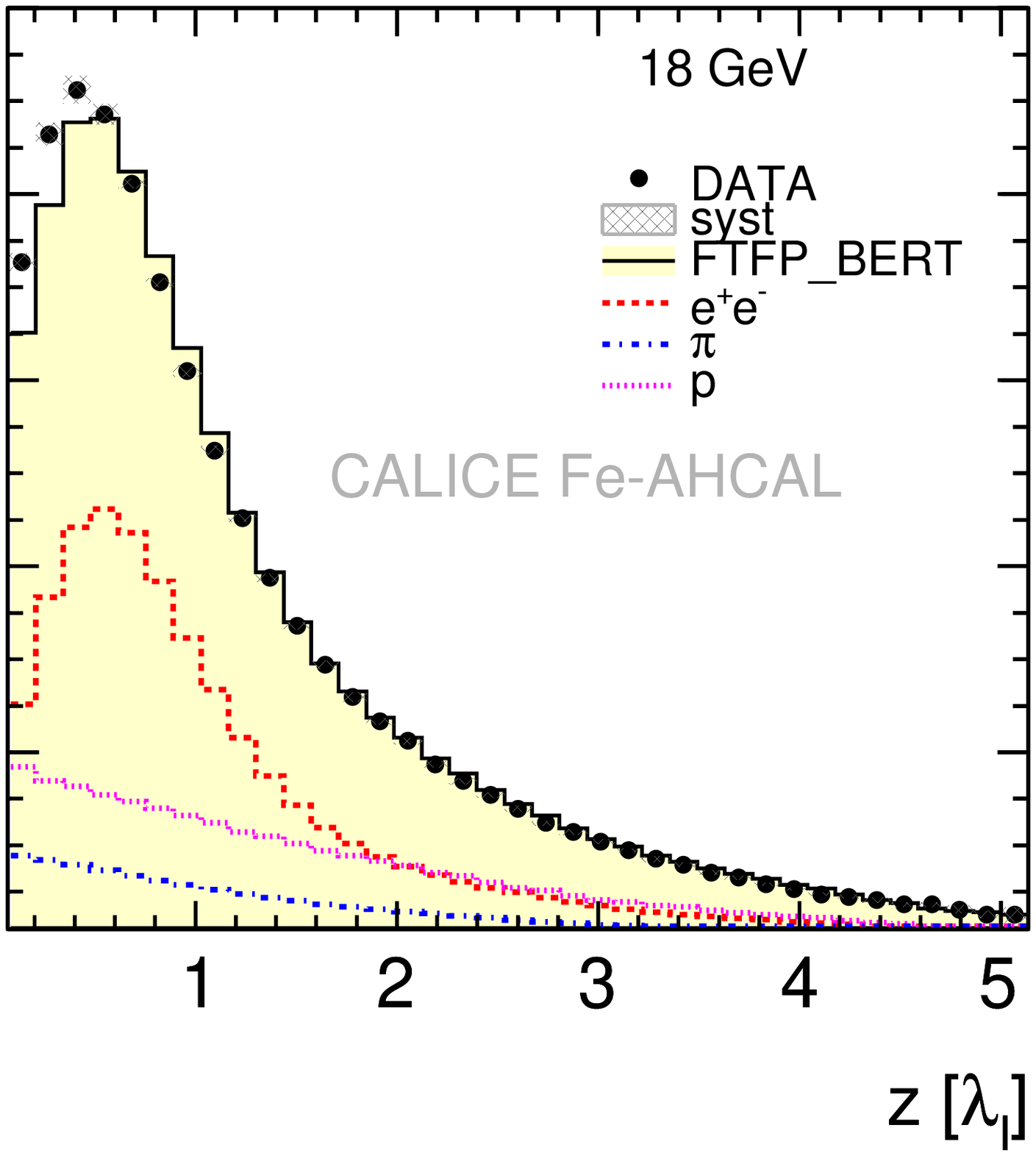}\hspace{-4.mm}
\includegraphics[trim=80 0 10 10, clip, height=6cm]{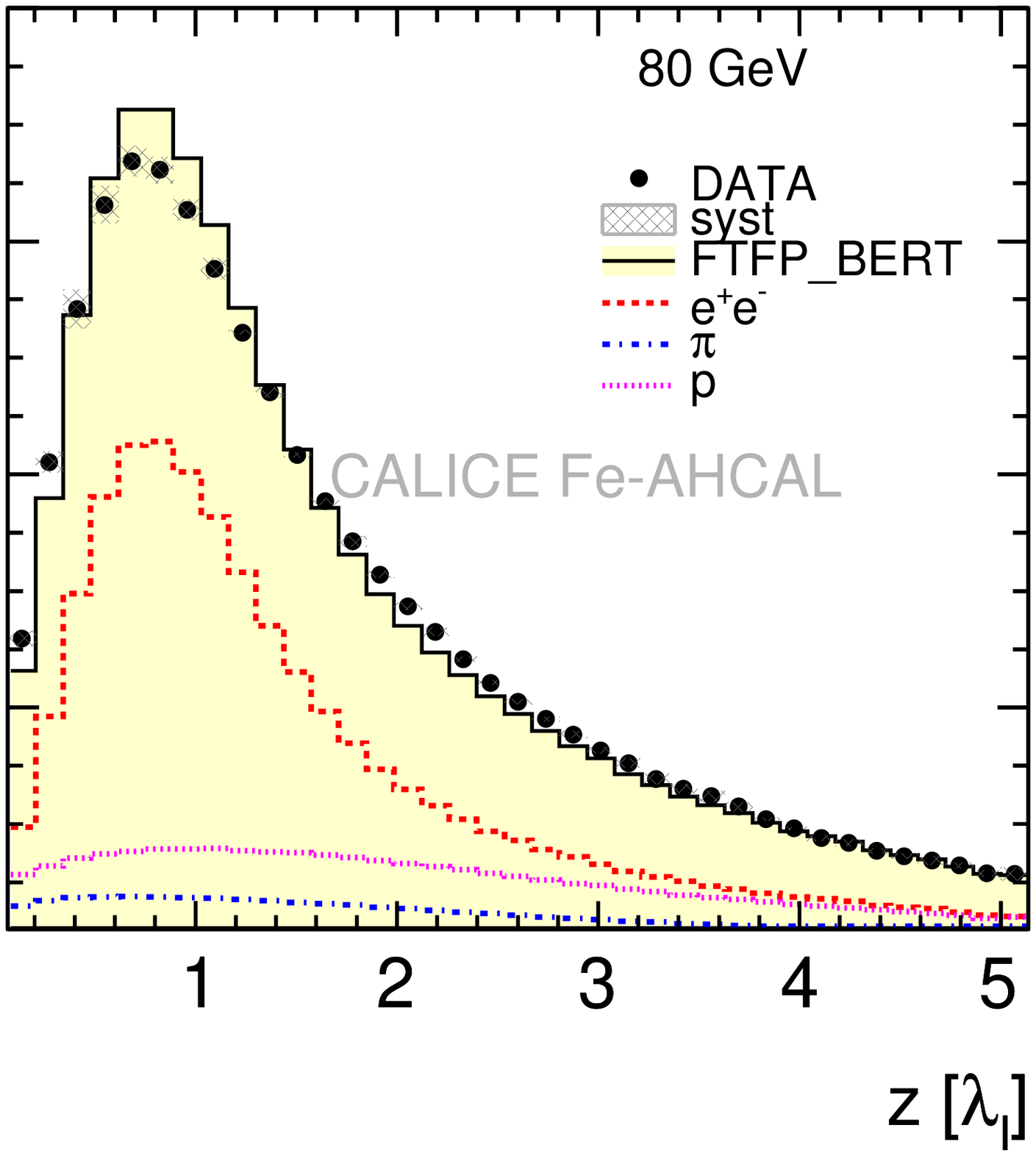}
}
\vspace{0mm}
\centerline{
\hspace{-1.5mm}
\includegraphics[trim=0 0 10 10, clip, height=6cm]{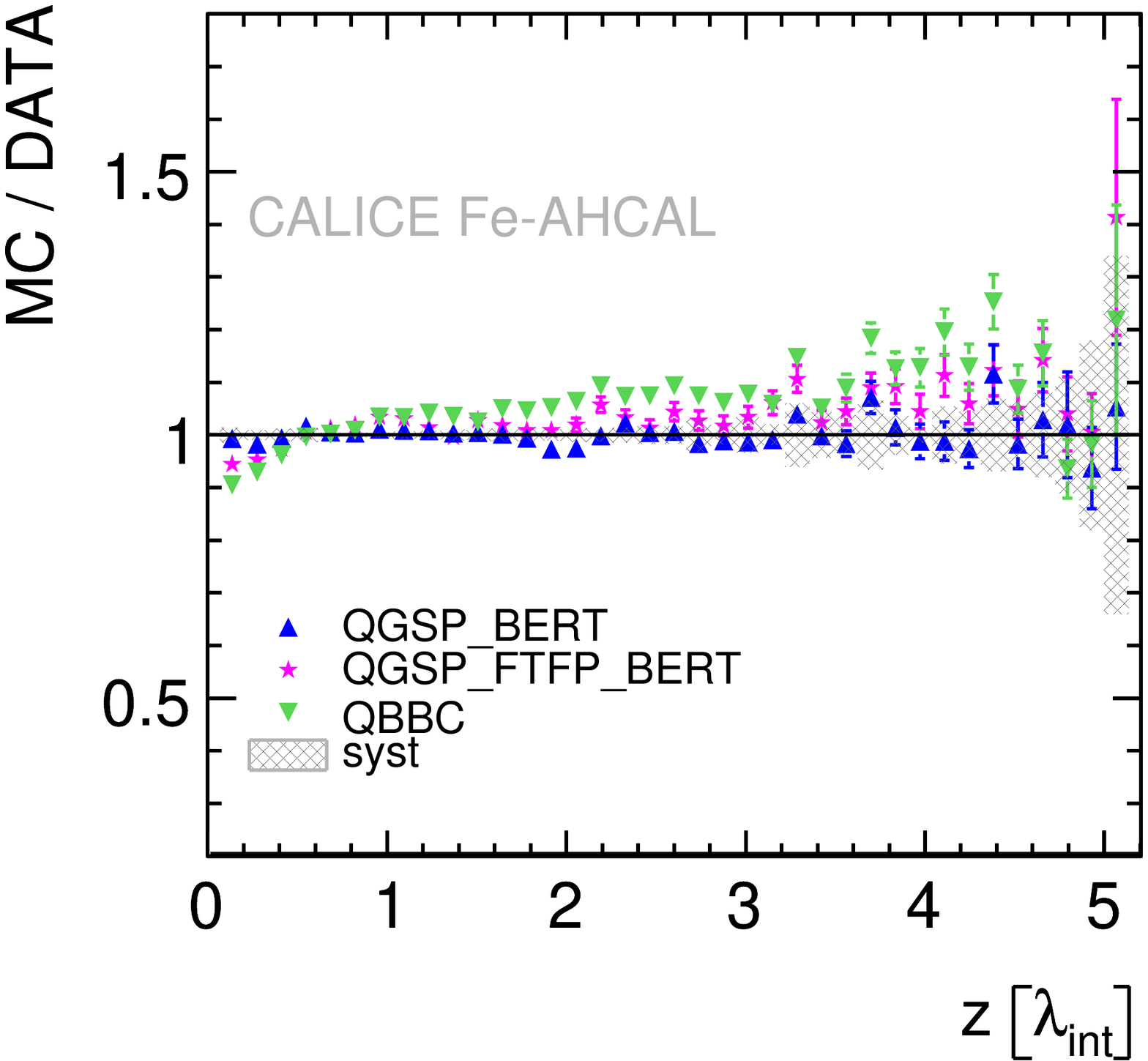}  \hspace{-5mm}
\includegraphics[trim=80 0 10 10, clip, height=6cm]{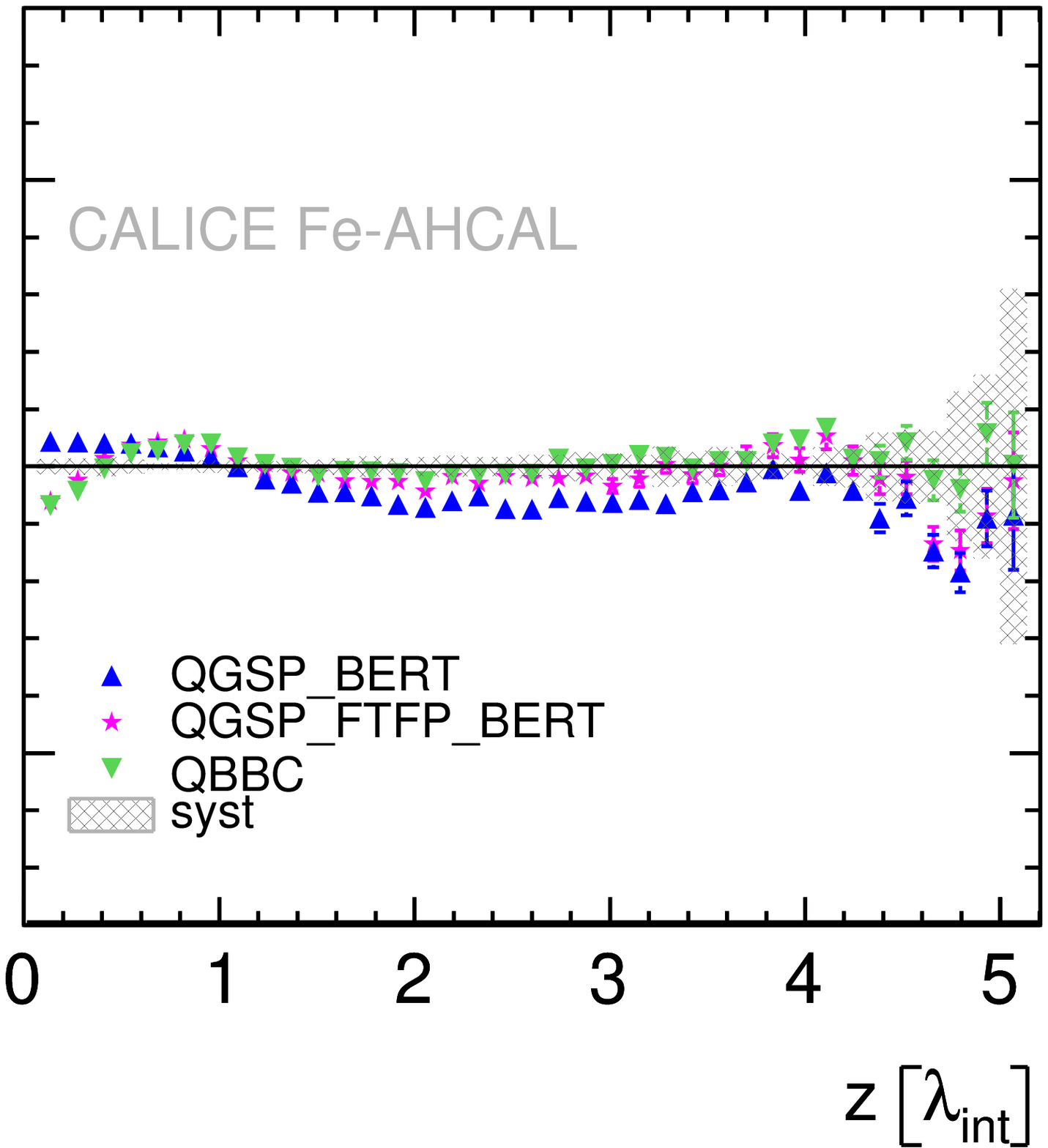}  \hspace{-5.3mm}
\includegraphics[trim=80 0 10 10, clip, height=6cm]{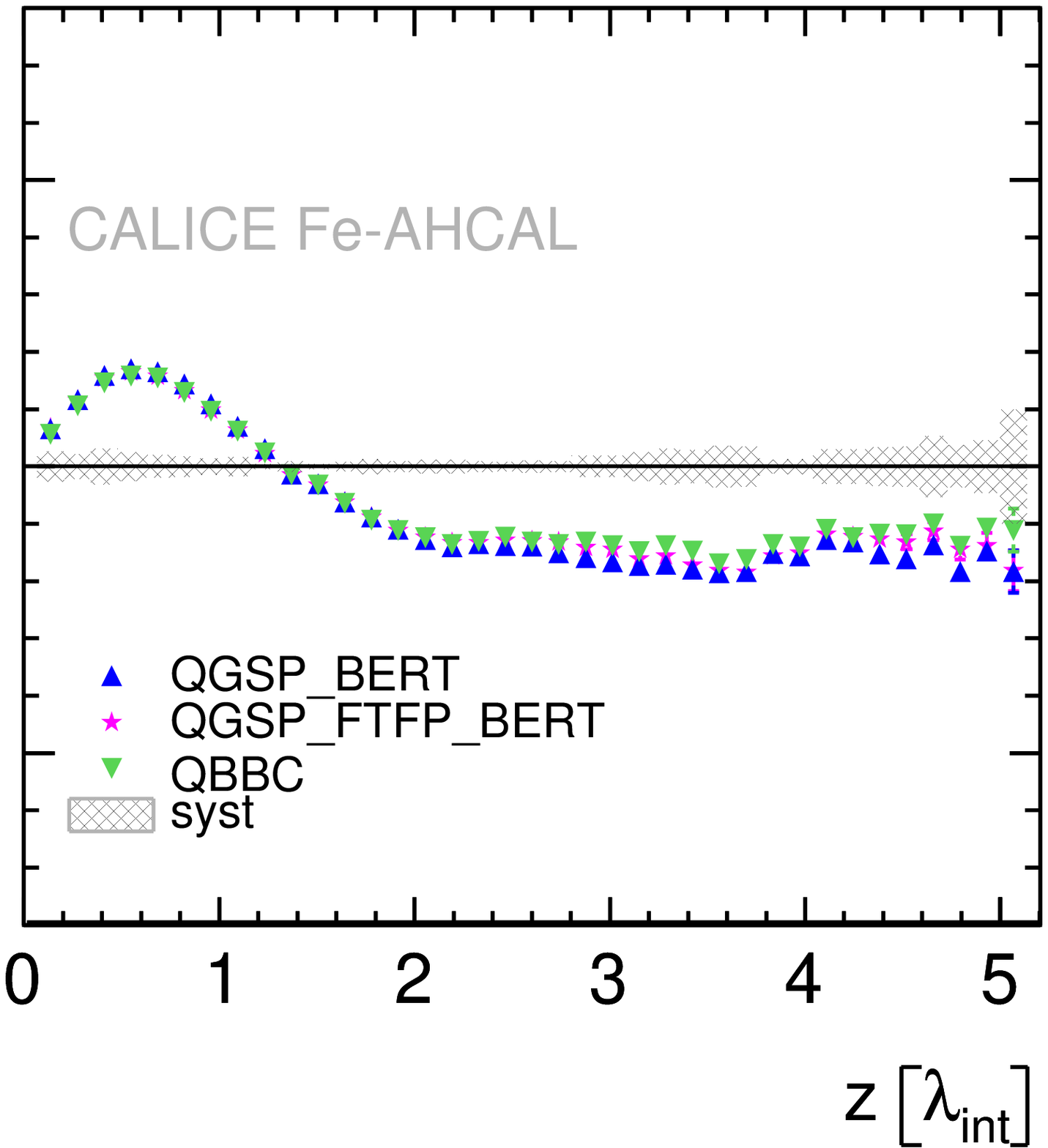} 
\vspace{-10mm}
}
\centerline{
\hspace{-1.5mm}
\includegraphics[trim=0 0 10 10, clip, height=6cm]{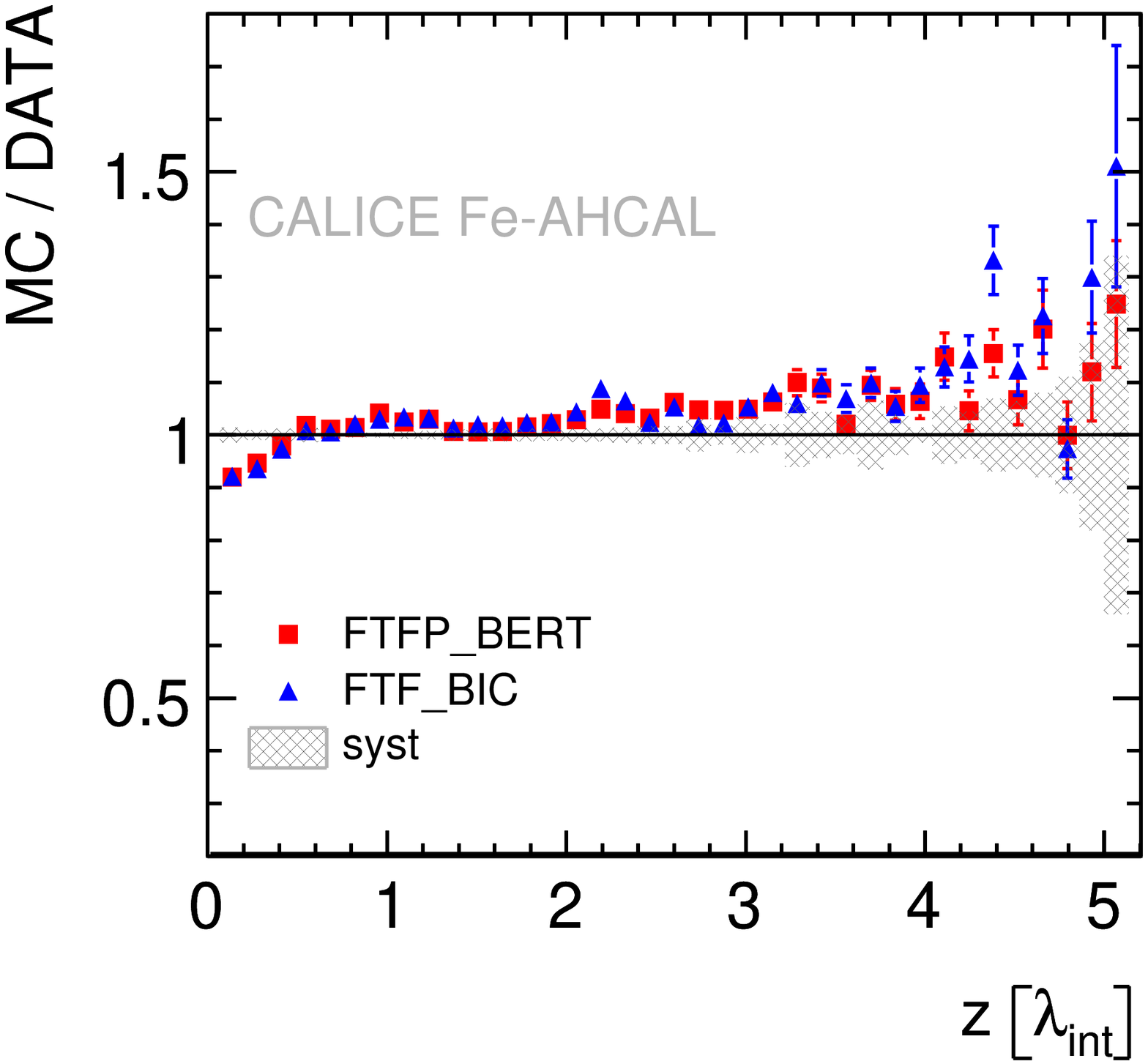}  \hspace{-5.mm}
\includegraphics[trim=80 0 10 10, clip, height=6cm]{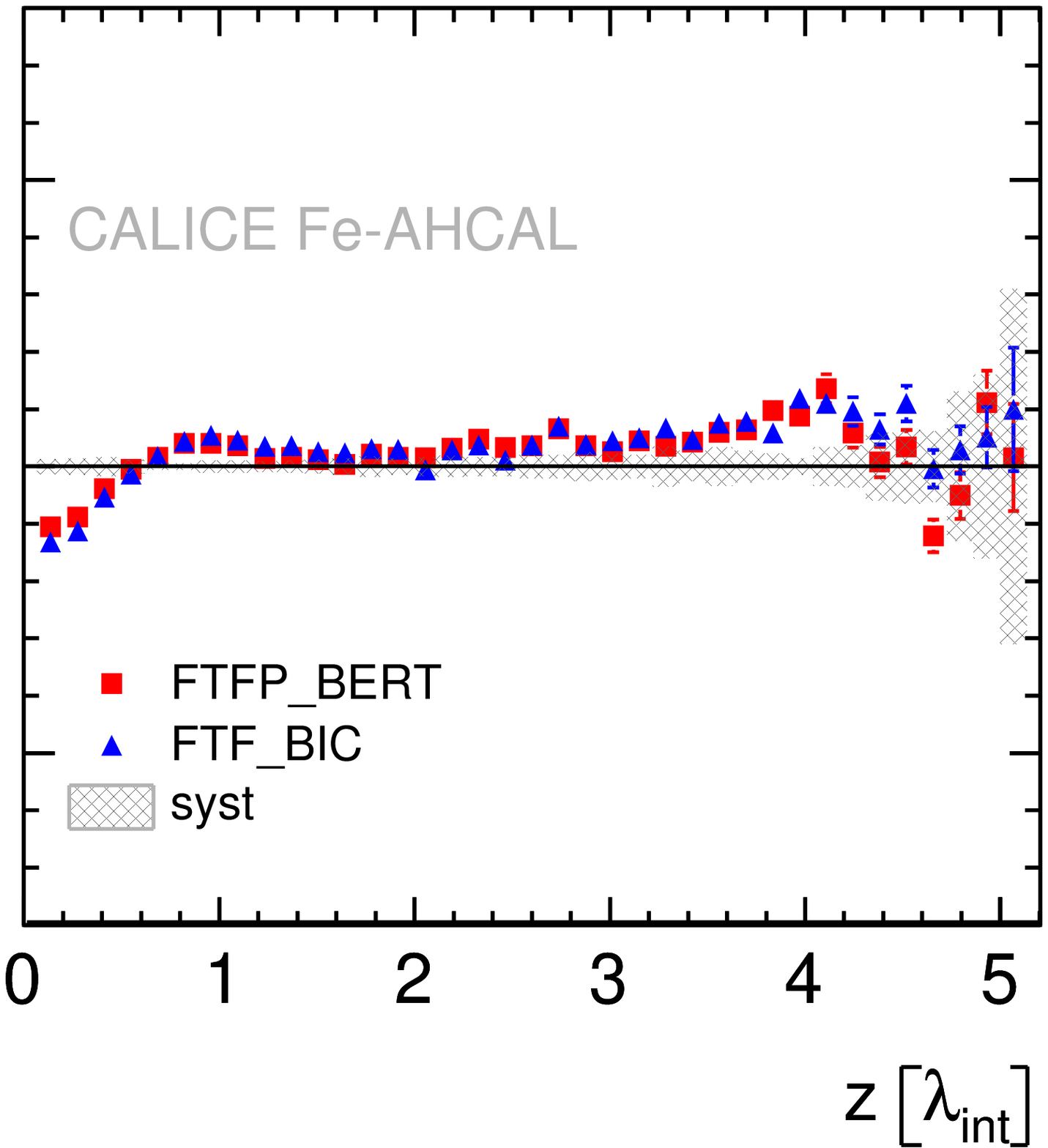}  \hspace{-5.3mm}
\includegraphics[trim=80 0 10 10, clip, height=6cm]{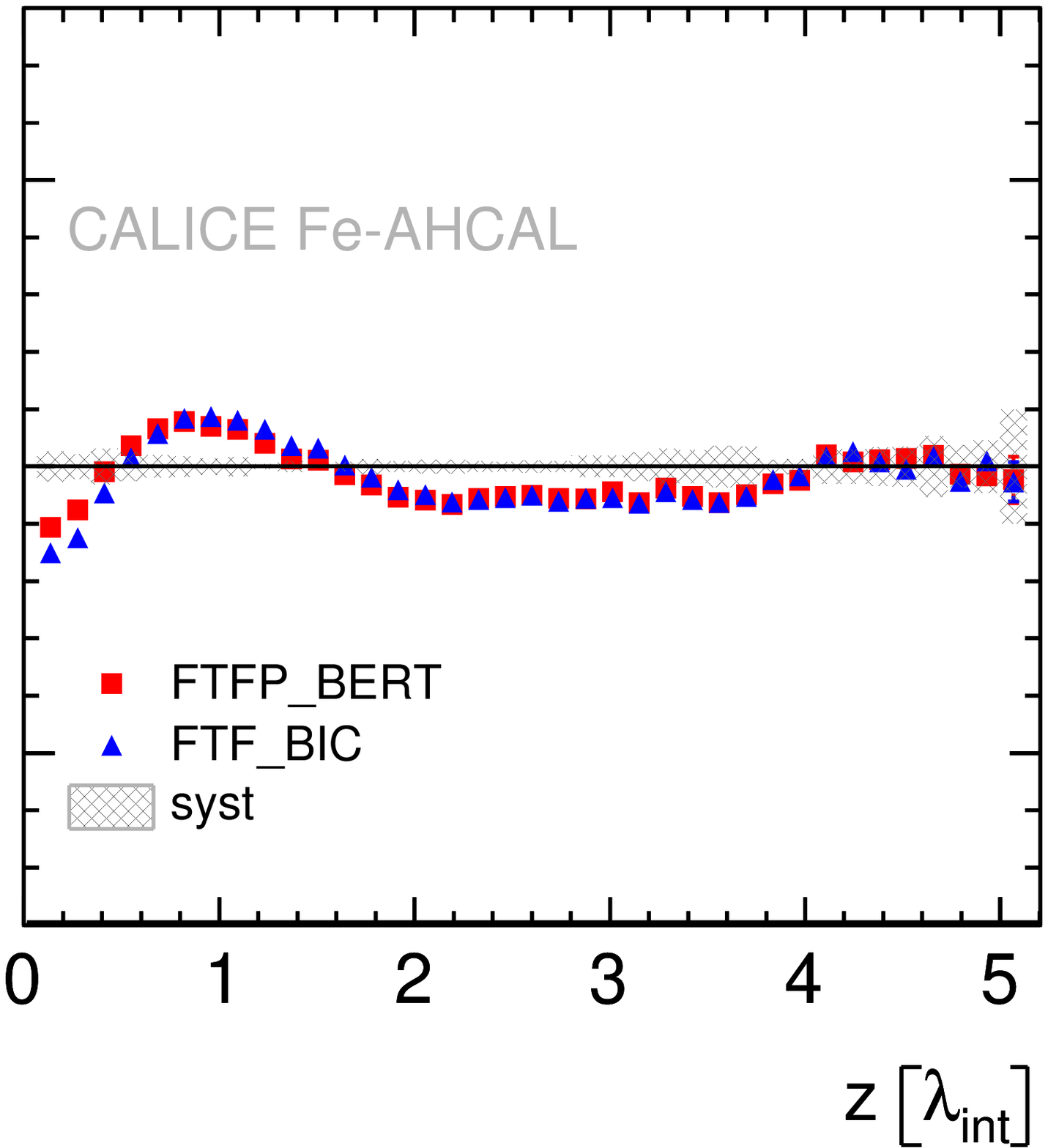}
\vspace{-10mm}
}
\centerline{
\hspace{-1.5mm}
\includegraphics[trim=0 0 10 10, clip, height=6cm]{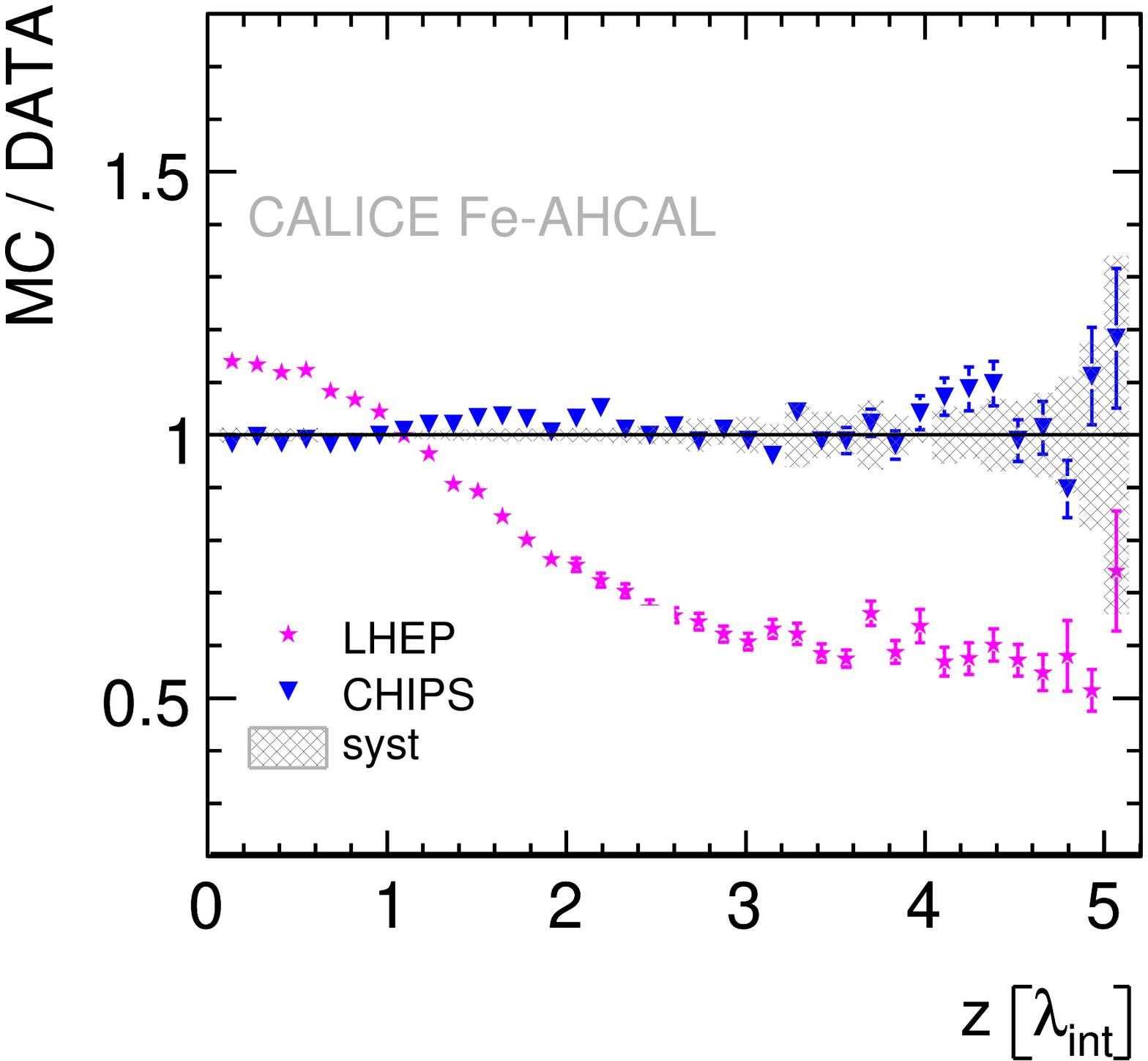}  \hspace{-5.mm}
\includegraphics[trim=80 0 10 10, clip, height=6cm]{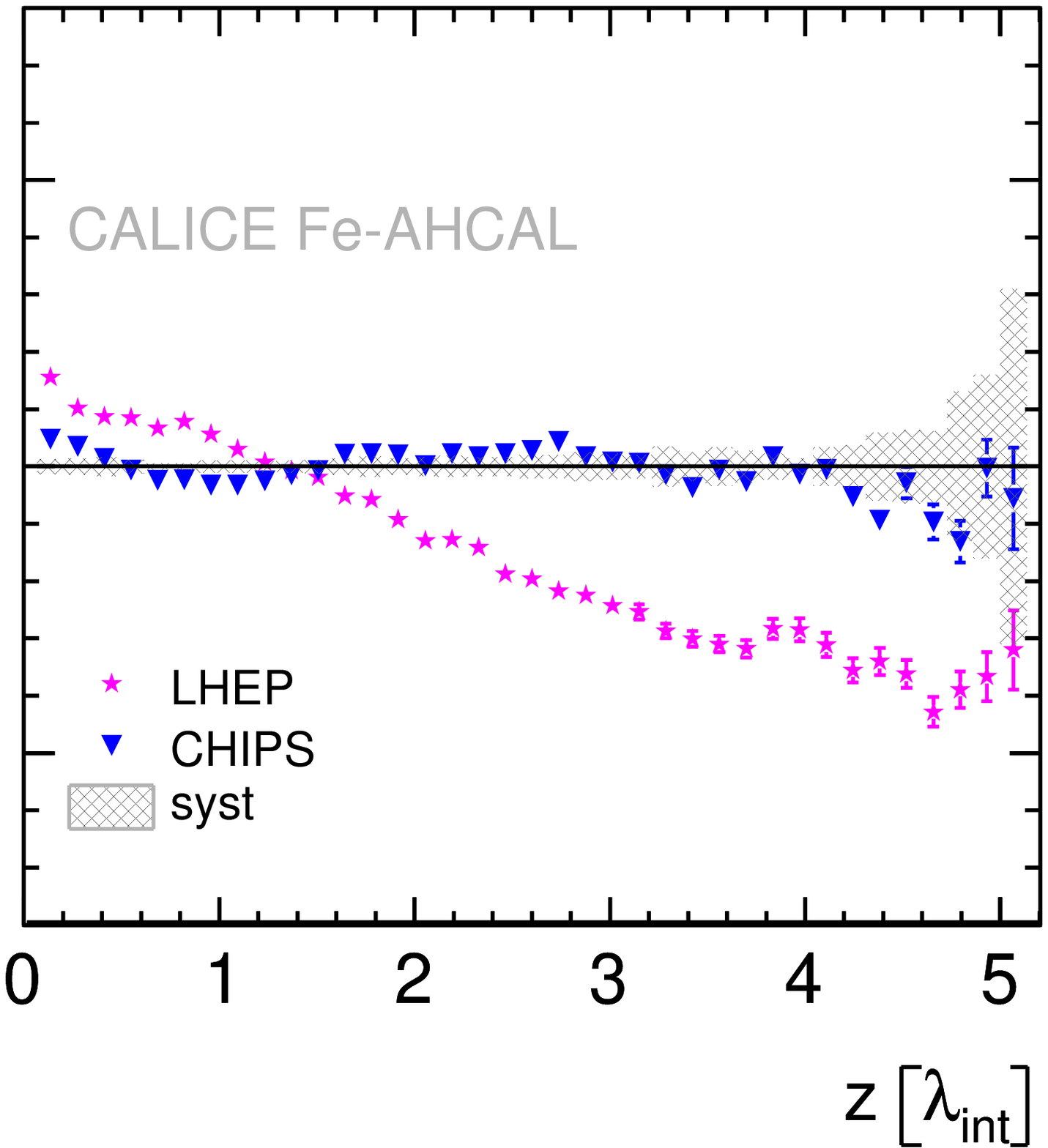}  \hspace{-5.3mm}
\includegraphics[trim=80 0 10 10, clip, height=6cm]{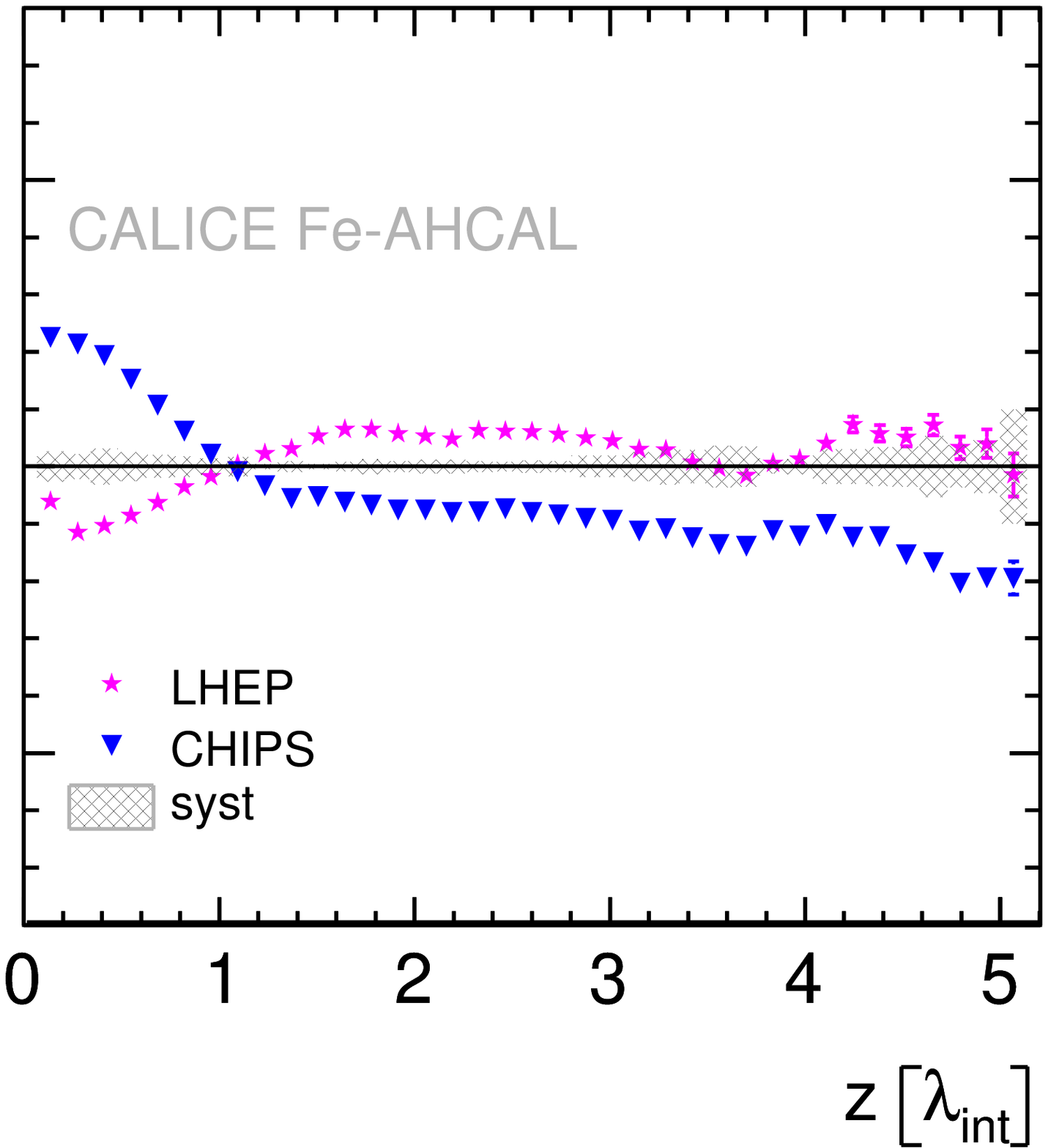} 
}
\caption[Longitudinal shower profiles from shower starting point in data and Monte Carlo]{\sl Mean longitudinal shower profiles from shower starting point for 8\,GeV (left column), 18\,GeV (center column) and 80\,GeV (right column) pions. First row: For data (circles) and for the \texttt{FTFP\_BERT} physics list (histogram). Second to fourth rows: Ratio between Monte Carlo and data for several physics lists. All profiles are normalized to unity. The grey area indicates the systematic uncertainty on data. $\langle \mathrm{E}_{\mathrm{rec}} \rangle / \Delta \lambda_{\mathrm{I}}$ is the average deposited energy in a $\Delta \lambda_{\mathrm{I}}$ thick transverse section of the calorimeter. $\mathrm{z}$ is the longitudinal coordinate, expressed in units of $\lambda_\mathrm{I}$.}
\label{figure:longitudinalprofilestart}
\end{figure}

Thanks to the fine longitudinal segmentation of the AHCAL into 38 layers, the longitudinal profile of hadronic showers can be investigated with an excellent accuracy. In particular, the fluctuations in the shower starting point, which are large since they are due to the statistical behavior of one single particle, can be separated from the intrinsic longitudinal shower development. This is shown in Fig.~\ref{figure:showerstartdistr} (right), where the shower profile relative to the calorimeter front face and the shower profile relative to the measured shower starting point are compared. The latter is shorter, since only the effective length of the hadronic shower is taken into account, excluding the path of the primary beam particle before the first hard interaction. In addition, the layer-to-layer fluctuations due to calibration uncertainties are smeared out into a smoother distribution, as different showers extend over different regions of the calorimeter. In the following, the results are therefore discussed for profiles relative to the measured shower starting point, shown in Fig.~\ref{figure:longitudinalprofilestart}. However, the trend of the agreement with data of the physics lists considered is independent from the definition of the profiles. The unit chosen to express the longitudinal development of showers is the nuclear interaction length $\lambda_{\rm I}$. For the AHCAL it has been estimated to correspond to 231.1\,mm.

Fig.~\ref{figure:longitudinalprofilestart} shows the longitudinal shower profiles for pions of 8\,GeV, 18\,GeV and 80\,GeV, in data and Monte Carlo. The shown distributions are average distributions over all selected events. The profiles are normalized to unity. The normalization allows the decoupling of the pure description of the spatial development of showers from the energy response already discussed above. The three energy points have been chosen in order to partially disentangle the different contributions from low, medium and high energy models (Fig.~\ref{figure:physicslists}). The energy contribution from electrons, positrons, pions and protons produced during the showers can be determined for the simulated events and is also shown in the graphs. The technique used for this decomposition is described in~\cite{Alexthesis}. This additional information helps to understand which physics processes contribute most in which phase of the shower development. In the first layers electrons and positrons contribute about equally to hadrons to the energy deposition. In the region around the shower maximum the deposition of electromagnetic energy dominates while in the tails the energy depositions by hadrons and electrons are again about equal.

The uncertainty on the measurement of the profiles has been evaluated creating 6 profiles, for showers starting in 6 different layers of the AHCAL, namely the first 6 layers. Due to the exponential distribution of the measured layer of the first hard interaction (Fig.~\ref{figure:showerstartdistr}), the 6 profiles considered dominate the final measurement of the longitudinal shower profiles, which has been performed without any selection on the shower starting point. The first (second, ...) bin of the profiles for showers starting in the first layer of the AHCAL, corresponds to the first (second, ...) layer of the AHCAL. The first (second, ...) bin of the profiles for showers starting in the second layer of the AHCAL, corresponds to the second (third, ...) layer of the AHCAL. An analogous correspondence between the bins of the profiles and the AHCAL layers is true for the remaining profiles measured for showers starting in layers 3-6 of the AHCAL. The variance $var_j$ of the energy desposited in the bin $j$ of the profiles has been evaluated as the variance between the 6 measurements. Some bins in the tails are not covered by all the 6 measurements. For instance, showers starting in the second layer of the AHCAL extend over at most 37 AHCAL layers and give no contribution to the bin 38 of the shower profiles. 

The development of showers does not depend on the measured layer of the first interaction and with an ideal calorimeter the 6 considered profiles would not differ. The measured variance is an inclusive measurement of the uncertainties associated to the energy measurement, such as bad calibrations and saturation effects. Thanks to the high statistics of several $10^4$ events used to create the profiles, the variance is dominated by systematic uncertainties, while statistical fluctuations are expected to be negligible.

For the final measurement of the longitudinal shower profiles no selection on the measured layer of the first hard interaction has been applied and $n_j$ different profiles are averaged to calculate the average energy deposition in the $j$-th layer, namely 37 for the first layer, 36 for the second, and so on for the remaining layers. The final uncertainty to be associated to the energy measurement in the layer $j$ is given by:
\begin{equation}
\sigma_{E_j} = \sqrt{\frac{var_j}{n_j}}.
\end{equation}
This uncertainty is shown with a grey area in Fig.~\ref{figure:longitudinalprofilestart}.

The position of the shower maximum is in general quite well reproduced by Monte Carlo models. In the first part of the showers \verb=QGSP_BERT=, \verb=QGSP_FTFP_BERT= and \verb=QBBC= agree with data at the 5-10\% level at 8\,GeV and 18\,GeV. The same is true for the Fritiof-based physics lists and for \verb=CHIPS=. Larger deviations are present in the tails of the showers, but they remain within systematic uncertainties. At 80\,GeV the energy deposition in the shower maximum is generally underestimated by Monte Carlo simulations, with discrepancies up to 20\%. At 80\,GeV \verb=CHIPS= shows a moderate tendency to prefer compact showers, overestimating the energy deposition in the first part of the shower and underestimating the tail. A similar trend is exhibited by \verb=LHEP= at 8\,GeV and 18\,GeV, though with a more pronounced disagreement with data, up to 50\% at 8\,GeV.

\begin{figure}
\centerline{	
     \subfigure{ \label{} \includegraphics[trim=0 0 45 0, clip, height=8cm]{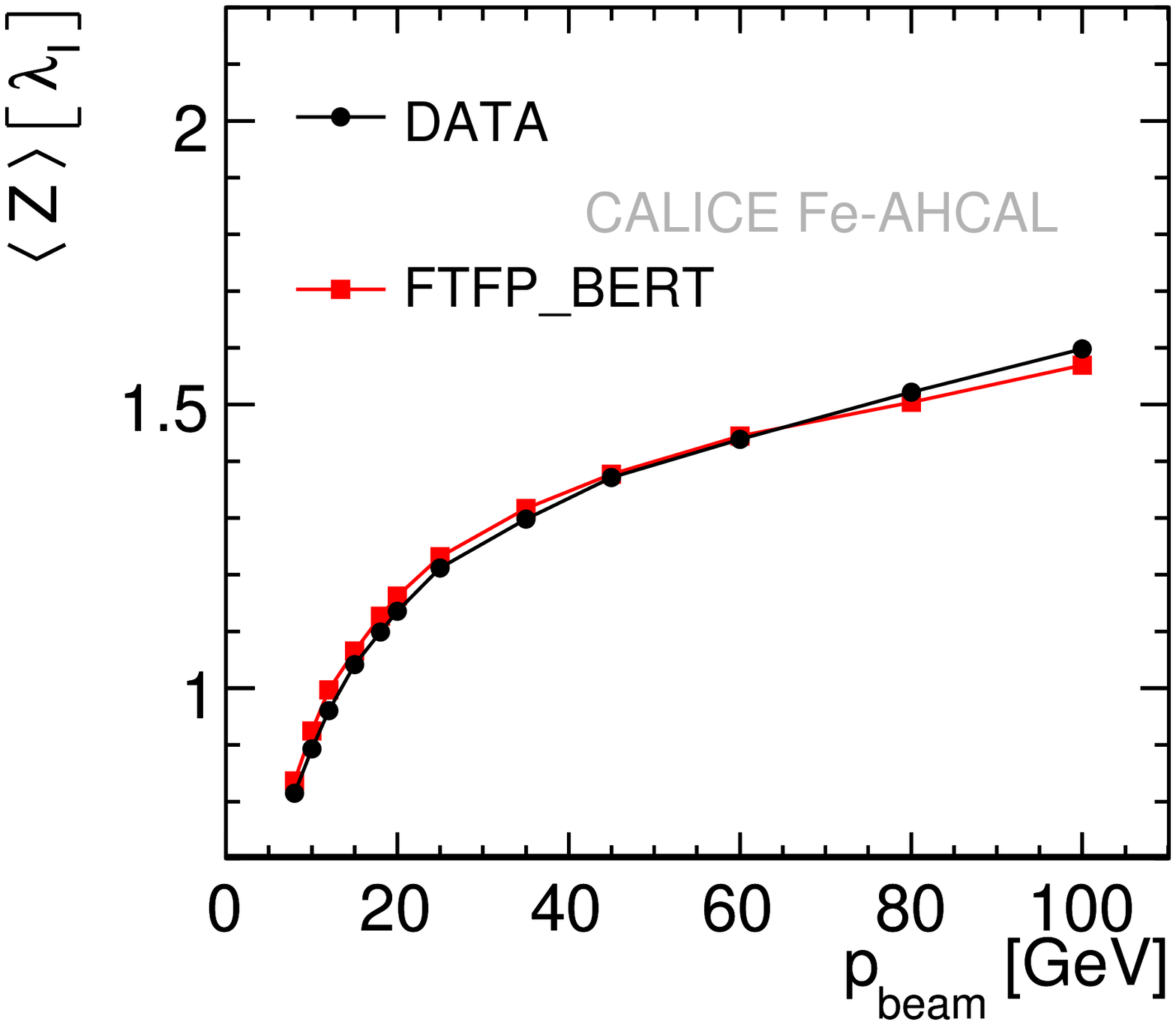} } \hspace{-4.5mm}
     \subfigure{ \label{} \includegraphics[trim=0 0 45 0, clip, height=8cm]{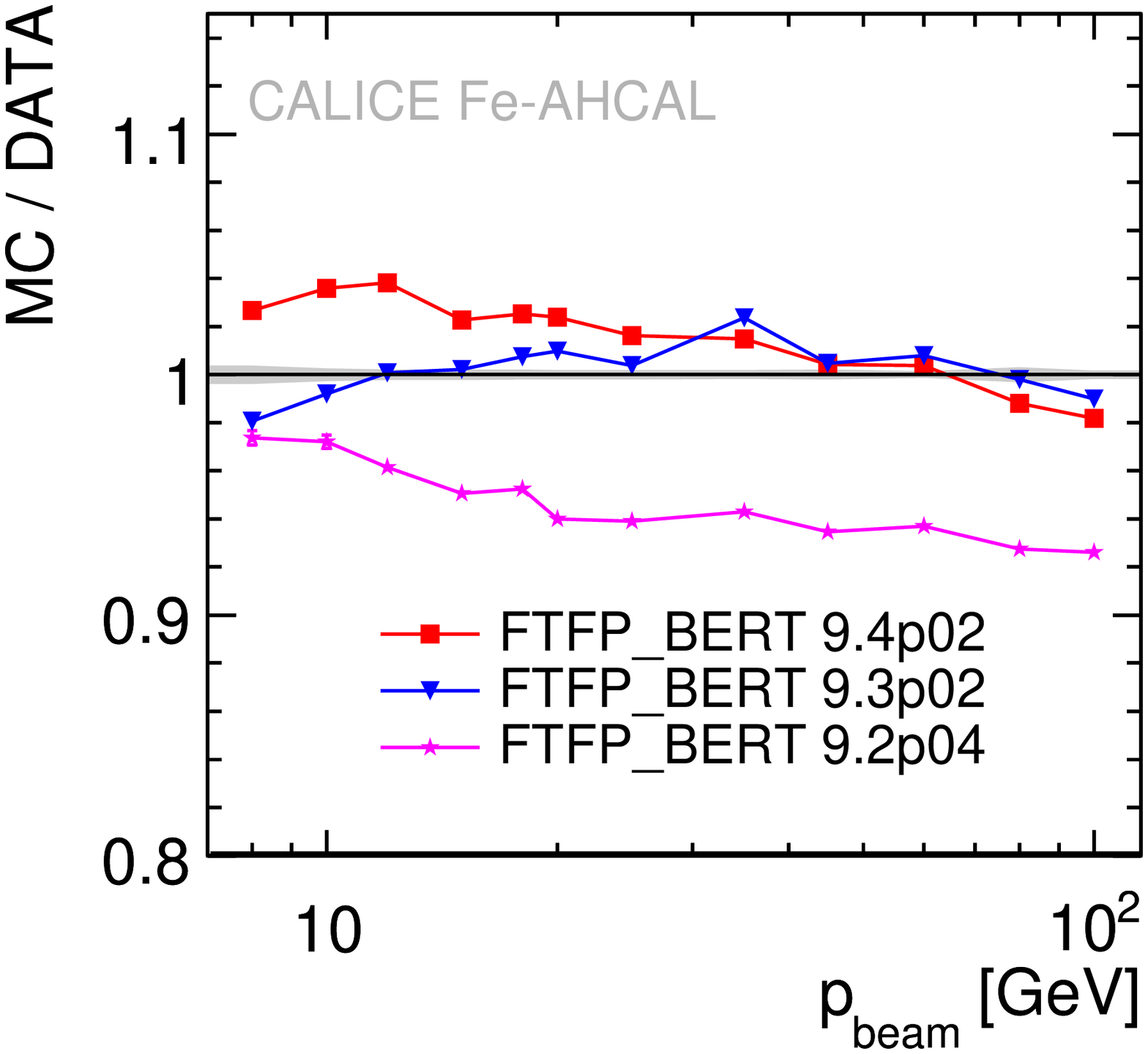} } \hspace{5mm}	
}
\centerline{
     \subfigure{ \label{} \includegraphics[trim=0 0 45 0, clip, height=6cm]{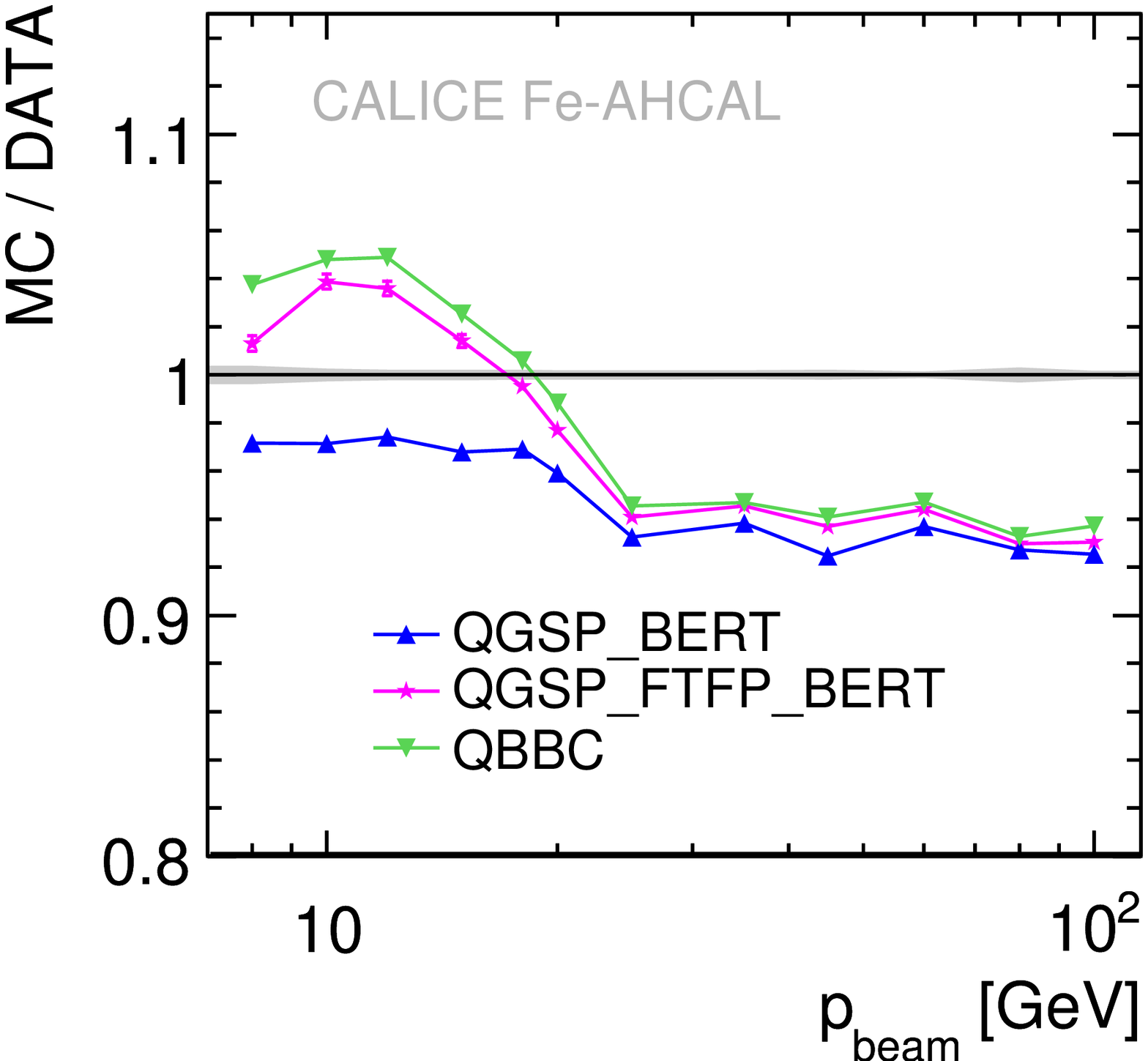} } \hspace{-4.5mm}
     \subfigure{ \label{} \includegraphics[trim=100 0 45 0, clip, height=6cm]{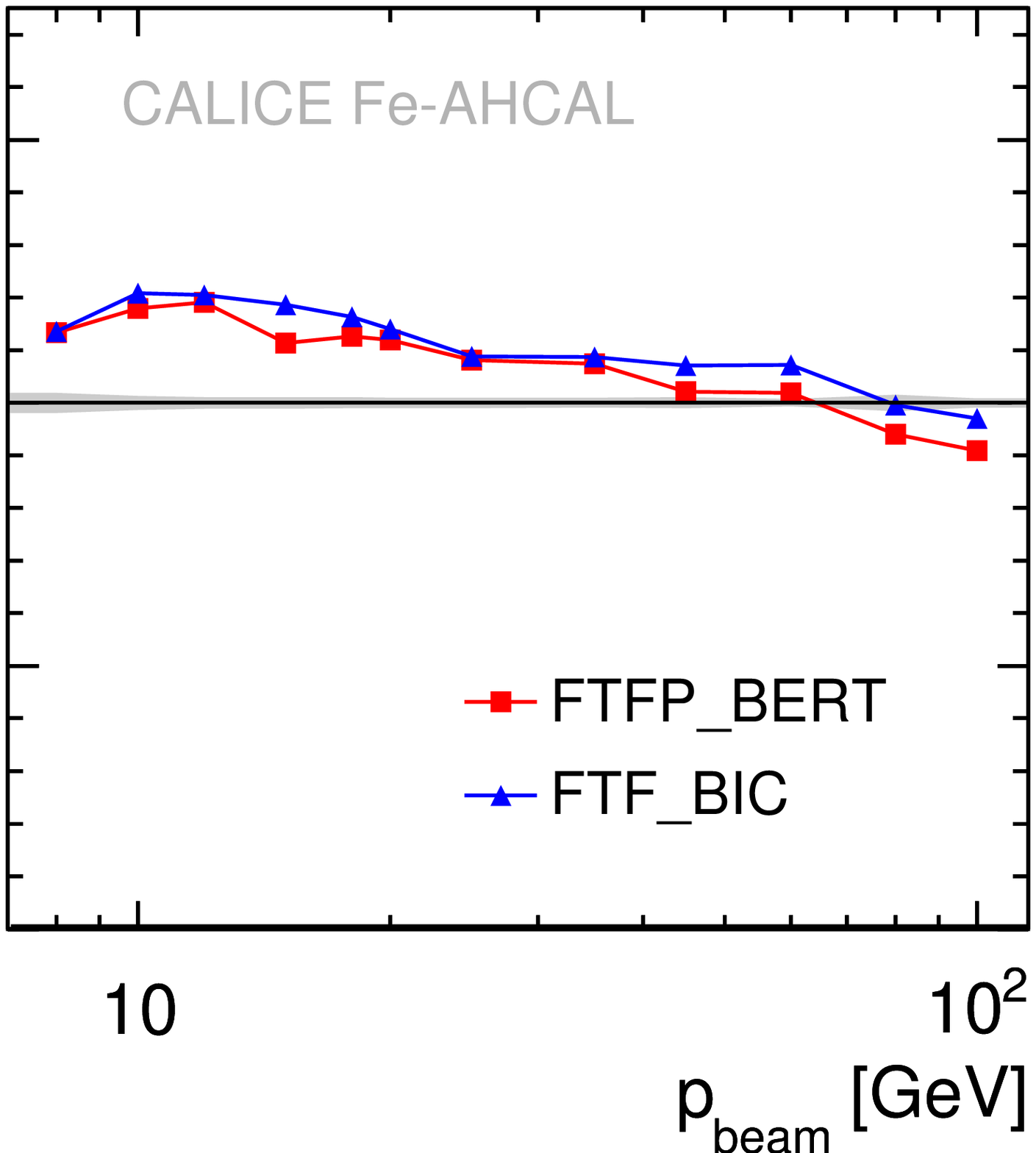} } \hspace{-4.5mm}
     \subfigure{ \label{} \includegraphics[trim=100 0 45 0, clip, height=6cm]{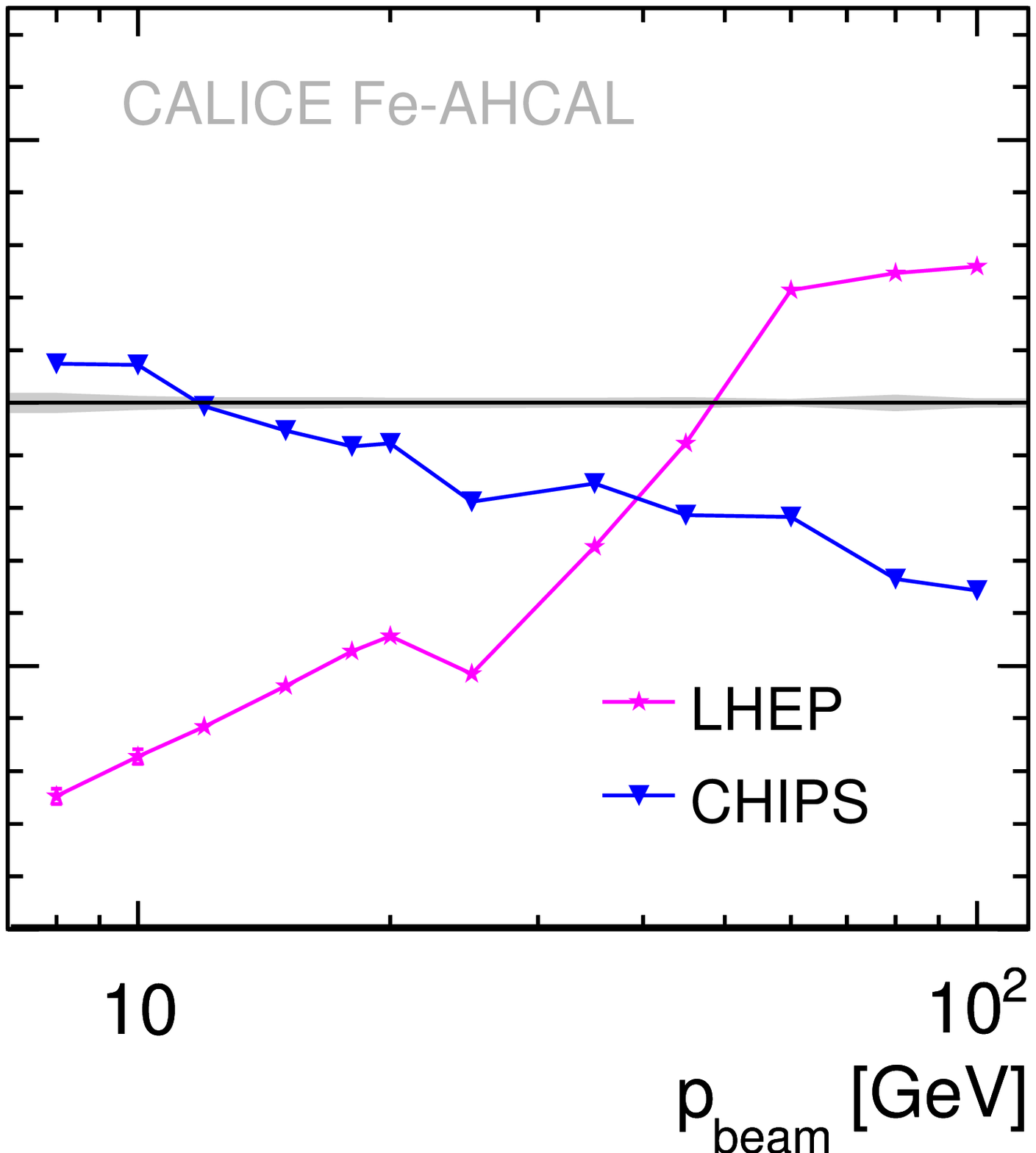} }
}
\caption[Center of gravity in the longitudinal direction]{\sl Summary of the measurement of the center of gravity in the longitudinal direction, for pions in the AHCAL. Top, left: For data and for the \texttt{FTFP\_BERT} physics list. Top, right: Ratio between Monte Carlo and data using the \texttt{FTFP\_BERT} physics list with different versions of \textsc{Geant4}. Bottom: Ratio between Monte Carlo and data for several physics lists. The gray band in the ratios represents the statistical uncertainty on data.}
\label{fig:Z}
\end{figure}

\begin{figure}
\centerline{	
     \subfigure{ \label{} \includegraphics[trim=0 0 45 0, clip, height=8cm]{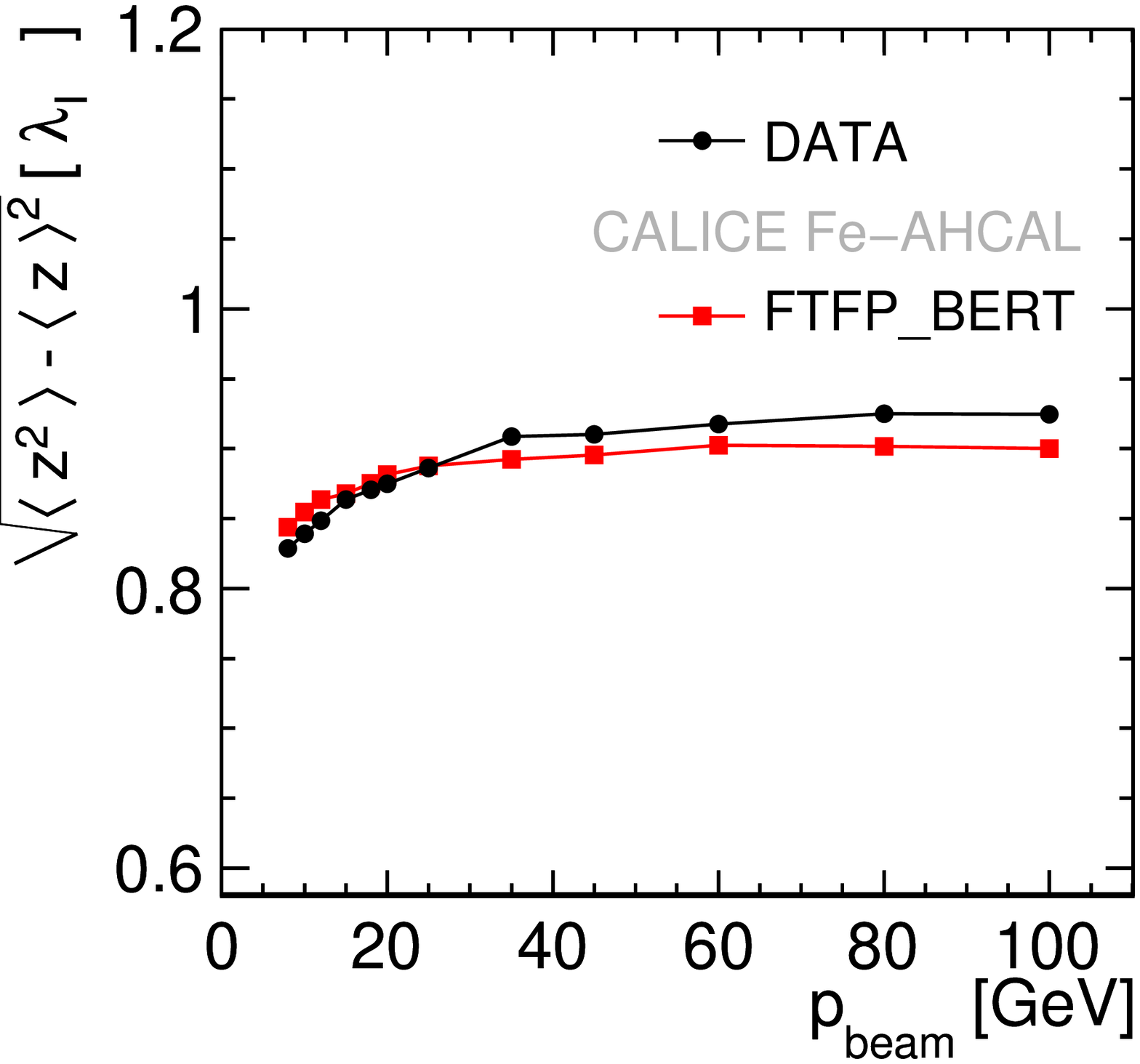} } \hspace{-4.5mm}
     \subfigure{ \label{} \includegraphics[trim=0 0 45 0, clip, height=8cm]{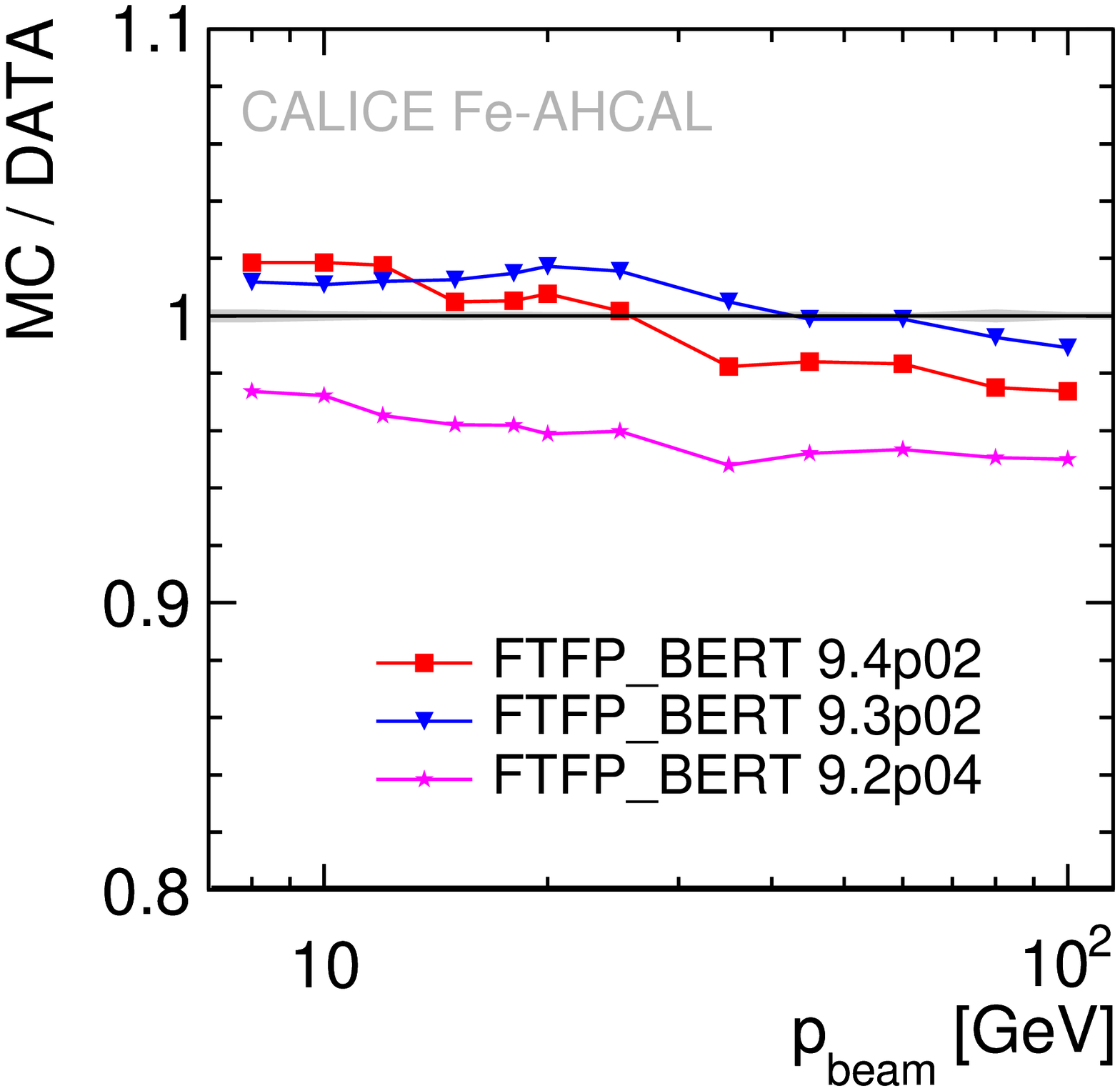} } \hspace{5mm}	
}
\centerline{
     \subfigure{ \label{} \includegraphics[trim=0 0 45 0, clip, height=6cm]{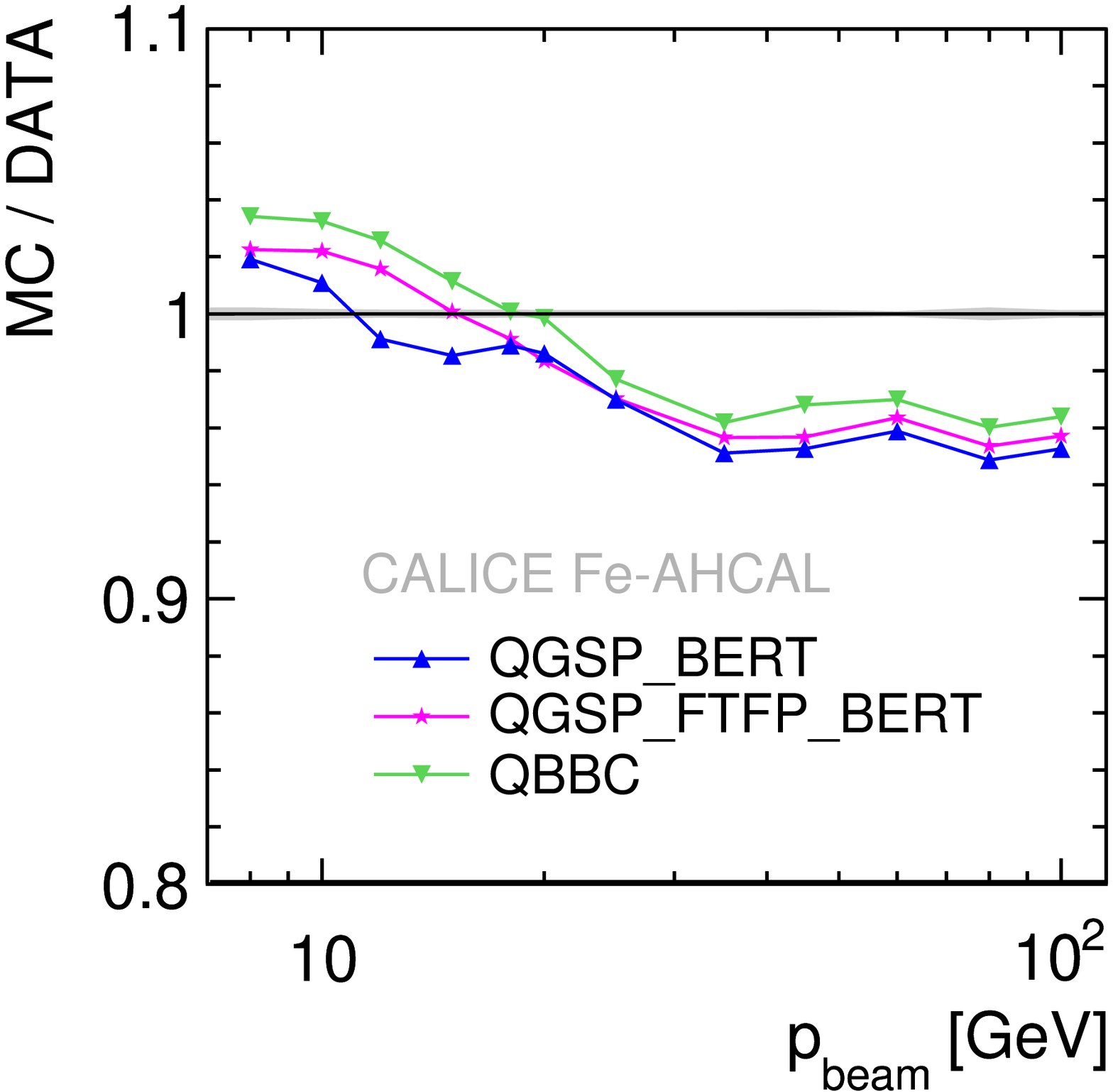} } \hspace{-4.5mm}
     \subfigure{ \label{} \includegraphics[trim=100 0 45 0, clip, height=6cm]{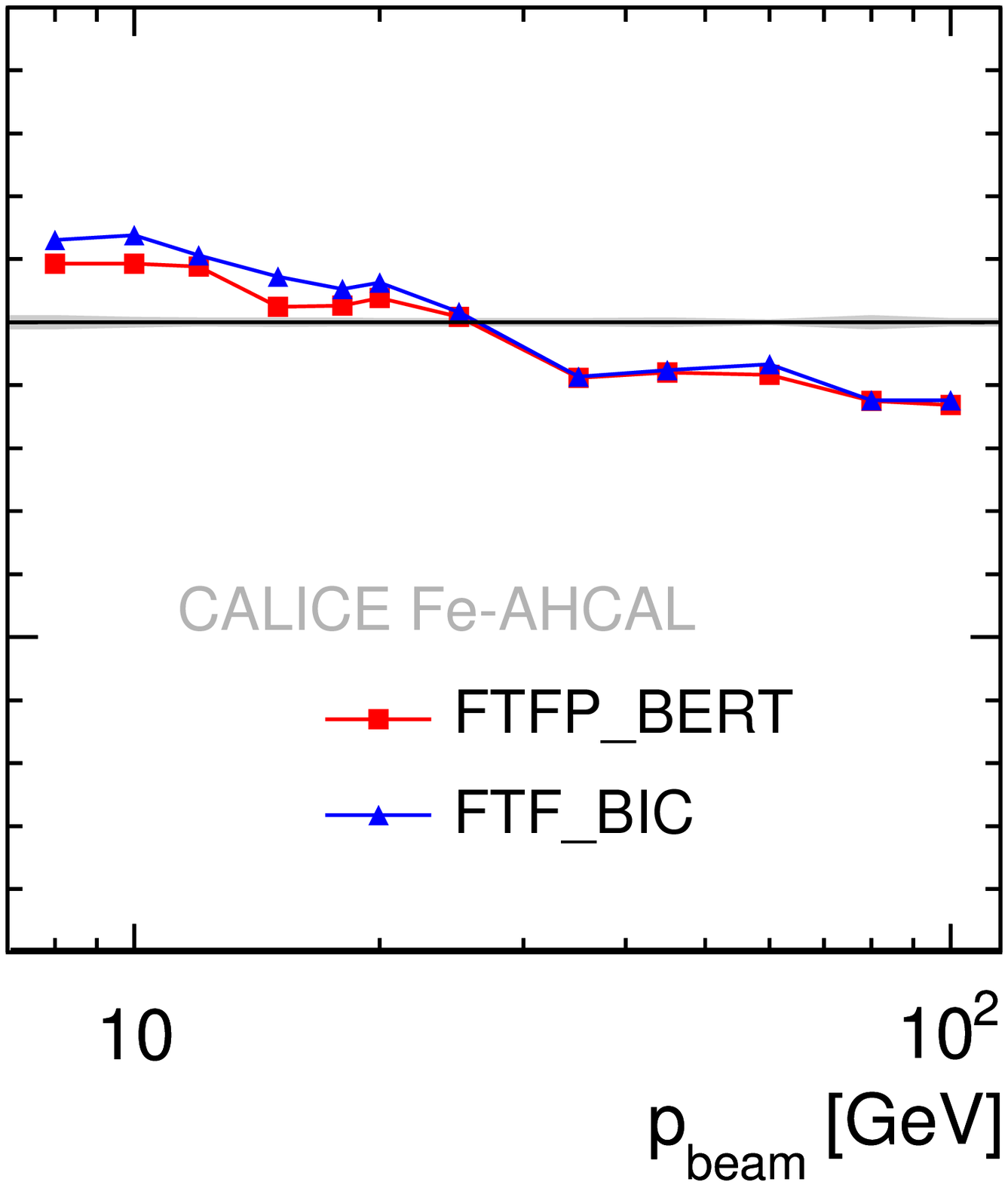} } \hspace{-4.5mm}
     \subfigure{ \label{} \includegraphics[trim=100 0 45 0, clip, height=6cm]{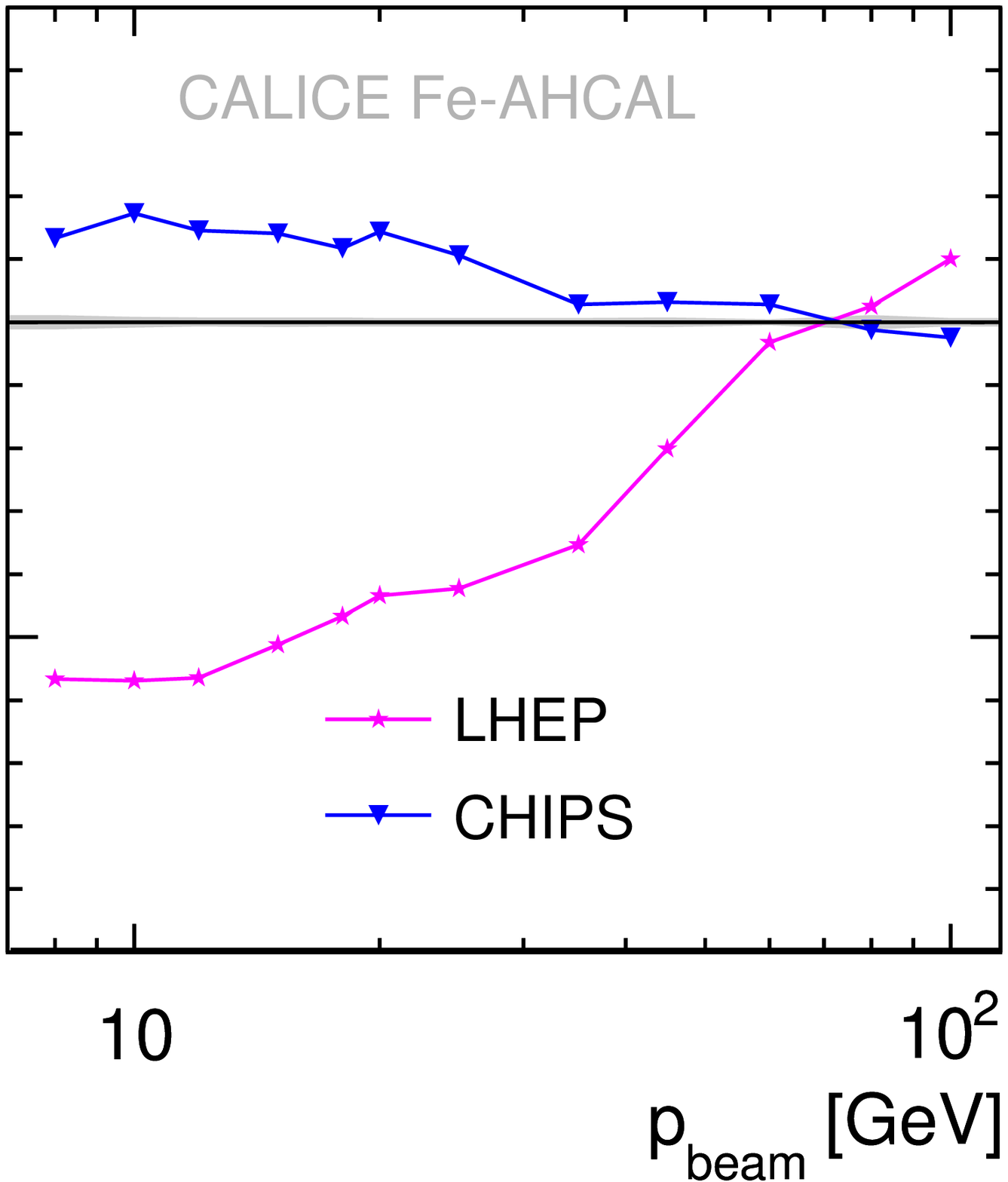} }
}
\caption[The standard deviation of the longitudinal shower profile]{\sl Summary of the measurement of the standard deviation of the longitudinal shower profile, for pions in the AHCAL. Top, left: For data and for the \texttt{FTFP\_BERT} physics list. Top, right: Ratio between Monte Carlo and data using the \texttt{FTFP\_BERT} physics list with different versions of \textsc{Geant4}. Bottom: Ratio between Monte Carlo and data for several physics lists. The gray band in the ratios represents the statistical uncertainty on data.}
\label{fig:L2}
\end{figure}

The energy dependence of the center of gravity and of the standard deviation of the longitudinal shower profiles are shown in Fig.~\ref{fig:Z} and Fig.~\ref{fig:L2}, respectively. The center of gravity is defined as the energy weighted mean of the hit longitudinal coordinate along the shower axis:
\begin{equation}
\langle z \rangle = \frac{\sum_i E_i \cdot z_i}{\sum_i E_i},
\end{equation}
while the standard deviation is defined as:
\begin{equation}
\sqrt{\langle z^2 \rangle - \langle z \rangle^2} = \sqrt{\frac{\sum_i E_i \cdot \left( z_i - \langle z \rangle \right)^2}{\sum_i E_i}},
\end{equation}
where $z_i$ is the longitudinal coordinate of the cell $i$ and $E_i$ is the energy measured in that cell.

The center of gravity is located between 0.8 and 1.6 interaction lengths and exhibits the expected logarithmic increase with energy. The standard deviation only mildly increases with energy from about 0.825 to about 0.925\,$\lambda_I$.

All physics lists except \verb=LHEP= show a similar energy-dependent behavior in the description of  the center of gravity $\langle z \rangle$ of showers. The ratio Monte Carlo over data decreases with energy.  Overall the best behavior is shown by the physics lists based on the Fritiof model, which agree with data at about the 4\% level.  The \verb=LHEP= parameterized models show also for this observable the worst agreement with data. The disagreement is up to 15\% at 8\,GeV. A similar message is given by the standard deviation $\sqrt{\langle z^2 \rangle - \langle z \rangle^2}$. 

The changes in \texttt{FTFP\_BERT} with different \textsc{Geant4} versions are significant for both longitudinal observables and the agreement with data changes by a few percent for different versions. The ratio Monte Carlo over data for the center of gravity $\langle z \rangle$ gets generally worse when comparing versions 9.3 and 9.4, apart from the intermediate energy points between 30\,GeV and 50\,GeV, where the agreement is comparable. Both versions of the physics list show an improvement with respect to the older version 9.2.  The agreement with data for the standard deviation $\sqrt{\langle z^2 \rangle - \langle z \rangle^2}$ improves in the version 9.3 of  \textsc{Geant4}, with respect to the older version. The most recent version has a worse performance than the version 9.3 at energies greater than 20\,GeV, but is still significantly better than version 9.2.

\section{Radial Development}
\label{section:radial}

An accurate modeling of the transverse shower profile is particularly important for a successful development of particle flow algorithms, since it affects the degree of overlap between showers and therefore the efficiency in disentangling single particles within jets. Using the AHCAL, it is possible to reconstruct the radial development of showers with high precision, thanks to the fine lateral segmentation of the sensitive layers.

For each cell of the AHCAL, the radial distance to the incoming particle trajectory is determined as:
\begin{equation}
r_i = \sqrt{(x_i - x_0)^2 + (y_i - y_0)^2},
\label{equation:radialco}
\end{equation}
where $(x_i,y_i)$ are the coordinates of the center of the cell $i$ and $(x_0, y_0)$ is the position of the energy weighted shower center:
\begin{equation}
x_0 = \frac{\sum_i E_i \cdot x_i}{\sum_i E_i}\;\;\mathrm{and}\;\;y_0 = \frac{\sum_i E_i \cdot y_i}{\sum_i E_i},
\end{equation}
$E_i$ being the energy measured in the cell $i$.

\begin{figure}
\centerline{
\includegraphics[trim=0 0 10 0, clip, height=6cm]{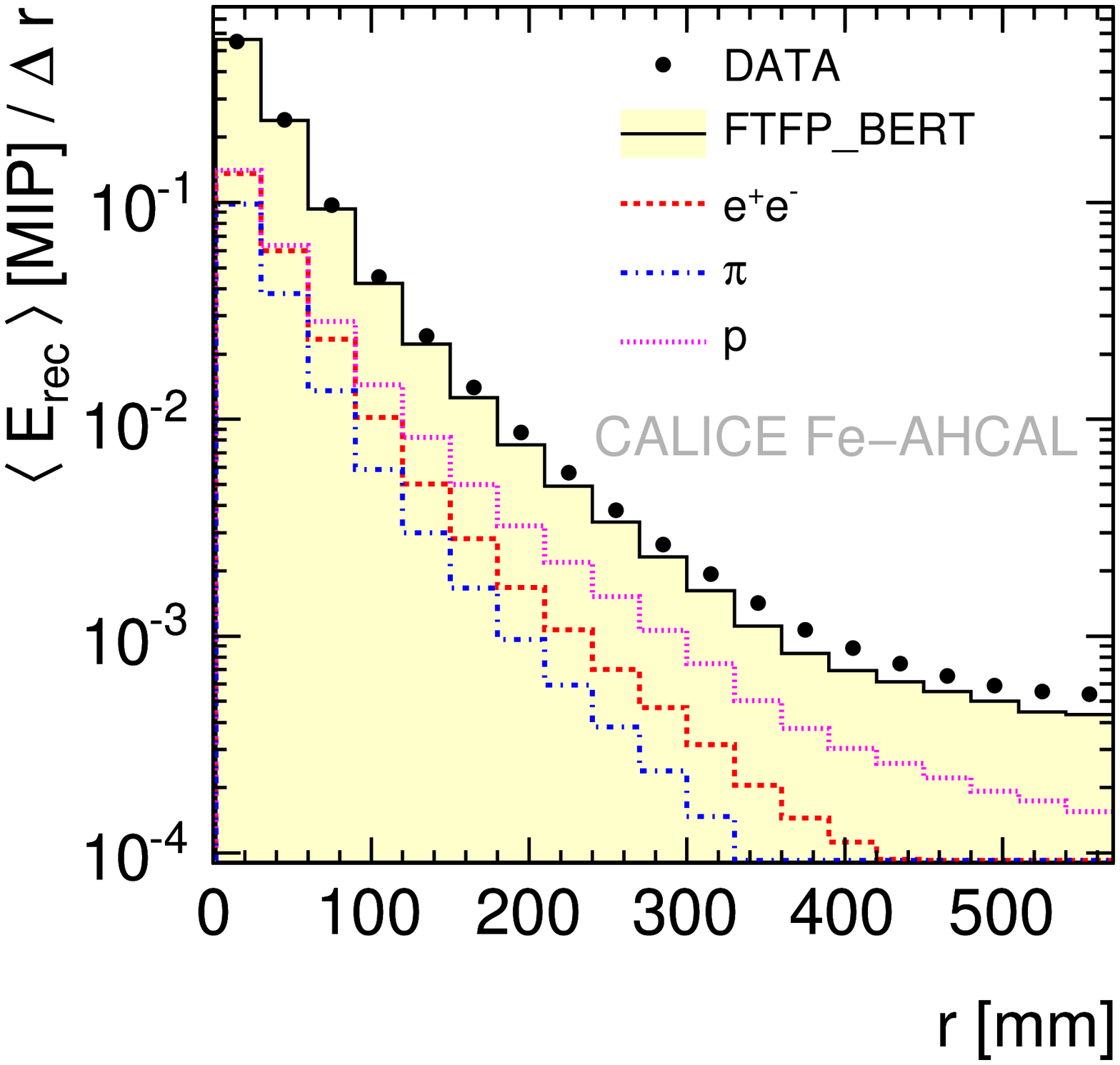}\hspace{-4.mm}
\includegraphics[trim=80 0 10 0, clip, height=6cm]{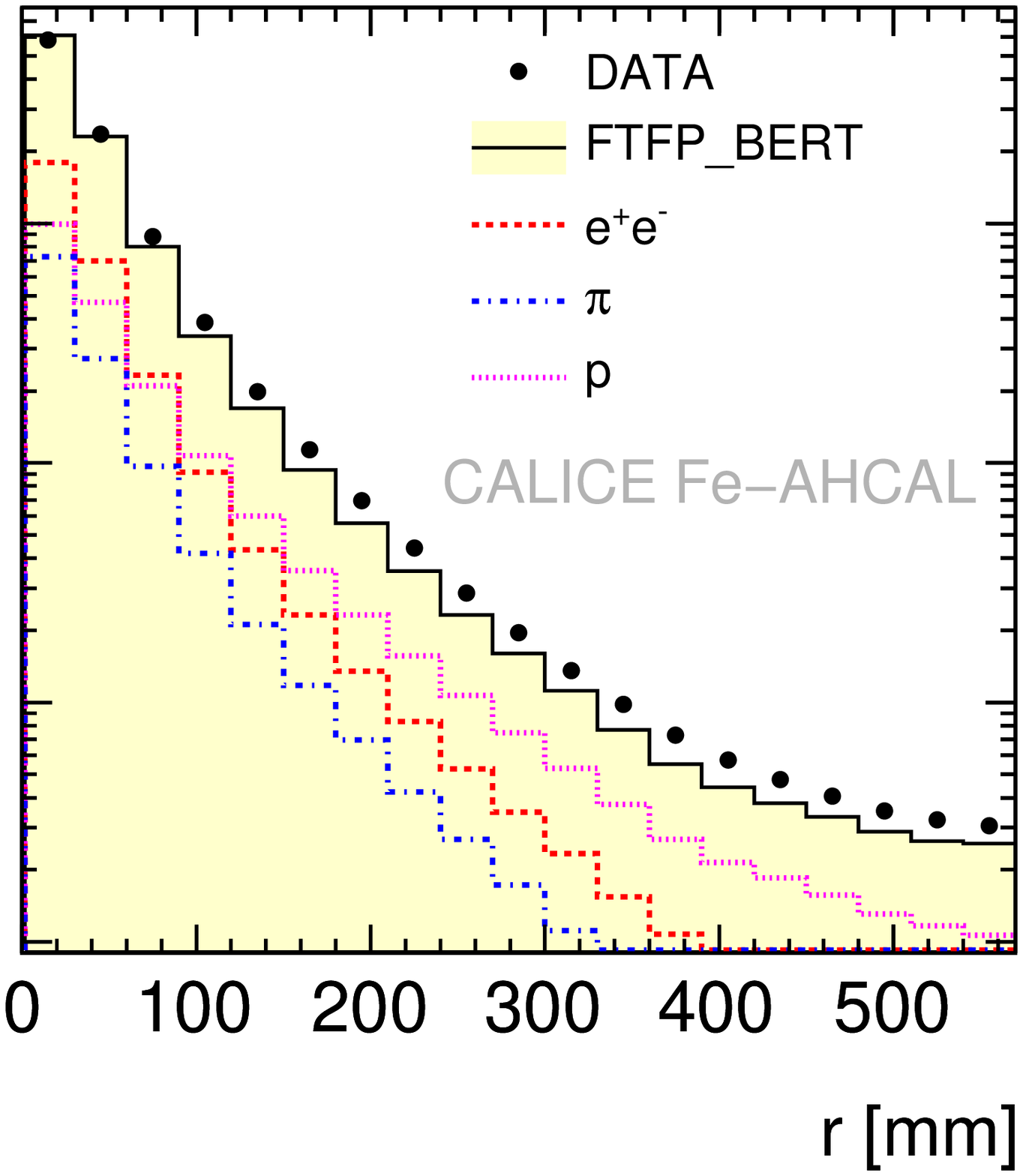}\hspace{-4.mm}
\includegraphics[trim=80 0 10 0, clip, height=6cm]{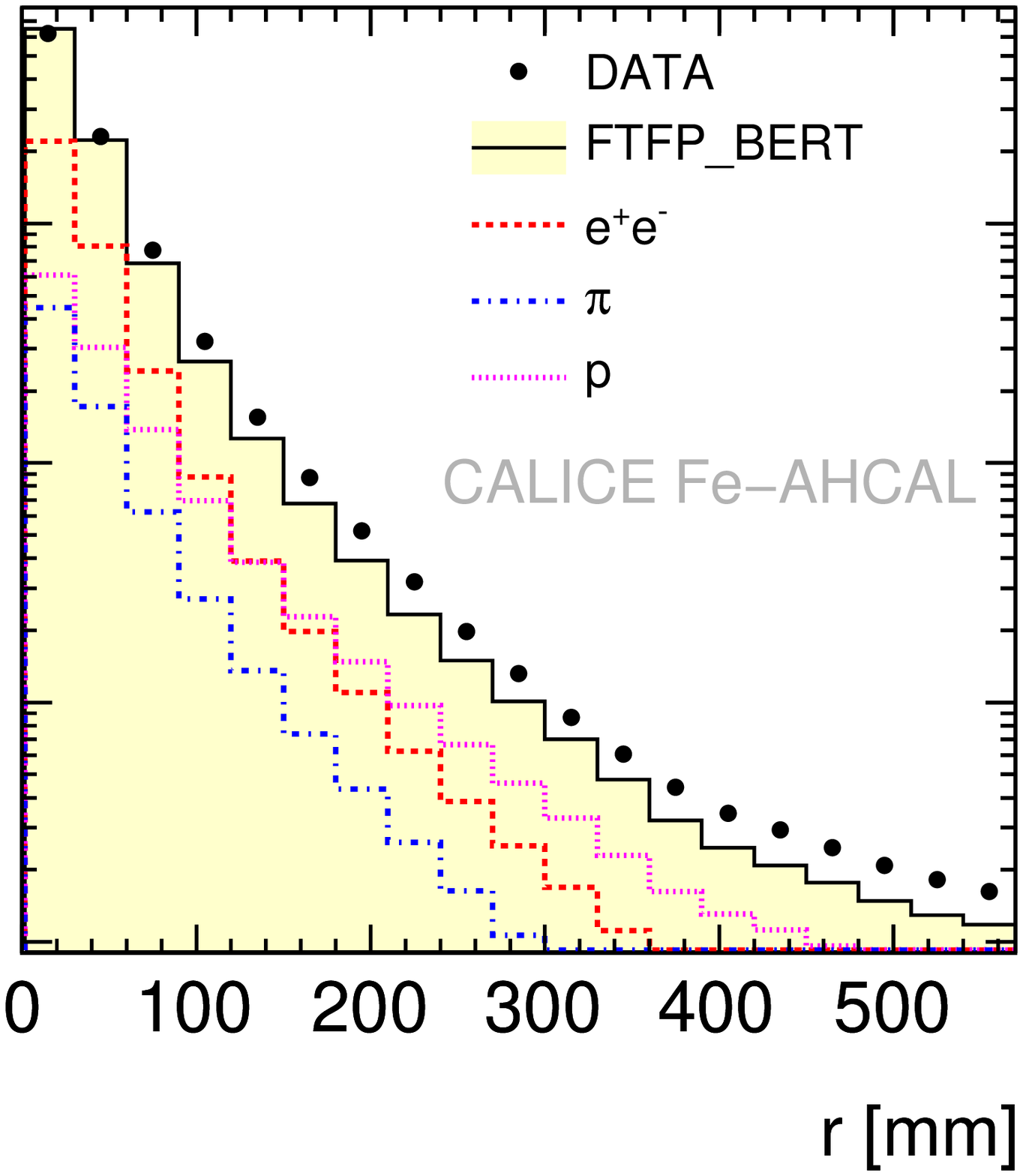}
}
\vspace{0mm}
\centerline{
\hspace{-1.5mm}
\includegraphics[trim=0 0 10 0, clip, height=6cm]{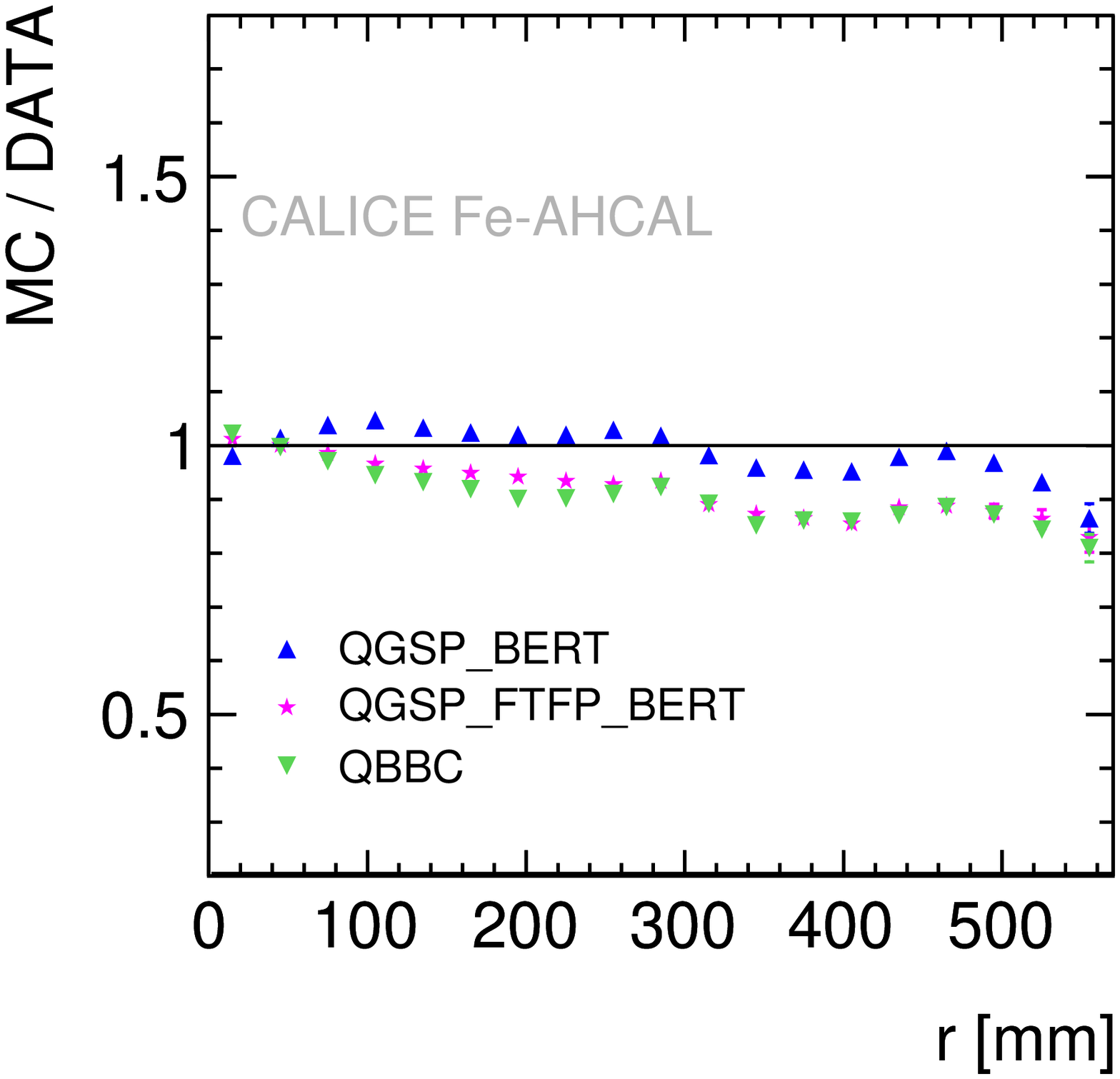}  \hspace{-5.mm}
\includegraphics[trim=80 0 10 0, clip, height=6cm]{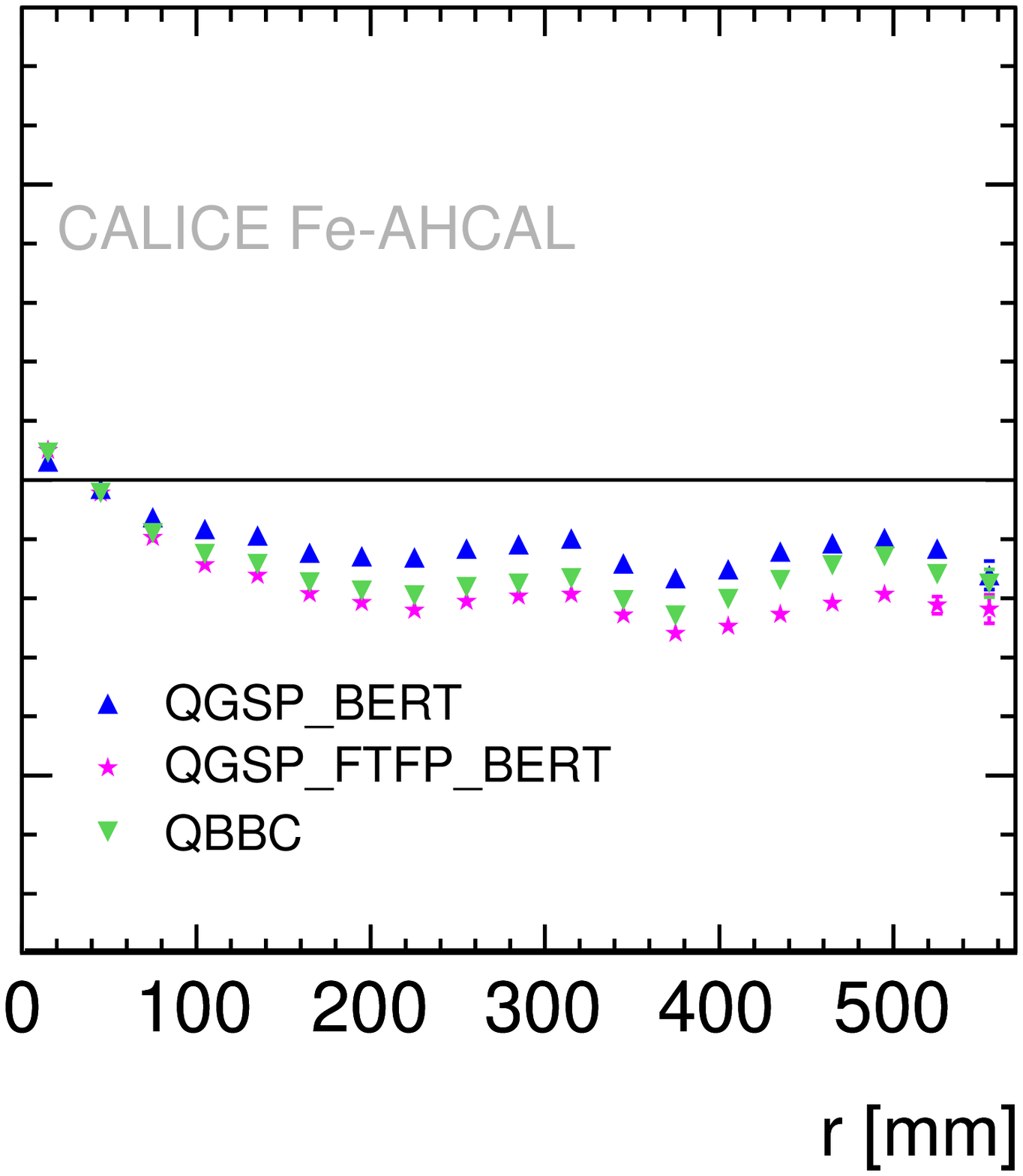}  \hspace{-5.3mm}
\includegraphics[trim=80 0 10 0, clip, height=6cm]{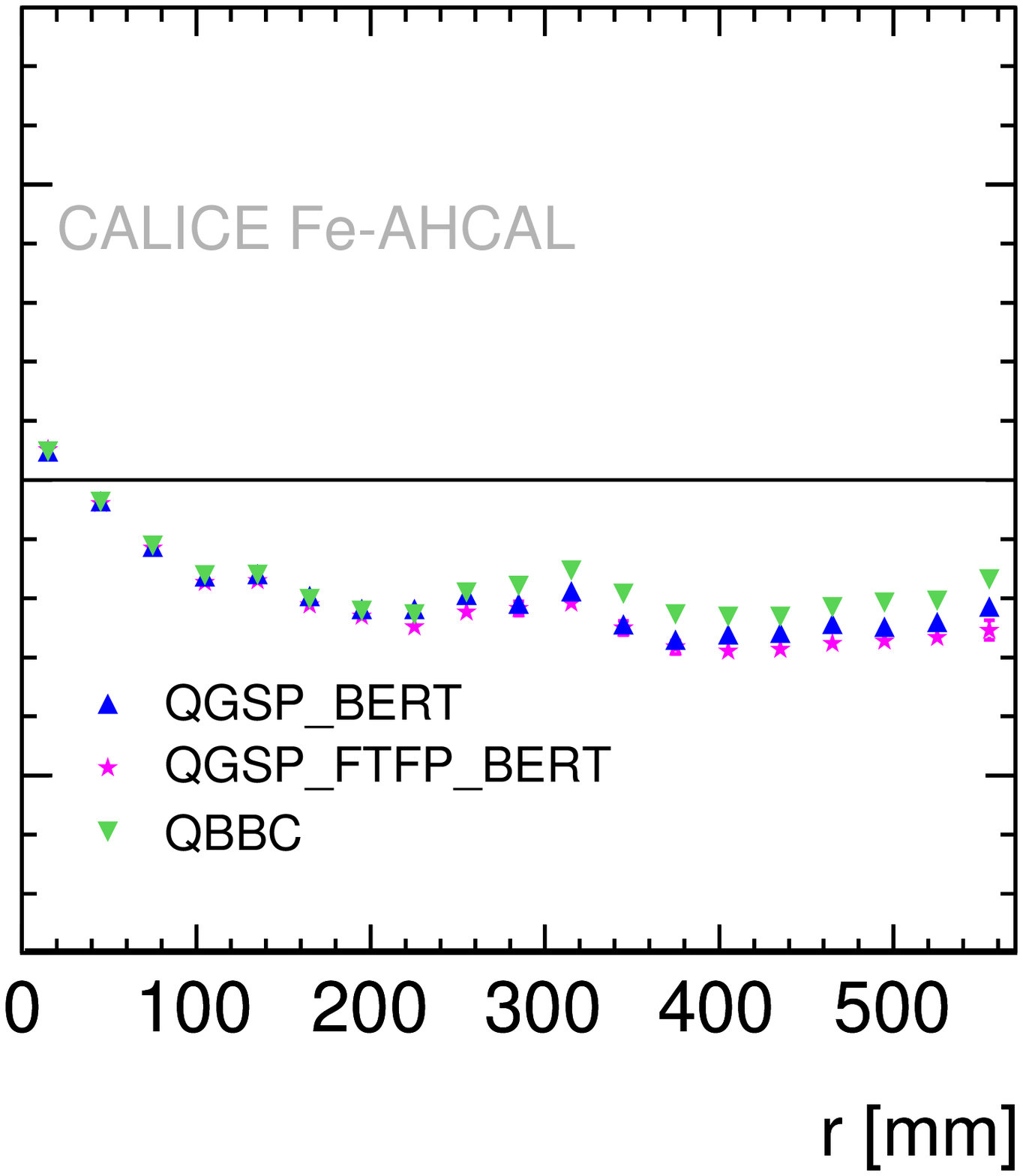} 
\vspace{-10mm}
}
\centerline{
\hspace{-1.5mm}
\includegraphics[trim=0 0 10 0, clip, height=6cm]{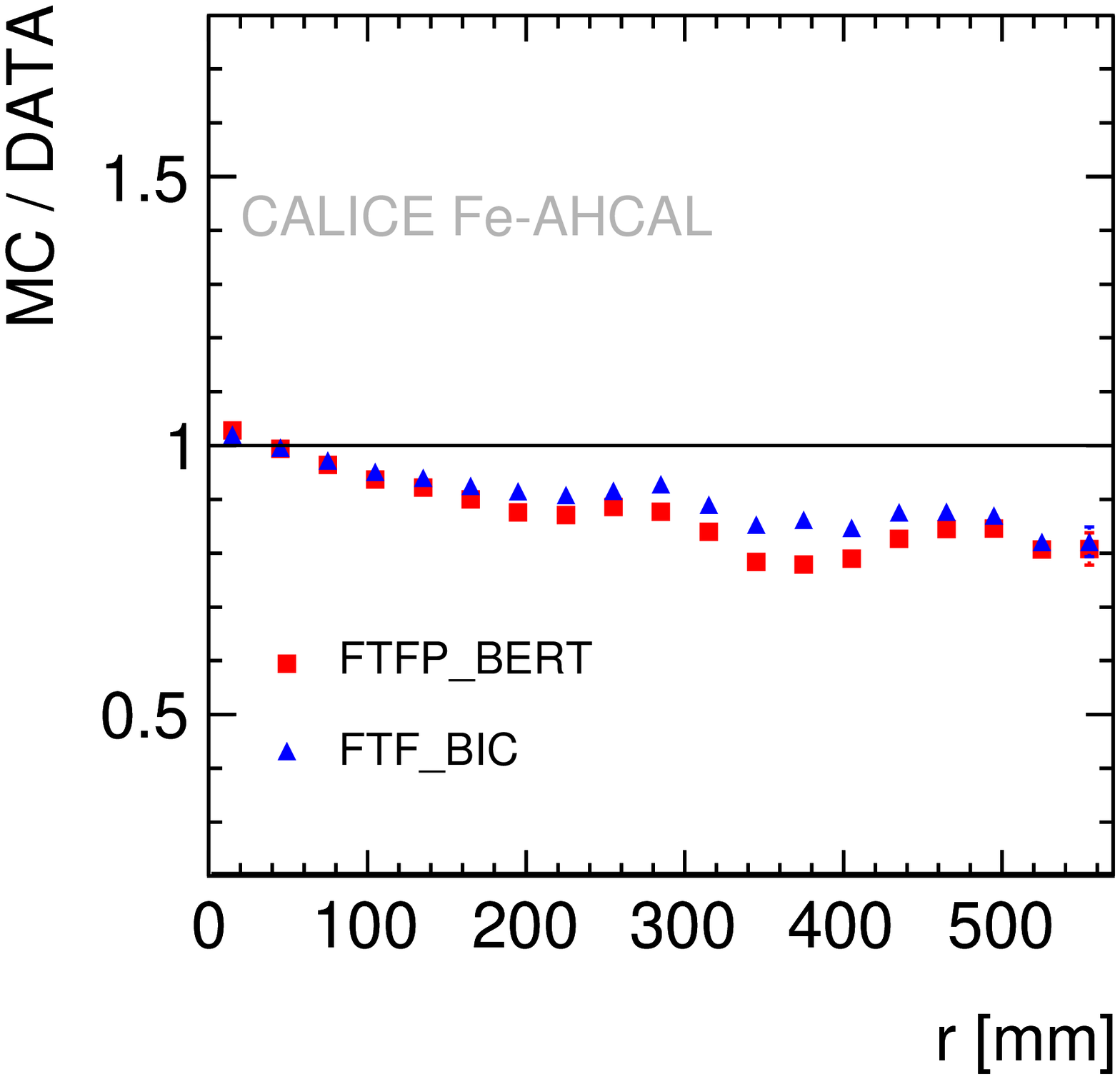}  \hspace{-5.mm}
\includegraphics[trim=80 0 10 0, clip, height=6cm]{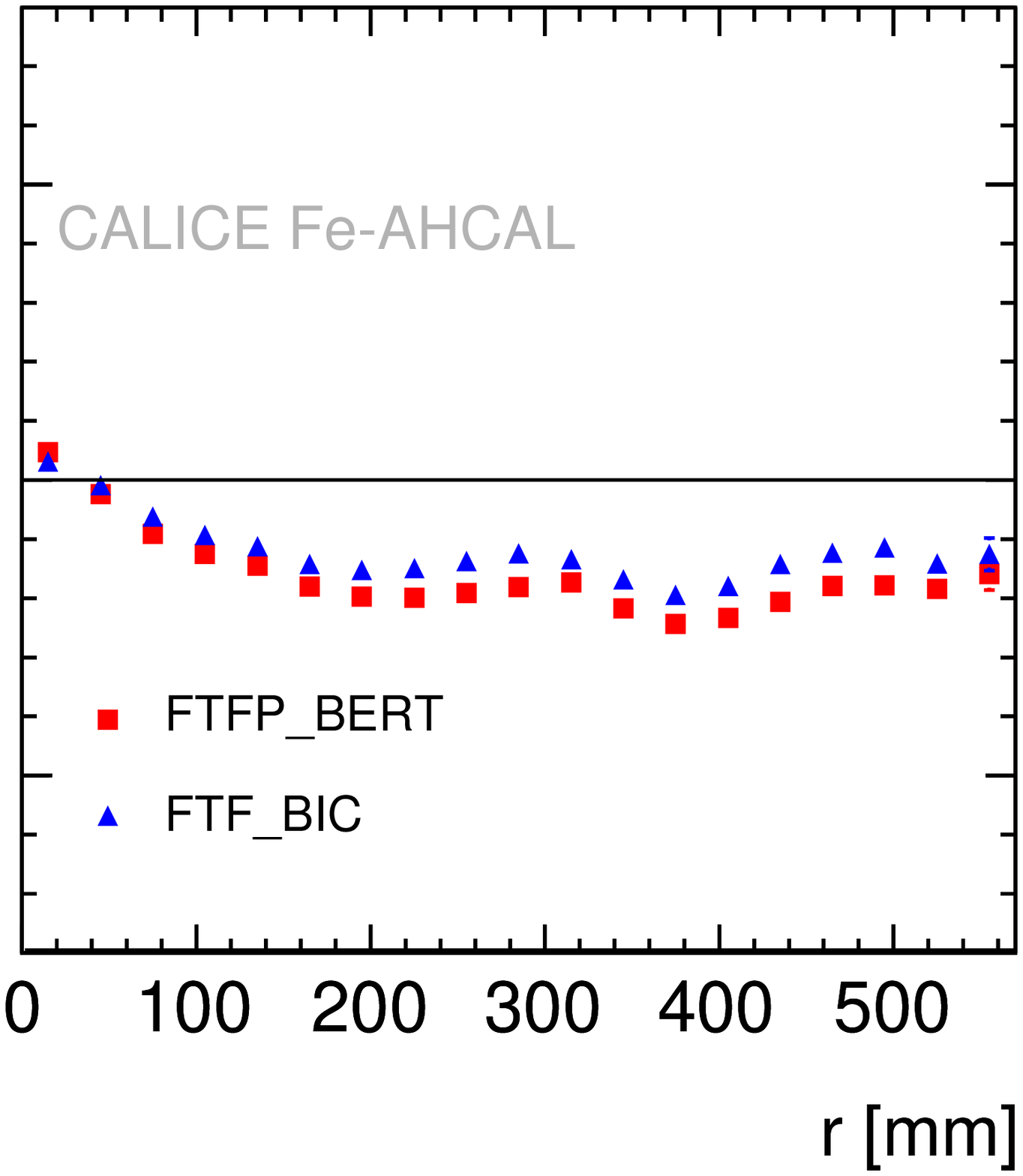}  \hspace{-5.3mm}
\includegraphics[trim=80 0 10 0, clip, height=6cm]{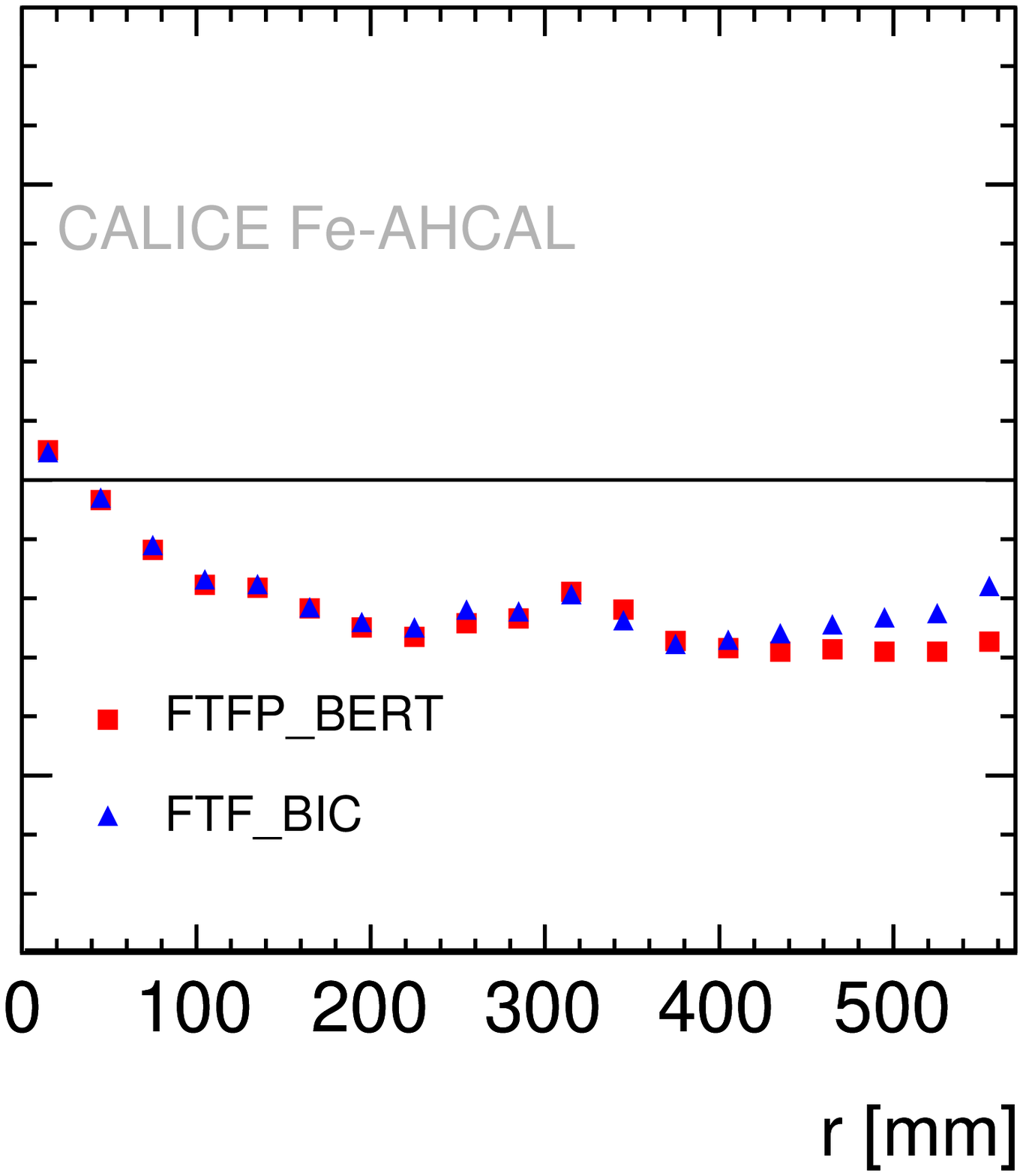}
\vspace{-10mm}
}
\centerline{
\hspace{-1.5mm}
\includegraphics[trim=0 0 10 0, clip, height=6cm]{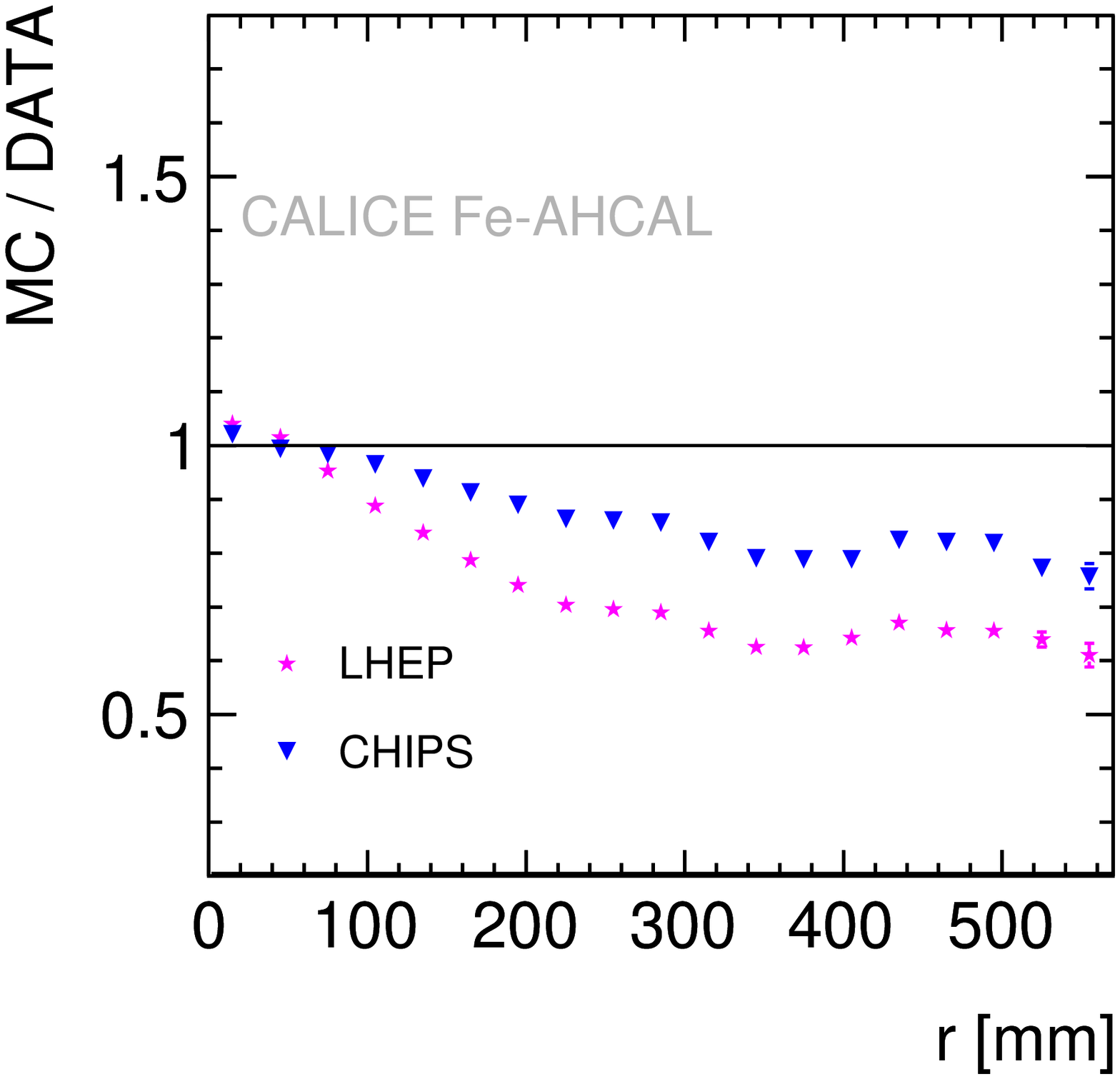}  \hspace{-5.mm}
\includegraphics[trim=80 0 10 0, clip, height=6cm]{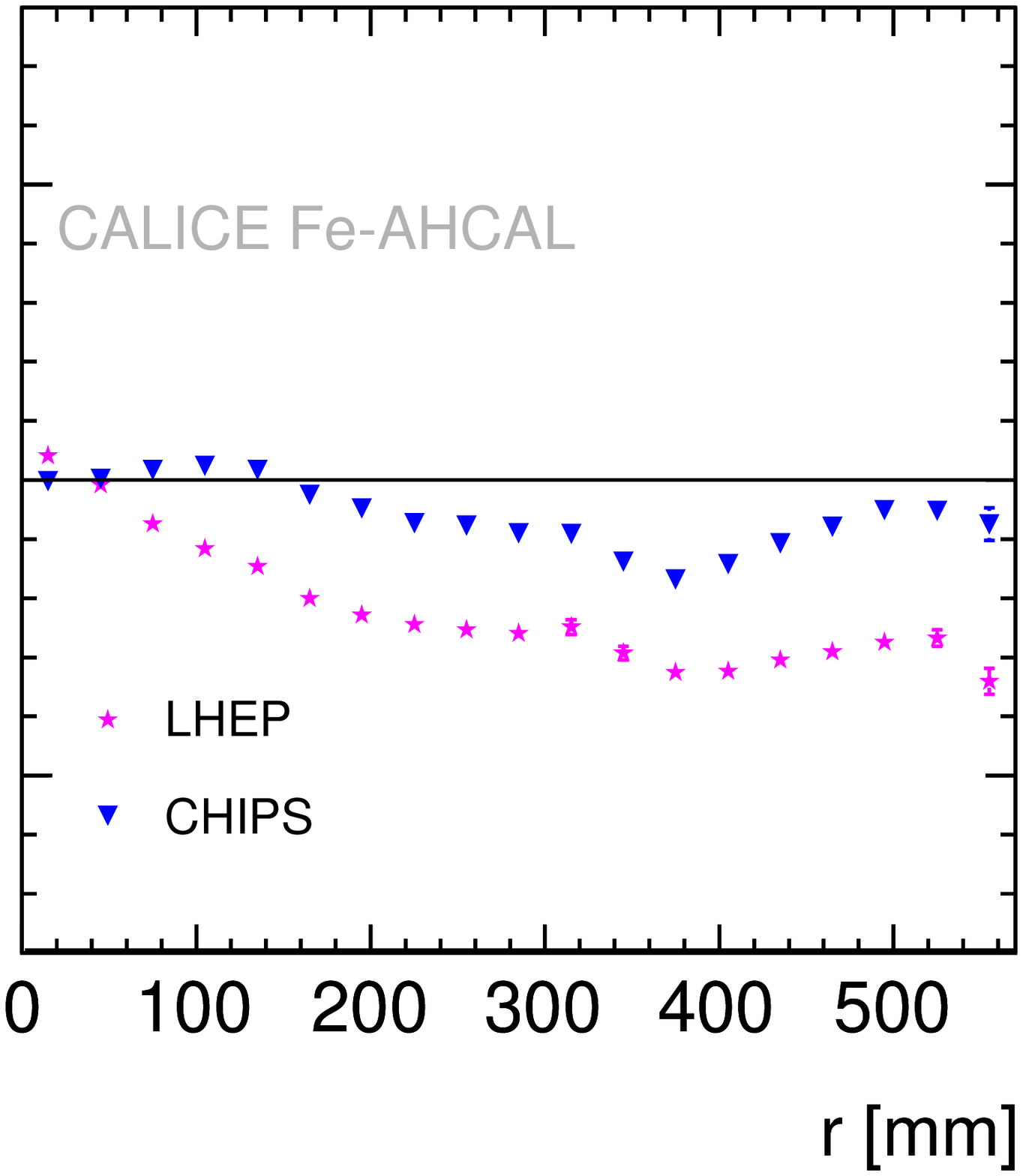}  \hspace{-5.3mm}
\includegraphics[trim=80 0 10 0, clip, height=6cm]{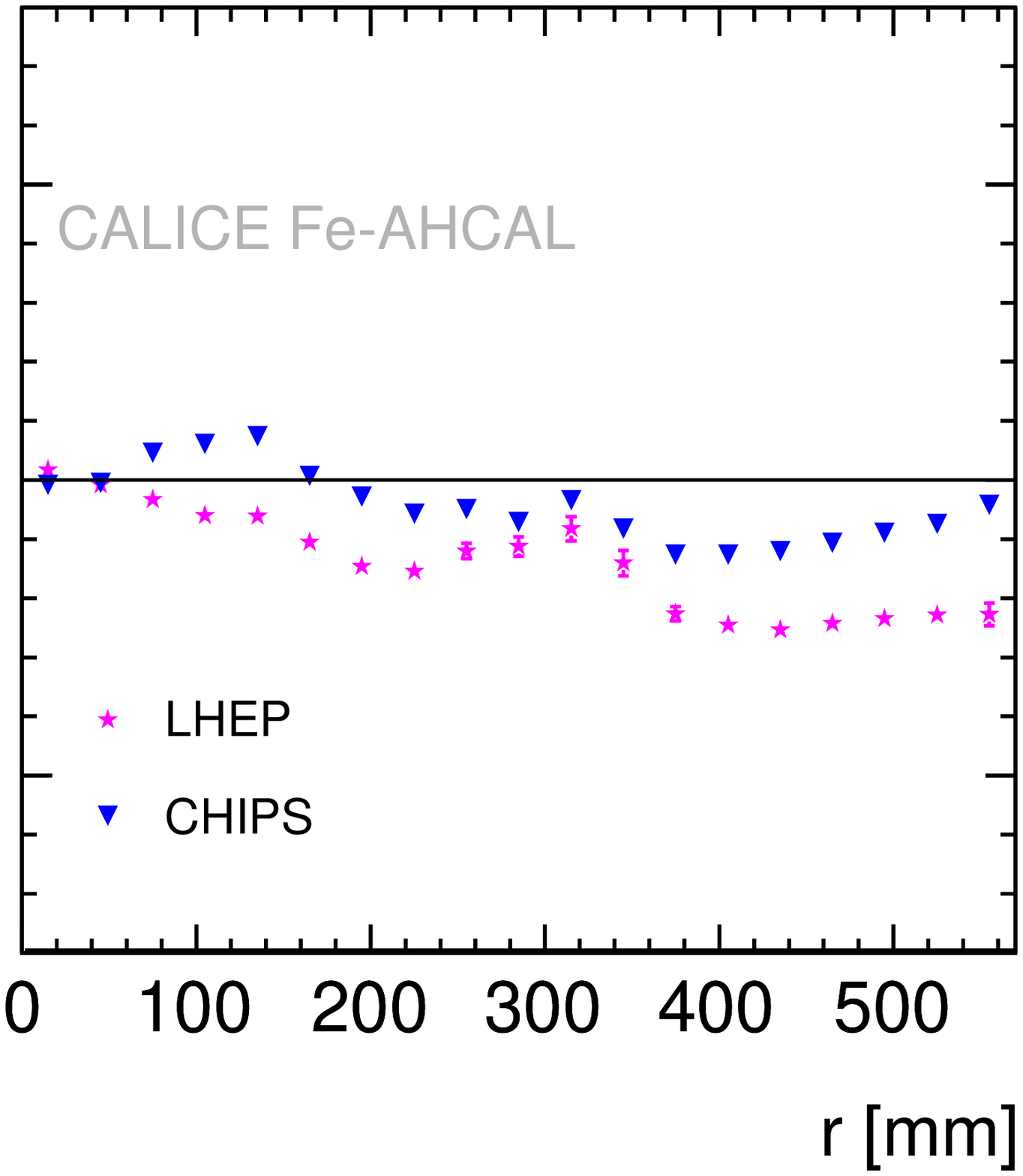} 
}
\caption[Radial shower profiles in data and Monte Carlo]{\sl \sl Mean radial shower profiles for 8\,GeV (left column), 18\,GeV (center column) and 80\,GeV (right column) pions. First row: For data (circles) and for the \texttt{FTFP\_BERT} physics list (histogram). Second to fourth rows: Ratio between Monte Carlo and data for several physics lists. All profiles are normalized to unity. $\langle \mathrm{E}_{\mathrm{rec}} \rangle / \Delta \mathrm{r}$ is the average deposited energy in $\Delta \mathrm{r}$, where $\mathrm{r}$ is the radial coordinate.}
\label{fig:rp}
\end{figure}

For this study, all physical AHCAL cells are subdivided into virtual cells of $1 \times 1\,\mathrm{cm}^{2}$~\cite{Lutz:2010zz}. The energy deposited in the physical cells is equally distributed over the virtual cells covering its area. The dimension of the smallest AHCAL tiles, i.e. $3 \times 3\,\mathrm{cm}^{2}$, is chosen as natural bin width for the radial shower profiles, which show the average energy deposited in the AHCAL as a function of the coordinate $r$. The radial shower profiles are shown in Fig.~\ref{fig:rp}, for the same set of beam energies and physics lists considered for the longitudinal profiles.

All physics lists show a similar behavior and tend to underestimate the radial extent of showers, showing a relatively higher energy deposition in the core of the showers. The disagreement is more pronounced at high energies. \verb=QGSP_BERT=, \verb=QGSP_FTFP_BERT= and \verb=QBBC= agree better with data at low energies, in particular \verb=QGSP_BERT=, which is consistent with data within 5-10\%. \verb=CHIPS= has the best behavior at 80\,GeV, describing the data at the 5-10\% level. The ratios between data and Monte Carlo generally show a discontinuity at about 300-350\,mm, corresponding to the transition between the tiles with a granularity of 6\,$\times$\,6\,cm$^2$ and the outer tiles with a granularity of 12\,$\times$\,12\,cm$^2$ in the active layers.

As expected, the contributions from electrons, pions and protons show that the electromagnetic component of the showers is concentrated in the core, while in the tails the energy is mostly deposited by protons.

\begin{figure}
\centerline{	
     \subfigure{ \label{} \includegraphics[trim=0 0 45 0, clip, height=8cm]{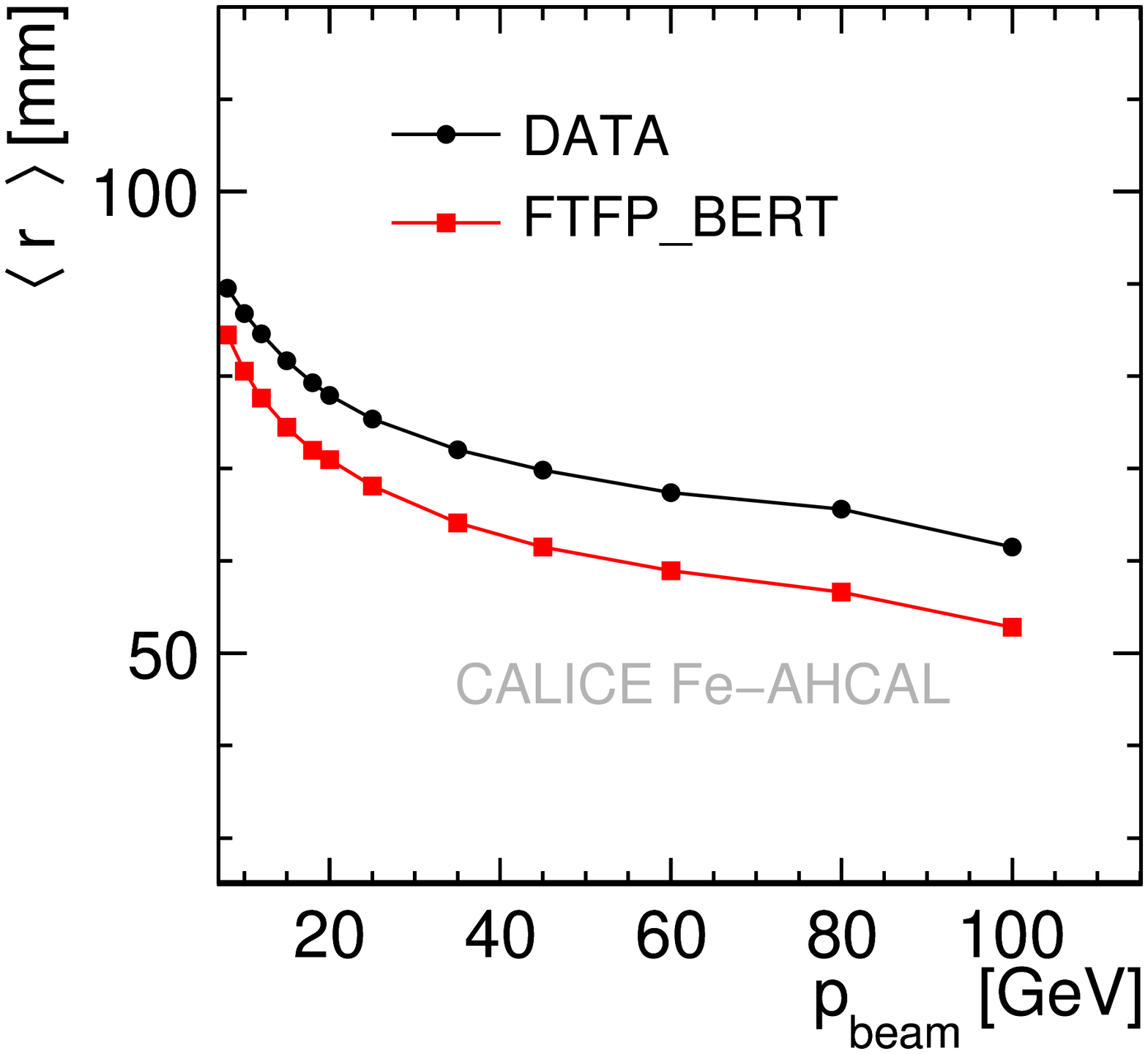} } \hspace{-4.5mm}
     \subfigure{ \label{} \includegraphics[trim=0 0 45 0, clip, height=8cm]{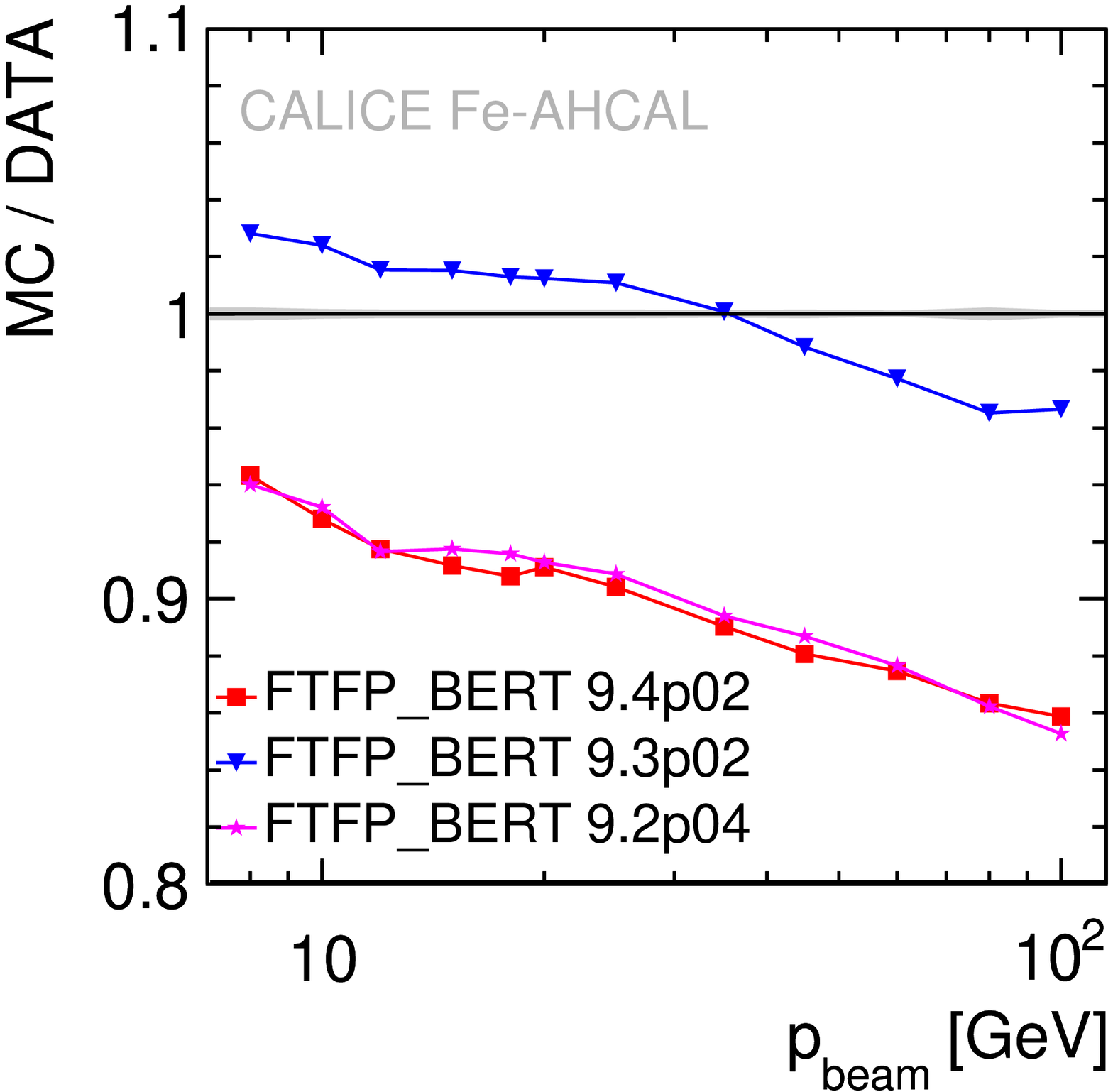} } \hspace{5mm}	
}
\centerline{
     \subfigure{ \label{} \includegraphics[trim=0 0 45 0, clip, height=6cm]{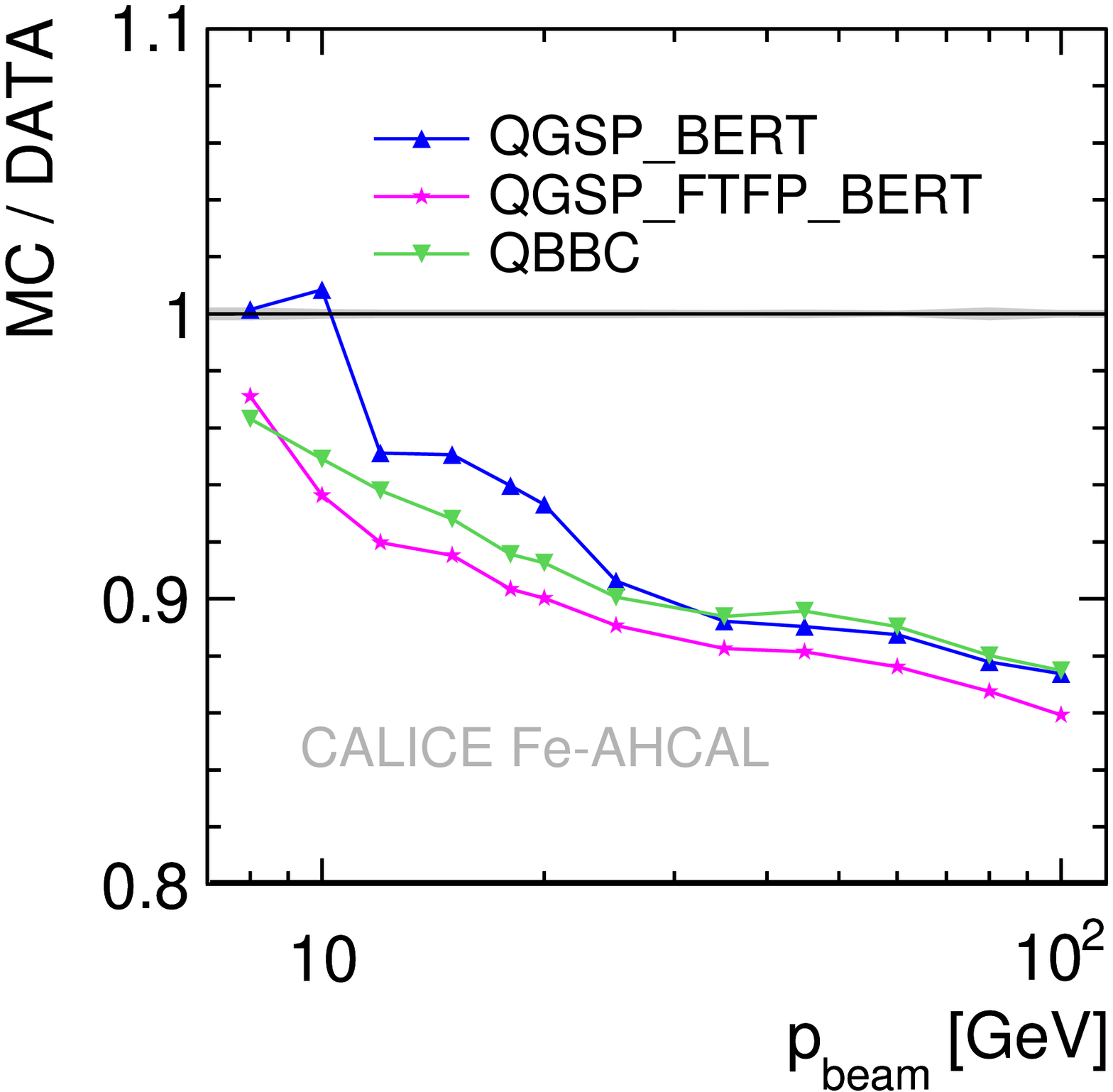} } \hspace{-4.5mm}
     \subfigure{ \label{} \includegraphics[trim=100 0 45 0, clip, height=6cm]{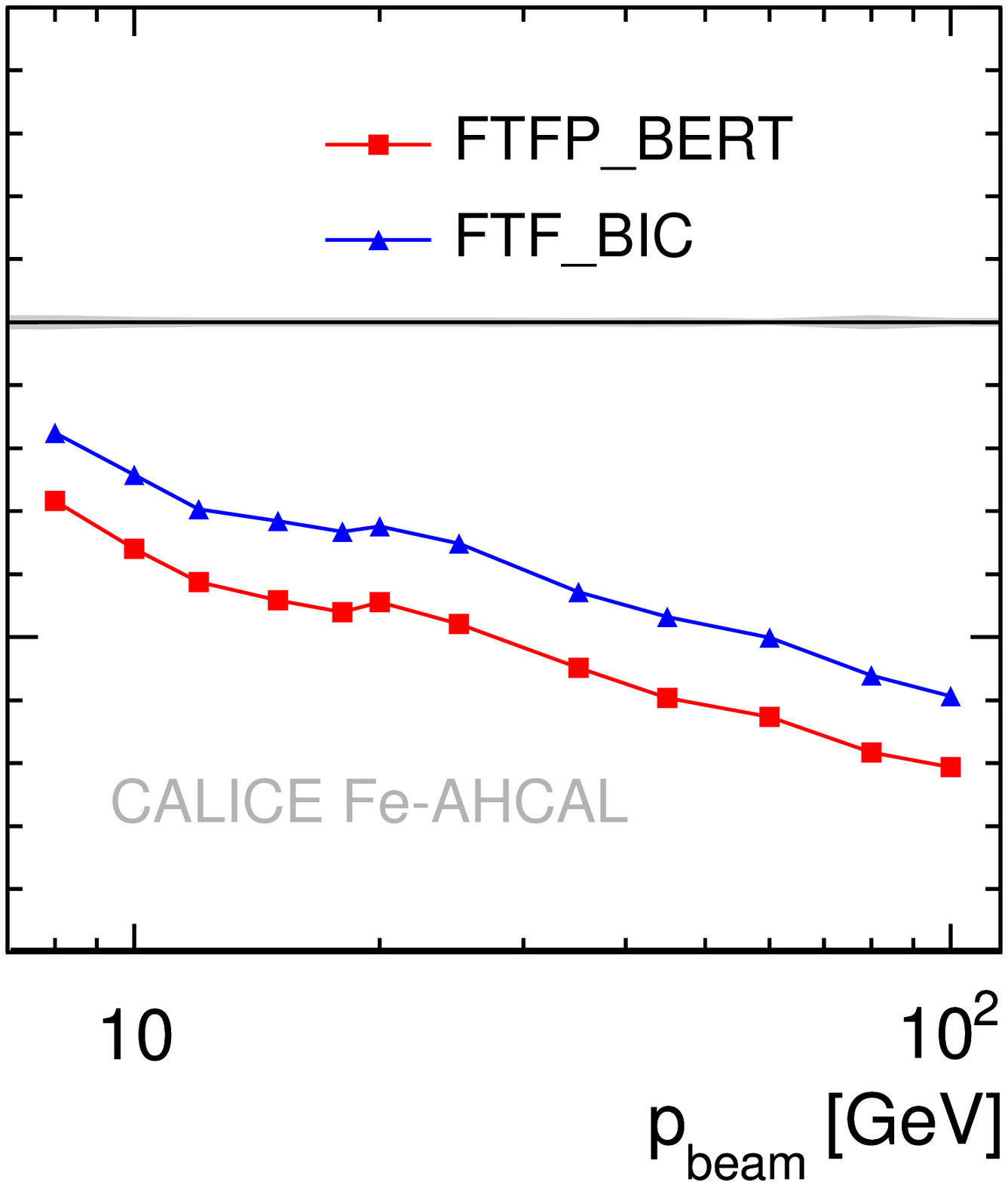} } \hspace{-4.5mm}
     \subfigure{ \label{} \includegraphics[trim=100 0 45 0, clip, height=6cm]{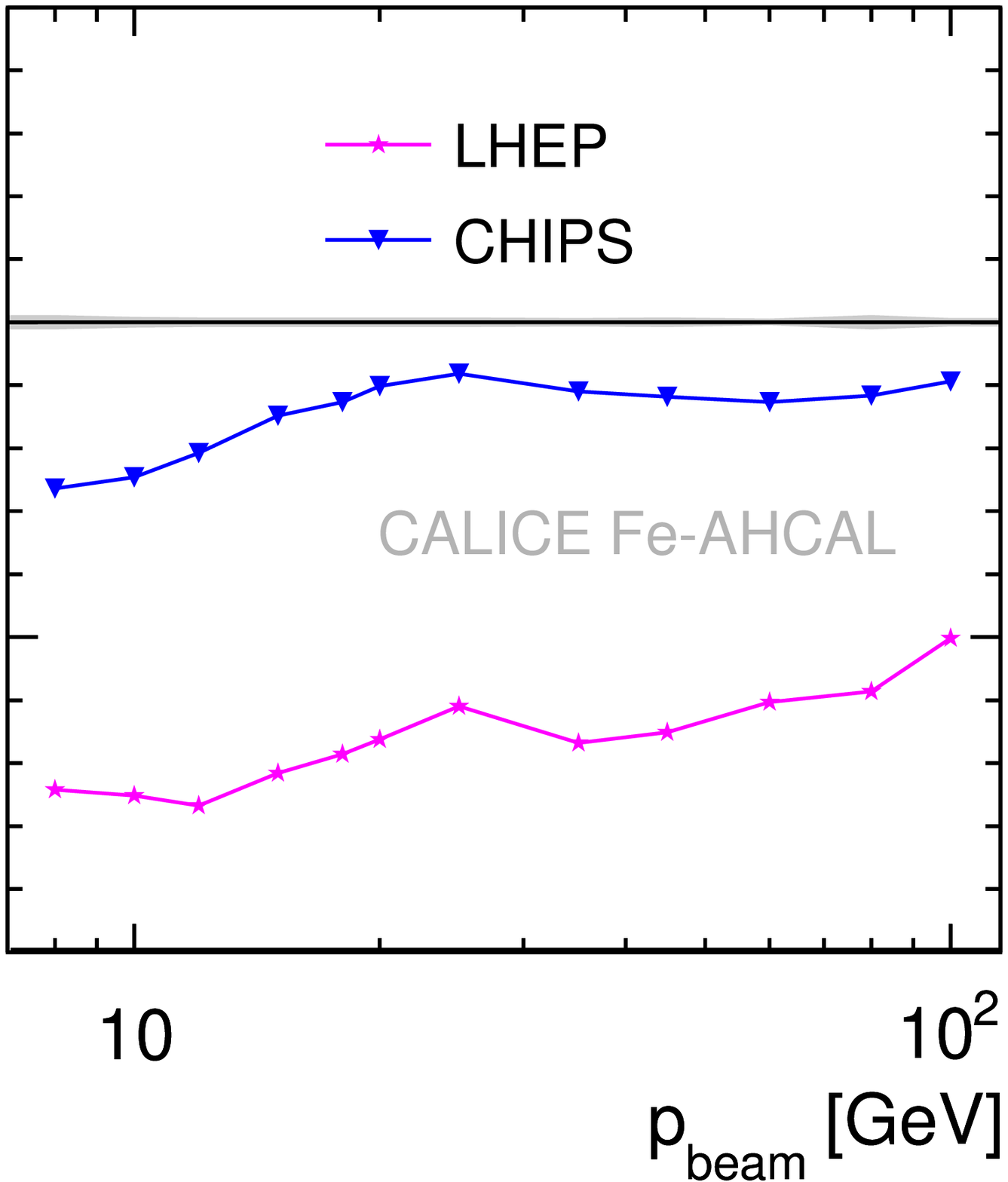} }
}
\caption[Center of gravity in the radial direction]{\sl Summary of the measurement of the average center of gravity $\langle \mathrm{r} \rangle$ in the radial direction, for pions in the AHCAL. Top, left: For data and for the \texttt{FTFP\_BERT} physics list. Top, right: Ratio between Monte Carlo and data using the \texttt{FTFP\_BERT} physics list with different versions of \textsc{Geant4}. Bottom: Ratio between Monte Carlo and data for several physics lists. The gray band in the ratios represents the statistical uncertainty on data.}
\label{fig:R}
\end{figure}

\begin{figure}
\centerline{	
     \subfigure{ \label{} \includegraphics[trim=0 0 45 0, clip, height=8cm]{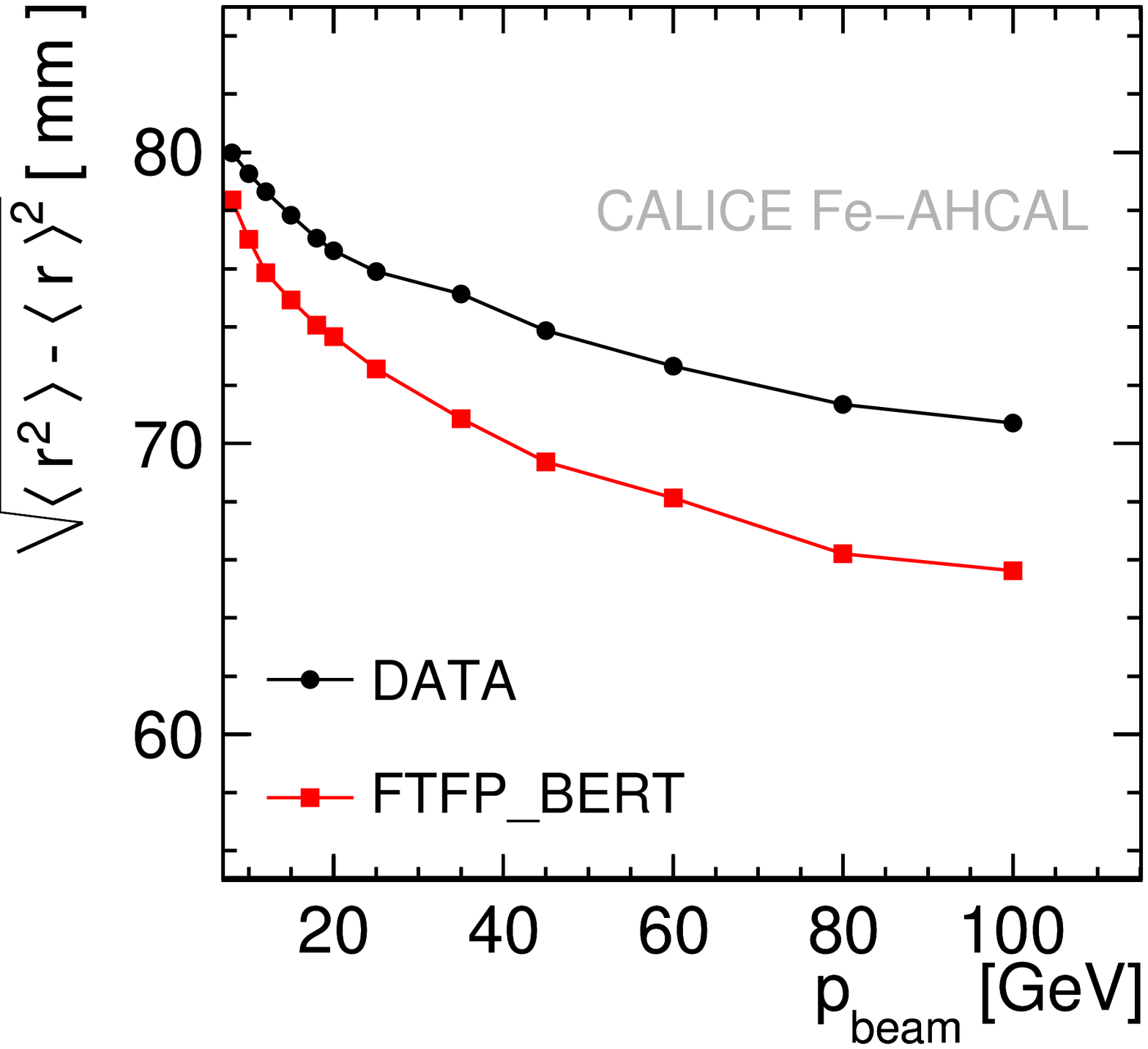} } \hspace{-4.5mm}
     \subfigure{ \label{} \includegraphics[trim=0 0 45 0, clip, height=8cm]{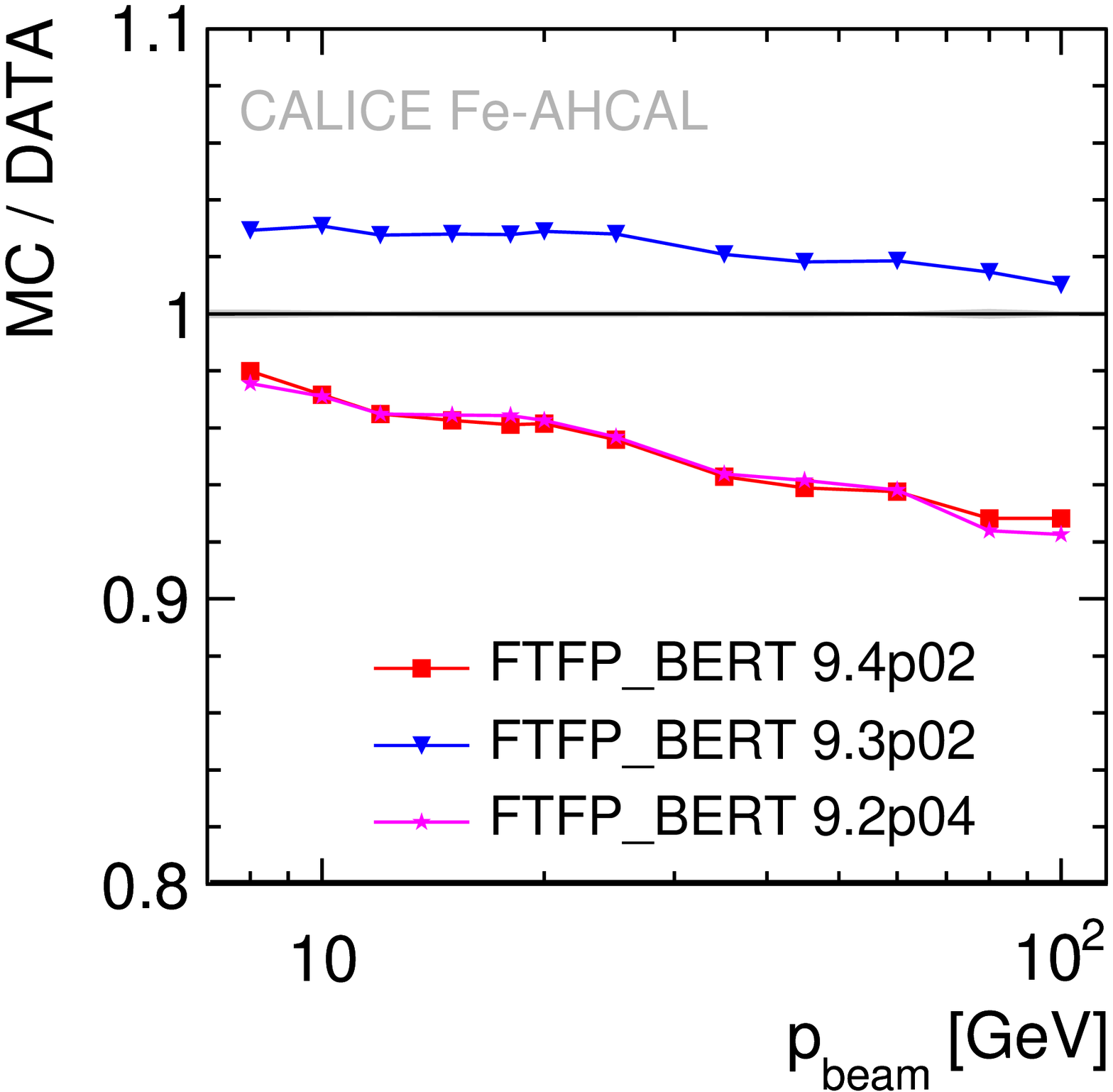} } \hspace{5mm}	
}
\centerline{
     \subfigure{ \label{} \includegraphics[trim=0 0 45 0, clip, height=6cm]{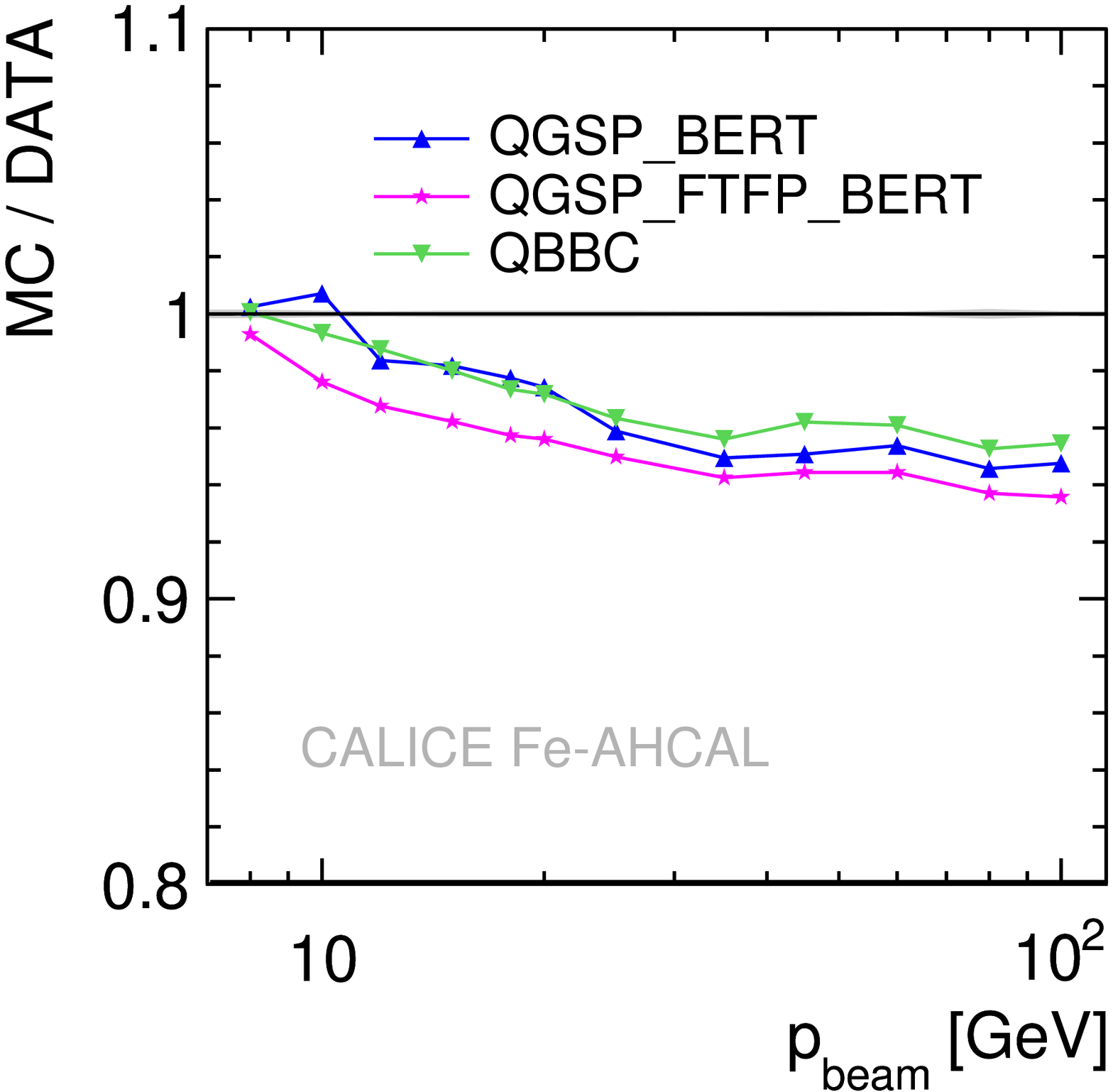} } \hspace{-4.5mm}
     \subfigure{ \label{} \includegraphics[trim=100 0 45 0, clip, height=6cm]{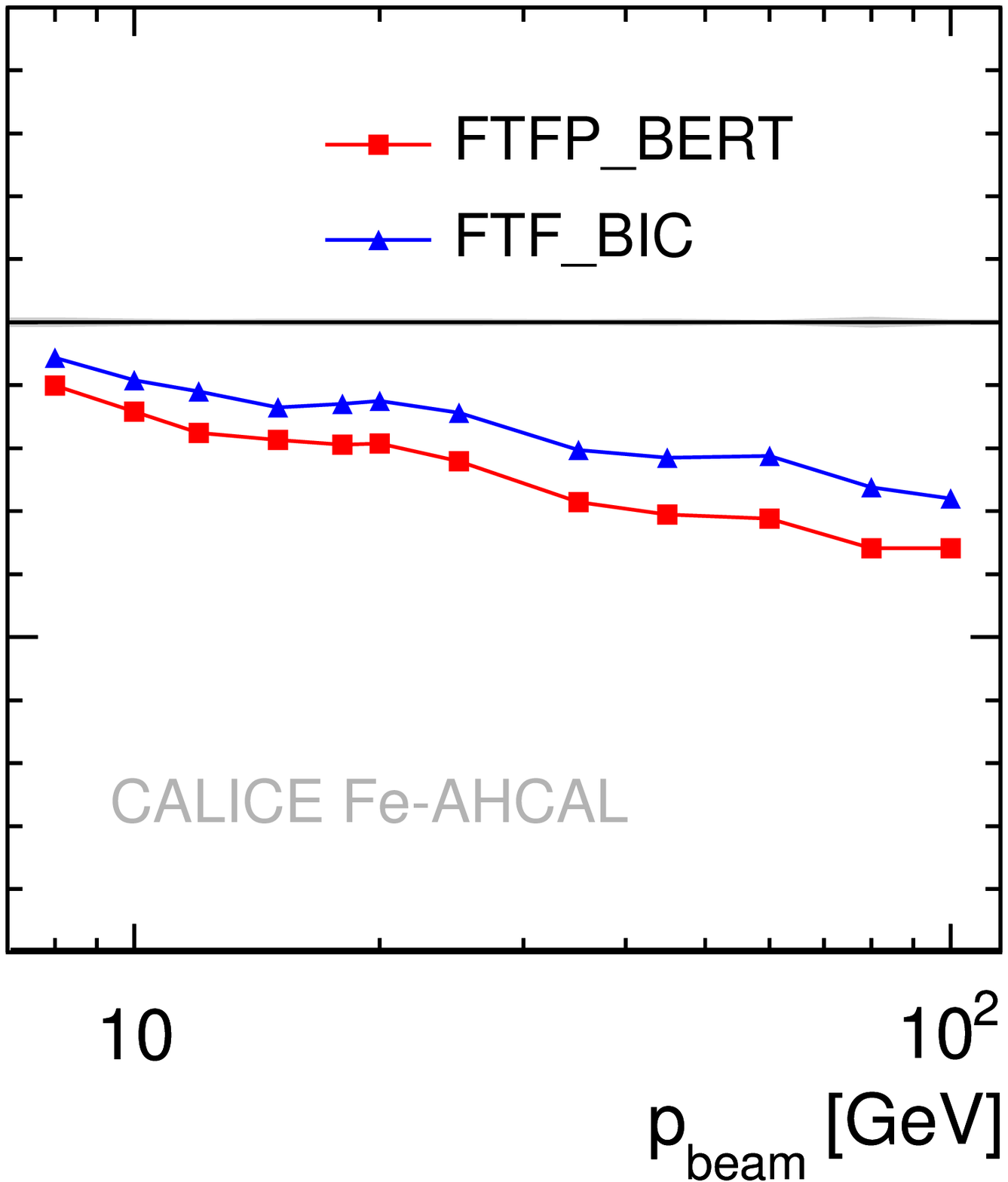} } \hspace{-4.5mm}
     \subfigure{ \label{} \includegraphics[trim=100 0 45 0, clip, height=6cm]{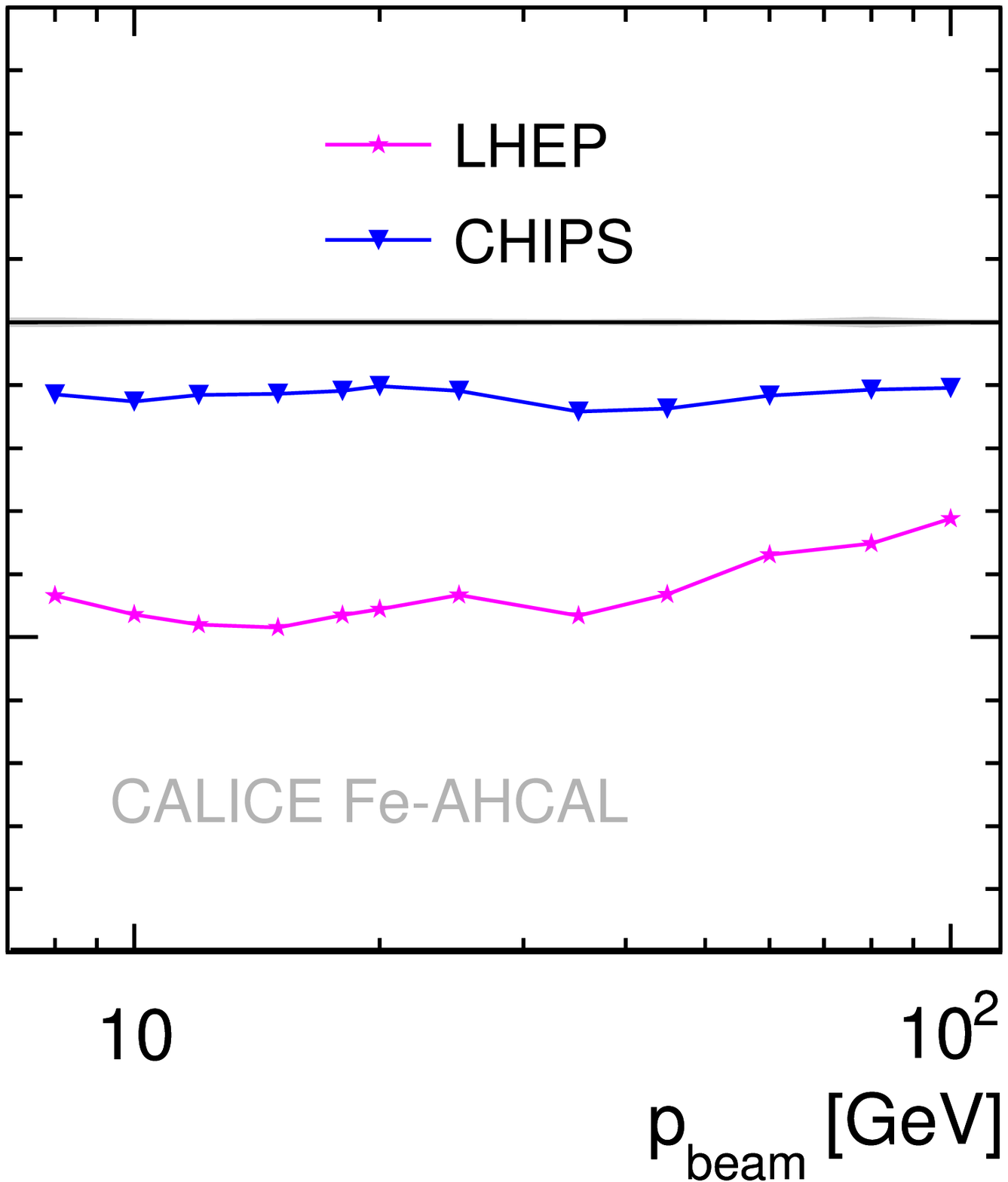} }
}
\caption[The standard deviation of the radial shower profile]{\sl Summary of the measurement of the average standard deviation of the radial shower profile $\sqrt{\langle r^2 \rangle - \langle r \rangle^2}$, for pions in the AHCAL. Top, left: For data and for the \texttt{FTFP\_BERT} physics list. Top, right: Ratio between Monte Carlo and data using the \texttt{FTFP\_BERT} physics list with different versions of \textsc{Geant4}. Bottom: Ratio between Monte Carlo and data for several physics lists. The gray band in the ratios represents the statistical uncertainty on data.}
\label{fig:R2}
\end{figure}

The mean energy-weighted shower radius and the standard deviation of the radial energy distribution are compared for all physics lists and all energies in Fig.~\ref{fig:R} and Fig.~\ref{fig:R2}, respectively. As for the longitudinal observables, the mean is defined as:
\begin{equation}
\langle r \rangle = \frac{\Sigma E_i \cdot r_i}{\Sigma E_i}, 
\end{equation}
and the standard deviation as:
\begin{equation}
\sqrt{\langle r^2 \rangle - \langle r \rangle^2},\;\;\;\langle r^2 \rangle = \frac{\Sigma E_i \cdot r_i^2}{\Sigma E_i},
\end{equation}
where $r_i$ is defined in Eq.~\ref{equation:radialco}.

The energy-weighted shower radius ranges between 90\,mm and 60\,mm and exhibits the expected exponential decrease with energy~\cite{Leroy:2000mj}. The standard deviation only mildly decreases with energy from about 80\,mm to about 70\,mm.

Monte Carlo models underestimate the radial extent of showers, both in terms of radius and of standard deviation. The majority of simulations exhibit an energy-dependent behavior and the disagreement with data increases with energy. \verb=QGSP_BERT= reproduces the mean radius of showers at 8-10\,GeV, but disagrees with data by up to about 12\% at high energies. Similar results are achieved by \verb=QGSP_FTFP_BERT= and \verb=QBBC=, though they give a worse agreement with data at low energies. \verb=FTF_BIC= underestimates the mean shower radius by about 3\% at 8\,GeV and by up to $\sim$12\% at high energies. A similar behaviour is shown by \texttt{FTFP\_BERT}, though the disagreement between data and Monte Carlo is about 2\% worse at all energies. Above 10 GeV, the best results are achieved by \verb=CHIPS=, which shows the least energy dependence and underestimates the mean radial shower expansion by 2-4\%. At lower energies the performance degrades and the disagreement increases by up to 5\% at 8\,GeV. \verb=LHEP= shows a dramatic disagreement with data at all energies.

Similar trends are observed for the standard deviation of the radial energy distributions, though the agreement with data improves for all physics lists with respect to the mean radius of showers (Fig.~\ref{fig:R2}). 

The study of the evolution of \verb=FTFP_BERT= with different  \textsc{Geant4} versions (Fig.~\ref{fig:R} and~\ref{fig:R2}) shows that in the version 9.4 the simulation of the radial shower development went back to the performance achieved for the version 9.2. The intermediate version 9.3 is the closest to the data. This conclusion concerns both the observables considered.

It should be underlined that the residual discrepancies in radial profiles observed in electromagnetic showers~\cite{collaboration:2010rq}, discussed above, do not prevent conclusions from being drawn concerning these radial observables and the results shown remain an important input to the process of validation of Monte Carlo models. The absolute uncertainties in the description of electromagnetic radial profiles are of the order of 2\,mm, while the deviations observed in hadronic showers are of the order of 10\,mm. 

\section{Conclusions}
\label{section:conclusions}

The response of the CALICE analog hadron calorimeter to pions is measured for energies between 8\,GeV and 100\,GeV, using data collected at CERN in 2007. The high sampling frequency together with the fine segmentation of the sensitive layers of this detector allows the investigation of the properties of hadronic showers and the validation of Monte Carlo models with an unprecedented spacial accuracy.

The paper covers the measurement of several properties of hadronic showers, such as the global energy response and the longitudinal and radial development of showers. The physics lists based on the Fritiof model yield overall the best simulation of the energy response, with a particularly good agreement with data at energies between 10\,GeV and 30\,GeV. A good simulation of the energy response at low energies is achieved also by the physics lists based on the quark-gluon string model, such as \verb=QGSP_BERT= and \verb=QGSP_FTFP_BERT=, which agree with data at the 2-4\% level. At high energies these physics lists overestimate the energy response by up to 10\%. A similar trend is exhibited by the \verb=CHIPS= physics list, while the \verb=LHEP= physics list underestimates the deposited energy at all energies by up to 10\%. This list has generally the worst performance also with respect to the other observables considered.

The best description of the longitudinal development of showers is achieved by the physics lists based on the Fritiof model and by \verb=CHIPS=. The physics lists based on the quark-gluon string model agree with data at about the 6-8\% level.

Previous publications show limitations in the understanding of the radial development of electromagnetic showers~\cite{collaboration:2010rq}. However, the results published in~\cite{:2011ha} show that despite these limitations the ability to model the hadronic shower separation is not degraded. Moreover, the discrepancies at the electromagnetic level are one order of magnitude smaller than the average radial extension of hadronic showers and do not prevent conclusions from being drawn concerning radial observables of hadronic showers. Hadronic data indicate broader showers than expected from simulation. The best performing list is \verb=CHIPS=, which shows an almost energy-independent behavior and underestimates the shower mean radial expansion by less than about 2-5\% at all energies. All the other physics lists with the exception of \verb=LHEP= exhibit a disagreement with data increasing at high energies. The \verb=LHEP= physics list shows a moderate energy-dependence, but largely disagrees with data by up to 15\%. More detailed studies are needed, in order to improve the description of radial shower profiles. 

\verb=FTFP_BERT= and \verb=CHIPS= are overall the physics lists that best agree with the observables presented. \verb=CHIPS= is a very recent list, which is still under development. \verb=FTFP_BERT= is an older list and the presented time evolution reflects changes in both the Bertini and the Fritiof models and their combination. Most observables fluctuate by $\pm$2-6\% depending on the \textsc{Geant4} version considered and remain in acceptable agreement with data. The considered radial observables present more significant changes of the order of 10\% and show a better agreement with data for an older version of the physics list.

In the last decade the LHC experiments have also performed comparison studies of their test-beam data to \textsc{Geant4} v9.3.p01 models. The beam energies available in the LHC beam tests were either below 9\,GeV or above 20\,GeV. They concluded in~\cite{Adragna:2009zz,Abat:2009zz} that the physics list \verb=QGSP_BERT= was the closest to their pion test-beam data. The agreement was within 2-3\%, with \verb=QGSP_BERT= response higher in data than in simulations. During the ATLAS test beam the lateral spread of pion showers has also been quantified. Simulated showers were found significantly narrower than in data, with a disagreement of about 15\% for \verb=QGSP_BERT=~\cite{Adragna:2010zz}. Since these tests several improvements have been implemented in the simulation. The results presented in this paper have extended the comparison to more recent versions of \textsc{Geant4} over a similar energy range.

\section*{Acknowledgements}
We gratefully acknowledge the DESY and CERN managements for their support and hospitality, and their accelerator staff for the reliable and efficient beam operation. 
We would like to thank the HEP group of the University of Tsukuba for the loan of drift chambers for the DESY test beam. The authors would like to thank the RIMST (Zelenograd) group for their help and sensors manufacturing. This work was supported by the Bundesministerium f\"{u}r Bildung und Forschung, Germany; by the  the DFG cluster of excellence `Origin and Structure of the Universe' of Germany;  by the Helmholtz-Nachwuchsgruppen grant VH-NG-206; by the BMBF, grant no. 05HS6VH1; by the Alexander von Humboldt Foundation (Research Award IV, RUS1066839 GSA); by joint Helmholtz Foundation and RFBR grant HRJRG-002, SC Rosatom; by the Russian Ministry of Education and Science via grants 8174, 8411, 1366.2012.2, P220; by MICINN and CPAN, Spain; by CRI(MST) of MOST/KOSEF in Korea; by the US Department of Energy and the US National Science Foundation; by the Ministry of Education, Youth and Sports of the Czech Republic under the projects AV0 Z3407391, AV0 Z10100502, LC527  and LA09042 and by the Grant Agency of the Czech Republic under the project 202/05/0653; by the National Sciences and Engineering Research Council of Canada; and by the Science and Technology Facilities Council, UK.

\clearpage


\clearpage
\begin{footnotesize}

\bibliographystyle{ieeetr}\bibliography{main}{}

\end{footnotesize}


\end{document}